\documentclass[useAMS,usenatbib]{mn2e}
\usepackage{mn2ejour}

\usepackage{graphicx}

\DeclareSymbolFont{UPM}{U}{eur}{m}{n}
\DeclareMathSymbol{\uppartial}{0}{UPM}{"40}

\title[Trapping of giant-planet cores]{
Trapping of giant-planet cores - I. Vortex aided trapping at the outer dead zone edge
}

\author[Zs. Reg\'aly, Zs. S\'andor, P. Csom\'os, S. Ataiee]{Zs. Reg\'aly$^{1,2}$\thanks{E-mail:regaly@konkoly.hu}, Zs. S\'andor$^3$, P. Csom\'os$^4$, and S. Ataiee$^5$\\
$^{1}$Konkoly Observatory, Research Centre for Astronomy and Earth Sciences, Hungarian Academy of Sciences, \\\,\,\,\,\,PO Box 67, H-1525 Budapest, Hungary; regaly@konkoly.hu\\
$^2$ELTE Gothard-Lend\"ulet Research Group, H-9704 Szombathely, Szent Imre Herceg u. 112., Hungary\\
$^3$Department of Astrophysics, University of Vienna, A-1180 Vienna, Austria\\
$^4$Department of Mathematics, University of Innsbruck, 6020 Innsbruck, Austria\\
$^5$Institut f\"ur Theoretische Astrophysik, Universit\"at Heidelberg, 69120 Heidelberg, Germany}

\begin{document}

\date{Accepted 2013 May 24.  Received 2013 May 21; in original form 2013 April 18}

\pagerange{\pageref{firstpage}--\pageref{lastpage}} \pubyear{2012}
 
\maketitle
\label{firstpage}

\begin{abstract}
In this paper the migration of a $10\mathrm{M_\oplus}$ planetary core is investigated at the outer boundary of the `dead zone' of a protoplanetary disc by means of 2D hydrodynamic simulations done with the graphics processor unit version of the {\small FARGO} code. In the dead zone, the effective viscosity is greatly reduced due to the disc self-shielding against stellar UV radiation, X-rays from the stellar magnetosphere and interstellar cosmic rays. As a consequence, mass accumulation occurs near the outer dead zone edge, which is assumed to trap planetary cores enhancing the efficiency of the core-accretion scenario to form giant planets. Contrary to the perfect trapping of planetary cores in 1D models, our 2D numerical simulations show that the trapping effect is greatly dependent on the width of the region where viscosity reduction is taking place. Planet trapping happens exclusively if the viscosity reduction is sharp enough to allow the development of large-scale vortices due to the Rossby wave instability. The trapping is only temporarily, and its duration is inversely proportional to the width of the viscosity transition. However, if the Rossby wave instability is not excited, a ring-like axisymmetric density jump forms, which cannot trap the $\mathrm{10M_\oplus}$ planetary cores. We revealed that the stellar torque exerted on the planet plays an important role in the migration history as the barycentre of the system significantly shifts away from the star due to highly non-axisymmetric density distribution of the disc. Our results still support the idea of planet formation at density/pressure maximum, since the migration of cores is considerably slowed down enabling them further growth and runaway gas accretion in the vicinity of an overdense region. 

\end{abstract}


\section{Introduction}

It is well known that planetary objects, up to $20\mathrm{M_\oplus}$, embedded in accretion discs are subject to Type I migration, which is linearly proportional to their mass \citep{Ward1997}. The typical bodies undergoing Type I migration are the planetary cores, which play an important role in the core-accretion scenario of planet formation. In this scenario, first a solid core with mass between $5\textrm{--}15\mathrm{M_\oplus}$ is formed, which later on evolves to a giant planet by accreting a massive gaseous envelope. Type I migration of these bodies happens on such a short time-scale \citep{Tanakaetal2002} that they will inevitably be engulfed by the host star within the disc lifetime \citep{Haischetal2001,Sicilia-Aguilaretal2006}, or before growth to the critical mass ($\sim10\mathrm{M_\oplus}$) that is enough to accrete a massive atmosphere and form a giant planet \citep{Pollacketal1996}. Type I migration has therefore been thought to be quite problematic in explaining planet formation. In order to simulate the formation of giant planets, an artificial slow-down of Type I migration (by a factor of 100) was necessary in the planet synthesis models \citep{Alibertetal2005,Mordasinietal2009}. 

In recent years, however, the understanding of Type I migration has considerably changed. In several investigations that presented planet--disc interaction models with the inclusion of radiation transport, there has been found slowed-down or even reversed Type I migration; see \cite{KleyNelson2012} and references therein. However, fast migration (whether inwards or outwards) is still a danger to the core of a Jupiter-like planet. To overcome this issue, the concept of a \emph{planet trap} has been introduced. A planet trap can appear at density enhancements that could be formed in discs. These locations are thought to trap low-mass planets \citep{Masset2002,Massetetal2006b,Morbidellietal2008}. Another type of planet traps has been suggested by \citet{Lyraetal2010}, which appears when the thermal properties of a disc are changing suddenly, e.g. a change of the dust opacity may result in a transition from the adiabatic to the isothermal regime.

An overdense region in the protoplanetary disc can be formed if the gas accretion rate is reduced significantly. The gas accretion is driven by an effective viscosity, for which the main source is the magneto-rotational instability (MRI; \citealp{BalbusHawley1991}. Deep in the disc the ionization of the gas is presumably very low due to gas self-shielding against ionization radiation, therefore the MRI cannot be triggered there. As a result, there should exist a region, the \emph{dead zone}, in which the MRI-triggered gas accretion is practically absent. The existence of the dead zone, in which the gas accretion takes place only in an ionized upper layer of a protoplanetary disc, proposed originally by \citet{Gammie1996}, and later on by \citet{Glassgoldetal1997} and \citet{IgeaGlassgold1999}, is widely accepted in the literature. Considering various ionization processes for the MRI, \citet{TurnerDrake2009} have found that the big majority of discs certainly contain a dead zone. We note that inside the dead zone there might be some remnant viscosity. Hydrodynamic waves excited by the turbulence in the viscous surface layer can penetrate into the midplane causing non-vanishing Reynolds stress \citep{FlemingStone2003,BaiGoodman2009,OishiMacLow2009}. Turbulent mixing of ionized gas downwards to the mid-plane causes the non-vanishing ionization level that may result in suppressed MRI viscosity \citep{IlgnerNelson2008,TurnerSano2008}. Since shears can generate azimuthal magnetic field if the radial magnetic field penetrates sufficiently to the midplane, laminar Maxwell stress can be generated, which can transport angular momentum without turbulent motions \citep{Latteretal2010}.

The disc material is accumulated at the outer edge of the dead zone if the gas accretion is reduced there, forming a density maximum demonstrated by \citet{VarniereTagger2006} and \citet{Terquem2008}. These density enhancements are found to be unstable to the Rossby wave instability (RWI), which can result in the formation of anticyclonic vortices \citep{Lovelaceetal1999,Lietal2000}. RWI is excited if the width of the viscosity reduction at the dead zone edge is smaller than twice of the local disc pressure scale height \citep{Lyraetal2009}. For RWI unstable discs, the vortices are subject to a merging process resulting in the formation of one large-scale, long-lived vortex \citep{Regalyetal2012}.

The formation of vortices in protoplanetary discs has been investigated thoroughly in recent years \citep{LyraMacLow2012,Meheutetal2012a,Meheutetal2013}. Large-scale vortices can accelerate the formation of planetesimals and planetary embryos in the core-accretion paradigm \citep{Braccoetal1999}. \citet{BargeSommeria1995} and \citet{KlahrHenning1997} have already shown that particles tend to get trapped in anticyclonic vortices. \citet{Lyraetal2009} have demonstrated that particles of about centimetre to metre in size subsequently drift into the pressure maximum, i.e., to the vortices and form gravitationally bound clumps of solids, which can coalesce forming embryos between the masses of the Moon and Mars. The swarm of these embryos may evolve further by mutual {\it N-body} collisions forming massive cores ($\sim10\mathrm{M_{\oplus}}$) of giant planets in less than $5\times10^5\,\rm yr$ \citep{Sandoretal2011}. Recent 3D hydrodynamic simulations of \citet{Meheutetal2012b} clearly demonstrated the dust collecting property of the RWI-triggered vortices. It has been found that the dust-to-gas ratio has reached unity after a very short time.

According to the above results, planet formation at a density maximum (which serves as a pressure maximum too) is very favourable. First of all, coagulating dust particles can easily reach sizes of a centimetre to a metre inside the density enhancement. In the second place, apart from solving the infamous metre-sized barrier problem \citep{Braueretal2008}, the larger bodies might be trapped there, and the core of a giant planet can be saved from fast inward migration. However, laboratory experiments \citep{Guettleretal2010} and detailed Monte Carlo simulations \citep{Zsometal2010} show that the maximum particle size at 1\,au in the minimum mass solar nebula model might be limited to a millimetre due to bouncing collisions [however, see \citet{Windmarketal2012} for a possible solution to this bouncing barrier].

In this paper, we investigate what happens to a planetary core which has been formed at a density maximum. In the present study, we follow the orbital evolution of a $10\mathrm{M_\oplus}$ planetary object in a locally isothermal global 2D disc model that has only outer dead zone edge. Incorporating in our model the viscous evolution of gas at the density maximum (formed near the dead zone edge), we show that a planetary core grown to the critical size of $10\mathrm{M_\oplus}$ gets trapped at the density maximum only in certain conditions, i.e. for which case a large-scale vortex is developed.

The paper is structured as follows. In Sect. 1, numerical investigation of the evolution of pressure maximum formed and trapping of a $10\mathrm{M_\oplus}$ planetary core at the outer dead zone edge is presented by means of 1D simulations. A series of 2D hydrodynamical simulations are presented in Sect. 2, demonstrating the migration and trapping efficiency of a $10\mathrm{M_\oplus}$ planetary core in density maximum and large-scale vortices formed at the outer dead zone edge. In Section 3, we discuss our findings, while in Sect. 4 we present our conclusions to the giant-planet formation theories which might be affected by the efficiency of planetary trapping at dead zone edges.

\section{Planetary migration in the 1D model}

\subsection{1D hydrodynamical model}

\begin{figure}
	\centering
	\includegraphics[width=\columnwidth]{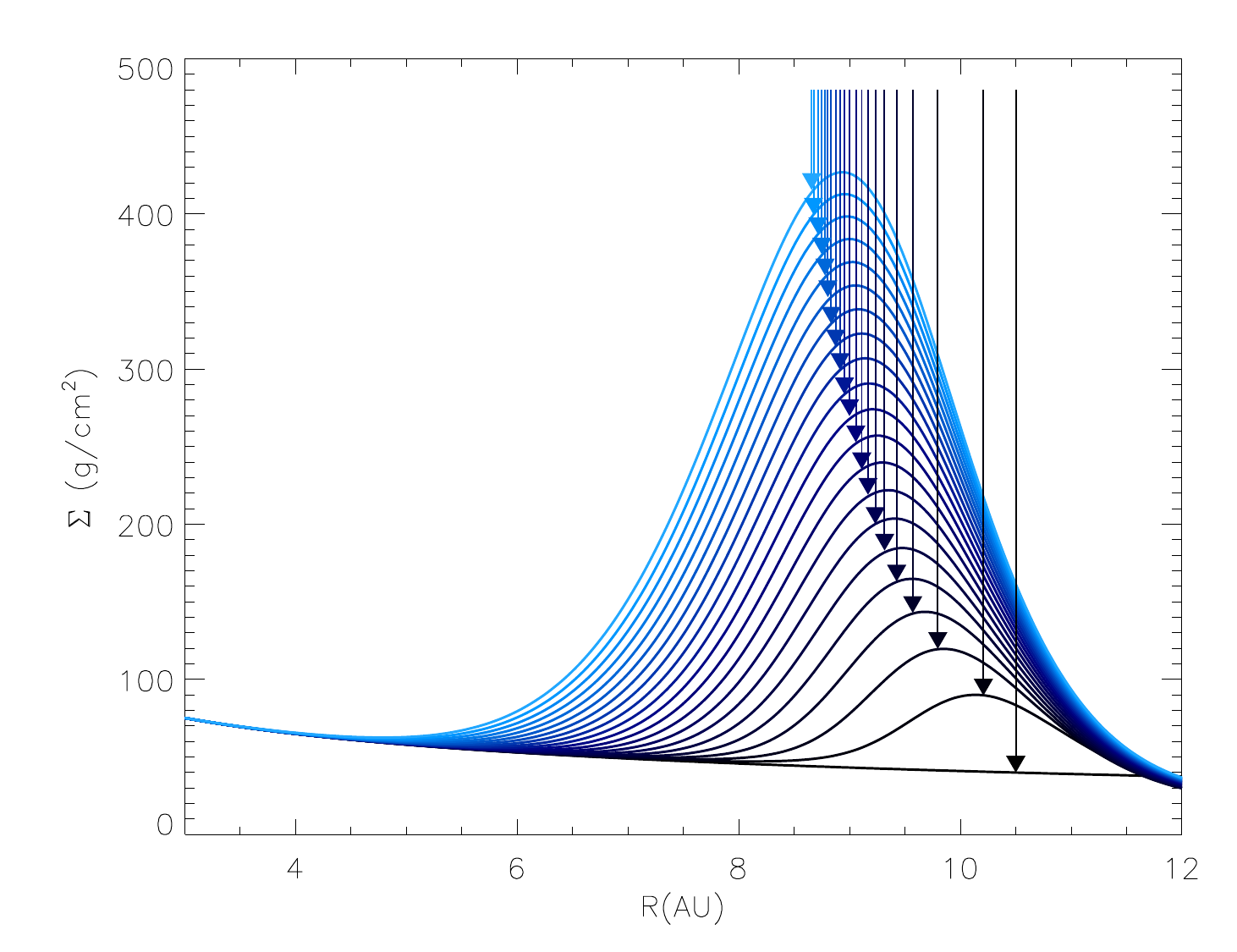}
	\includegraphics[width=\columnwidth]{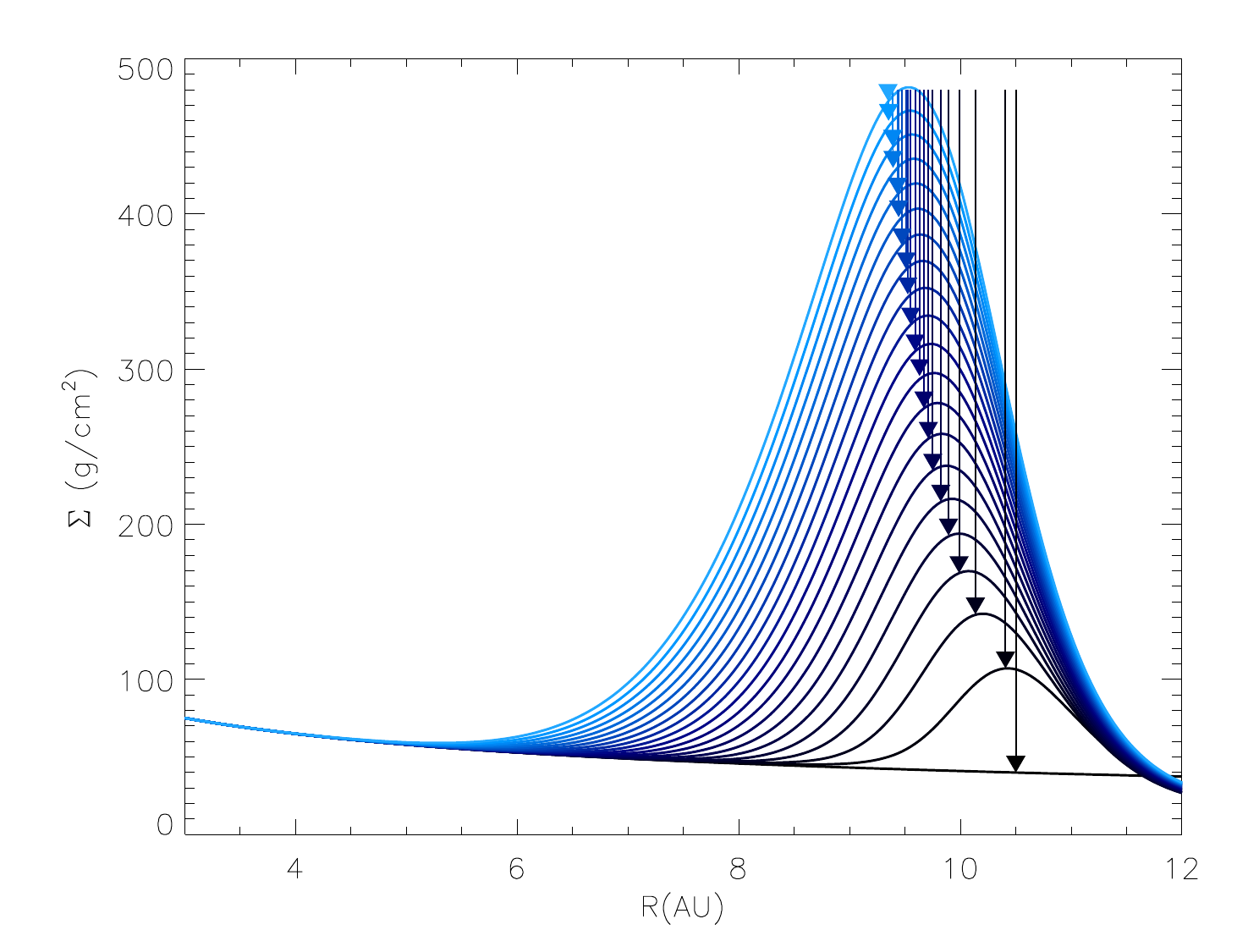}
	\caption{Surface mass density profile snapshots with a cadence of $5\times10^3$\,yr in model $\Delta R_\mathrm{dze}=2H$ (upper panel) and $\Delta R_\mathrm{dze}=1.5H$ (lower panel) calculated by the 1D code according to equations (\ref{eq:visc_evol})-(\ref{eq:nu}). The evolution of the semimajor axis of $10\mathrm{M_\oplus}$ planetary cores are shown with arrows calculated by equations (\ref{eq:mig-analithic-1}) and (\ref{eq:mig-analithic-3}), see details in Sect. 2.2.}
	\label{fig:1D}
\end{figure}

The viscous evolution of the density jump formed at the outer dead zone edge of the disc can be investigated by solving the vertically integrated continuity and Navier--Stokes equations. If the disc is assumed to be axisymmetric, one can use the 1D approximation, where the set of differential equations is significantly reduced. In this case, the evolution of the disc's surface mass density $\Sigma(R)$ can be described by the diffusion-like partial differential equation 
\begin{equation}
	\label{eq:visc_evol}
	\frac{\uppartial\Sigma(R)}{\uppartial t}=\frac{3}{R}\left[R^{1/2}\frac{\uppartial}{\uppartial R}\left(\nu(R)\Sigma(R)R^{1/2}\right)\right]
\end{equation}
\citep{Lynden-BellPringle1974}. In our model, the disc kinematic viscosity $\nu(R )$ is calculated by using the $\alpha$ prescription \citep{ShakuraSunyaev1973}:
\begin{equation}
	\nu(R)=\alpha\frac{c_\mathrm{s}^2(R)}{\Omega(R)},
	\label{eq:nu}
\end{equation}
where $c_\mathrm{s}(R)=\Omega(R) H(R)$ is the local sound speed, $H( R )$ is the pressure scale height, and $\Omega(R)$ is the angular velocity of the gas. Using the \citet{ShakuraSunyaev1973} approximation, the background viscosity parameter is set by the value of $\alpha$. The initial surface mass density is defined as $\Sigma(R)=\Sigma_0 R^{-0.5}$, where $\Sigma_0$ is the surface mass density at $R=1$\,au. We recall that using a power-law surface mass density profile with power-law index  $-0.5$ represents a steady-state solution for $\Sigma(R )$, if the disc viscosity is treated by the $\alpha$ prescription. In disc models presented in this paper, we set $\Sigma_0 = 1.45\times 10^{-5}\mathrm{M_{\odot}}/\mathrm{au}^2$ (in standard units $\Sigma_0 = 129.2\,\mathrm{g/cm^2}$). With this choice, similarly to the minimum mass solar nebula \citep{Weidenschilling1977}, our disc contains $0.01$ solar mass inside $30\,\rm au$. 

To model the decline of viscosity at the outer dead zone edge, the value of the parameter $\alpha$ is smoothly reduced to a value $\alpha\delta_\alpha$ at the outer dead zone edge, see more details in \citet{Regalyetal2012}. The global viscosity is set to the conventionally accepted value $\alpha=10^{-2}$. Inside the dead zone the viscosity is reduced to $\alpha\delta_\alpha=10^{-4}$, meaning that the viscosity is 1\% of its global value. Although the position of the dead zone edge is very uncertain, we set it to $R_\mathrm{dze}=12$\,au according to the theoretical calculations of \citet{MatsumuraPudritz2005}. The width of the dead zone edge ($\Delta R_\mathrm{dze}$) is the region where the viscosity is reduced to $\alpha\delta_\alpha$. $\Delta R_\mathrm{dze}$ is assumed to be in the range $1H_\mathrm{dze}\leq\Delta R_\mathrm{dze}\leq4H_\mathrm{dze}$, where $H_\mathrm{dze}$ is the disc pressure scale height at $R=R_\mathrm{dze}$. We assume a flat disc model; therefore, the local disc height is $H(R)=hR$, where the disc aspect ratio is set to the conventional value of $h=0.05$.
 
To solve equation (\ref{eq:visc_evol}) we applied the fourth-order Runge--Kutta method for time discretization, while the spatial derivatives of $\Sigma(R)$ have been calculated using the second-order central difference scheme. In order to provide the stability of the numerical scheme, an appropriate time step criterion has been applied. In our simulations, the disc extension is set to $0.1\rm{AU}\leq R \leq 20\rm{au}$, and the computational domain consists of 500 equidistant grid cells. At the boundaries we applied such a condition that the surface mass densities in the ghost cells (beyond the first and last grid cells) are the same as in the neighbouring cells, i.e. $\Sigma_{i=0}=\Sigma_{i=1}$ and $\Sigma_{i=N_R}=\Sigma_{i=N_R-1}$. The same condition is applied for the kinematic viscosity $\nu(R)$.

The time evolution of the density jump up to $10^5$\,yr is displayed in Fig.\,\ref{fig:1D} for two different models, in which the widths of the viscosity reduction at the dead zone edge are $\Delta R_\mathrm{dze}=2H_\mathrm{dze}$ and $1.5H_\mathrm{dze}$. In accordance with our recent 2D hydro results \citep{Regalyetal2012}, the peak of the density maximum is slowly evolving towards smaller radii. This inward drift is faster for blunt viscosity reduction. It is also remarkable that the strength and the width of the density jump are inversely proportional to $\Delta R_\mathrm{dze}$.

\subsection{Migration and trapping of a $10\mathrm{M_\oplus}$ planetary core in 1D}

In this section our results are presented on Type I migration of a $10\mathrm{M_\oplus}$ planetary core at a density maximum in a time-evolving locally isothermal 1D disc model. Type I migration of planets in 1D disc models can be investigated by applying analytic torque formulae. Such formulae are very useful, since they can be applied to follow the orbital dynamics of a body embedded in a protoplanetry disc (Lyra et al. 2010) or to investigate the {\it N-body} evolution of several planets embedded in a protoplanetary disc; see \citet{Sandoretal2011} and \citet{Hornetal2012} for two recent studies.

A recent analytic torque formula for Type I migration has been presented by \citet{Paardekooperetal2010}, assuming that the planet is in an inviscid and axisymmetric disc. The torque on the planet contains both the differential Lindblad and corotation torques. Although the differential Lindblad torque is in linear regime, \citet{PaardekooperPapaloizou2009} showed that the corotation torques are non-linear even for low-mass planets, unless the disc has high viscosity. Note that the viscosity is low in our simulations as the planet migrates in the disc dead zone, where the viscosity is greatly reduced. The total torque that \citet{Paardekooperetal2010} derived for a locally isothermal disc contains the linear differential Lindblad torque, linear entropy-related corotation torque and the non-linear horseshoe drag. The change of the semi major axis ($a$) of a planet undergoing Type I migration in a locally isothermal disc thus can be given as

\begin{equation}
	\frac{\mathrm{d}a}{\mathrm{d}t}=\frac{2\Gamma}{qa\Omega(a)},
	\label{eq:mig-analithic-1}
\end{equation}
where
\begin{eqnarray}
	\frac{\Gamma}{\Gamma_0}=&-&(2.5 - 0.5 \beta_\mathrm{p} - 0.1 \sigma_\mathrm{p}) \left(\frac{0.4}{\epsilon}\right)^{0.71}\nonumber\\
	&-&1.4 \beta_\mathrm{p} \left(\frac{0.4}{\epsilon}\right)^{1.26} + 1.1 \left(\frac{3}{2} - \sigma_\mathrm{p}\right) \frac{0.4}{\epsilon},
	\label{eq:mig-analithic-2}
\end{eqnarray}
\begin{equation}
	\Gamma_0=\left(\frac{q}{h}\right)^{2} \Sigma(a) a^{4} \Omega(a)^{2}.
	\label{eq:mig-analithic-3}
\end{equation}
Here, $q=M_\mathrm{p}/M_*$ is the mass ratio of the planet to the central star and $\Omega(a)$ is the circular Keplerian angular velocity at the location of the planet. Here $\sigma_\mathrm{p}$ and $\beta_\mathrm{p}$ are the negative of the local density and temperature gradients, respectively, which can be given by 
\begin{equation}
	\sigma_\mathrm{p}=-\frac{R}{\Sigma(R)}\left.\frac{\mathrm{d}\Sigma(R)}{\mathrm{d}R}\right|_{R=a}
\end{equation}
and  
\begin{equation}
	\beta_\mathrm{p}=-\frac{R}{T(R)}\left.\frac{\mathrm{d}T(R)}{\mathrm{d}R}\right|_{R=a}.
\end{equation}
The parameter $\epsilon$ in equation (\ref{eq:mig-analithic-2}) describes the planetary potential smoothing required for 2D calculations to take into account the disc thickness. To be consistent with our 2D simulations, the value of smoothing parameter is set to $\epsilon=0.6$, see details in Sect.\,\ref{sec:2D}. Using equations (\ref{eq:mig-analithic-2}) and (\ref{eq:mig-analithic-3}) yields
\begin{equation}
	\frac{\Gamma}{\Gamma_0}=-1.24 - 0.66 \sigma_\mathrm{p}
	\label{eq:ttorque}
\end{equation}
for a locally isothermal approximation where $\beta_\mathrm{p}=1$. Since the torque vanishes at $\sigma_\mathrm{p}=-1.88$, the planet should be trapped at a radial distance where the density gradient is positive, i.e. inside the density maximum.

\begin{figure}
	\centering
	\includegraphics[width=\columnwidth]{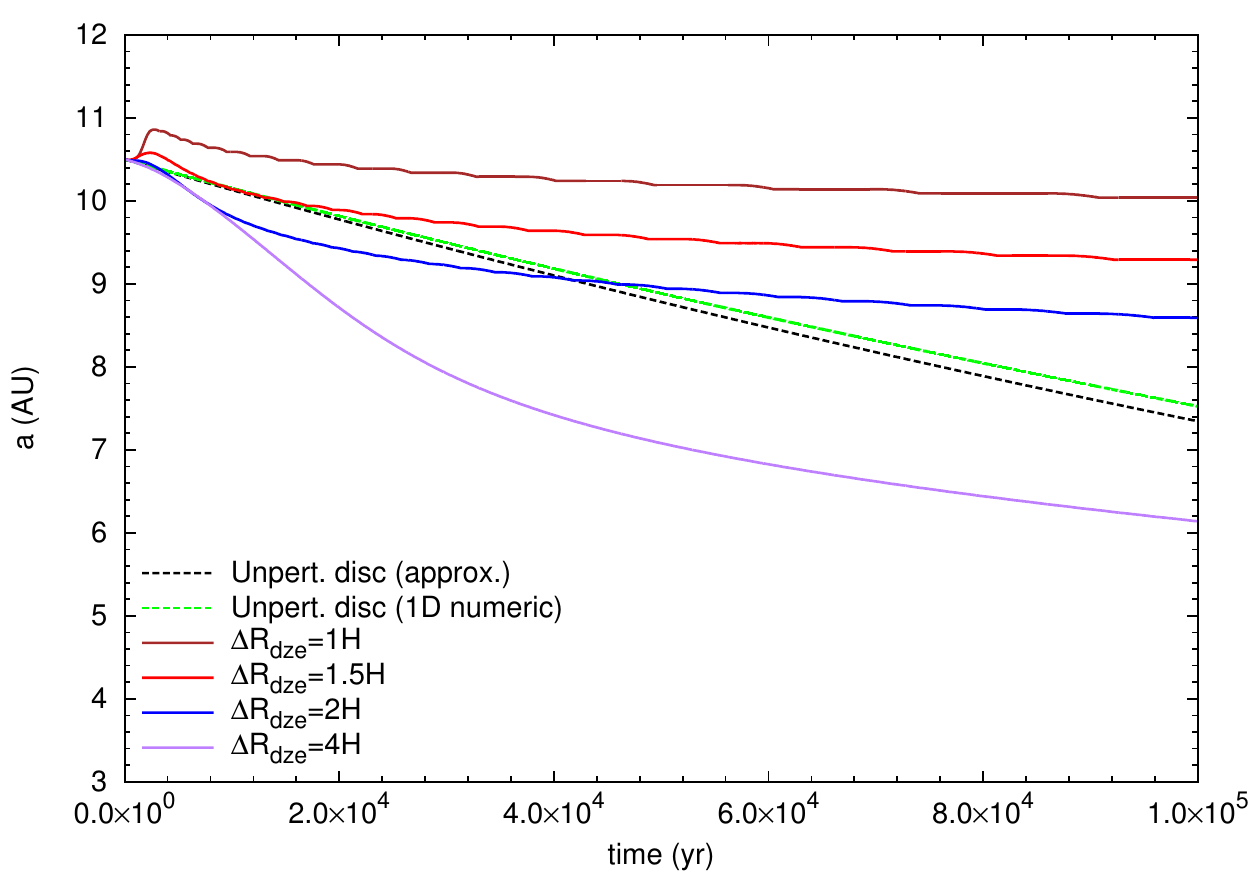}
	\caption{Migration of a $10\mathrm{M_\oplus}$ planetary core calculated by the 1D numerical code withs different width of the outer dead zone edge in the range $1H_\mathrm{dze}\leq\Delta R_\mathrm{dze}\leq4H_\mathrm{dze}$. Type I migration in an unperturbed disc is shown with green and black dashed curves calculated numerically and analytically, respectively.}
	\label{fig:1dmig}
\end{figure}

In order to investigate the trapping of a giant-planet core in different 1D disc models (including the unperturbed case, where no viscosity reduction is applied), we performed a series of 1D simulations in models where the width of the dead zone edge is in the range $1H_\mathrm{dze}\leq\Delta R_\mathrm{dze}\leq4H_\mathrm{dze}$. In all simulations we followed the time evolution of the semimajor axis of a $10\mathrm{M_\oplus}$ planetary core started at the initial position $a_0=10.5$\,au. Our results are shown in Fig.\,\ref{fig:1dmig}.

One can see that the planet migration rate is significantly reduced, if the density jump forms at the dead zone edge compared to the unperturbed case. However, due to the slow inward drifting of density maximum (being more significant for the $\Delta R_\mathrm{dze}=4H_\mathrm{dze}$ case), the planet slowly migrates inwards. As one can see in Fig.\,\ref{fig:1D}, the planet crosses the density maximum and stays close to the $\sigma_\mathrm{p}=-1.88$  where the torque given by equation (\ref{eq:ttorque}) vanishes. Since the density jump and therefore the zero torque position slowly drifts inwards with time, the planet migrates inwards with a much smaller rate than in the Type I regime; thus, we can say that the planet got trapped. Our 1D result confirms the findings of an earlier study of \citet{Matsumuraetal2007}.

The above 1D approach, however, is based on serious simplifications. The most important simplification is that it neglects the azimuthal asymmetry which is expected to develop in the disc's density distribution near the dead zone edge due to the RWI in certain conditions, such as  sharp viscosity transition. Moreover, it is also questionable whether one can use the above 1D torque approximation when the local density slope significantly changes in the vicinity of the planet. \citet{DAngeloLubow2010} stated that their 1D approximation of disc torques can be applied if the slope of density (as well as the slope of temperature) profile is estimated in terms of their mean slope over a radial region of about $3H(a)$. They also emphasize that this assumption may not be true for dead zone edges, where the density slope may change by large amounts over radial distances of the order of $H(a)$. We also  note that the 1D simulations do not take into account the possible back reaction of the planetary core to the surface density profile. Namely, if the viscosity is low enough, as is in our dead zone models, even a $10\mathrm{M_\oplus}$ planet can open a small gap, for which case the planetary migration is no more in the Type I regime. In order to study the migration in the vicinity of a density/pressure maximum in more realistic cases, we performed a series of 2D hydrodynamical simulations, which are presented in the following section.

\section{Planetary migration in the 2D model}

\subsection{2D hydrodynamical model}
\label{sec:2D}
We investigated the migration of a $10\mathrm{M_\oplus}$ planetary core near the disc's outer dead zone edge by 2D grid-based, global hydrodynamic disc simulations. To follow the disc evolution and the migration of the planetary core, we used the code {\small GFARGO}\footnote{http://fargo.in2p3.fr/spip.php?rubrique18}, which is the GPU\footnote{Graphics Processor Unit (GPU) constructed with nearly half thousand processors inside a single GPU device.} supported version of the code {\small FARGO} \citep{Masset2000}. {\small GFARGO} solves the vertically integrated continuity and Navier--Stokes equations numerically in 2D, using  isothermal equation of state for the gas.

The CUDA\footnote{http://www.nvidia.com/object/what\_is\_cuda\_new.html} kernel code of {\small GFARGO} was modified in order to handle the viscosity reduction according to equation (1) of \citet{Regalyetal2012}. For these investigations we used scientific grade GPU NVIDIA TESLA C2050 cards enabling us to study  the $1-2\times10^5$\,yr long viscous evolution of a 2D disc interacting with a migrating planetary core in a computational time of a couple of days.

We adopt as the unit of length $1\,\rm au$ and the unit of mass the solar mass $1\,\mathrm{M_\odot}$. Since the mass of the central star is taken to be $M_*=1\,\mathrm{M_\odot}$, and the gravitational constant has been set to unity, the unit of time is such that the orbital period of a planet at 1\,au is $2\mathrm{\upi}$. We use a frame that corotates with the planet in all simulations. 

In order to better resolve the disc compared to the 1D model, its radial extension is set to $3\,\mathrm{au}\leq R\leq 20\,\mathrm{au}$. In this way, the inner part of the disc ($R\leq3\,\mathrm{au}$) is excluded from the simulation, which can be done as the innermost region of the disc does not affect the viscous evolution (see Fig.\,\ref{fig:1D}).

For the viscosity we used the $\alpha$ prescription \citep{ShakuraSunyaev1973}. Outside the dead zone $\alpha = 10^{-2}$, inside the dead zone its value is reduced to $\alpha\delta_\alpha=10^{-4}$ similarly to the 1D case. The radial position of dead zone edge is set to $R_\mathrm{dze}=12\,\mathrm{au}$ as in 1D simulations. 

The planet initially orbits at $a_0=10.5\,\mathrm{au}$, which is the position, where the density jump appears at the onset of gas accumulation.

At the inner and outer boundaries, the so-called damping boundary condition is applied \citep{deValBorroetal2006mn}, where a solid wall is assumed with wave killing zones next to the boundaries.

The 2D computational domain consists of $N_R=512$ radial and $N_\phi=1024$ azimuthal grid cells. The radial grid cells are distributed logarithmically, while the azimuthal spacing is equidistant. This choice helps in resolving the inner disc better than using uniformly placed radial grid cells. With the above disc extension and grid resolution, the planetary Hill sphere is resolved by about 68 grid cells. Note also that the corotation region is resolved by 15 radial zones, meaning that the overall error in the calculation of the corotation torque due to finite numerical resolution is less than 5\%, according to \citet{Masset2002}. According to our case study presented in Appendix\,\ref{apx:numres}, our simulations are in the numerical convergent domain with the above numerical resolution. 

The planet is not allowed to accrete the gas accumulated inside its Hill sphere, so its mass remained constant. However, a large quantity of gas has been accumulated inside the planetary Hill sphere forming a circumplanetary disc. This might influence its migration, if the disc self-gravity is not taken into account \citep{Cridaetal2009b}. Although the effect of the circumplanetary disc on the migration of planet is found to be significant only for giant planets ($q\geq10^{-4}$) according to \citet{Cridaetal2009b}, it might be important for our case. Namely, as the planetary core gets through the density maximum, the mass confined inside the Hill sphere significantly increases due to the increase of the ambient disc's gas surface mass density. An appropriate solution to take into account the effect of circumplanetary disc without considering the disc self-gravity is the exclusion of half of the material inside the planetary Hill sphere \citep{Cridaetal2009b}. Moreover, during the calculation of the disc torque exerted on the planet, the planetary Hill sphere should be excluded as the circumplanetary disc is bound to the planet \citep{TerquemHeinemann2011}.

According to the above train of thoughts, the mass inside of the planetary Hill sphere is excluded in the torque calculation by applying the a torque cut-off with multiplication of the torque by  $(1-\exp(d/R_\mathrm{H}))$, where $d$ is the grid centre distance to the planet and $R_\mathrm{H}$ is the radius of the planetary Hill sphere. Note that our study regarding the inclusion or exclusion of the planetary Hill sphere in the torque calculation showed that the migration of planetary core is sensitive only in some special circumstances, see details in Appendix\,\ref{apx:Hill}. 

To take the disc thickness into account when calculating the gravitational interaction between the planet and the gas in 2D, we use $\epsilon H(a)$ as the smoothing of the gravitational potential of the planet, where $H(a)$ is the disc's pressure scaleheight at the radial distance $a$ of the planet. \citet{Kleyetal2012} have investigated the effect of the choice of the smoothing length on the low mass planet and disc interactions in 2D simulations. They compared 3D and 2D simulations in a disc with constant surface density and low kinematic viscosity ($\nu=10^{-8}$). According to their results, the 2D simulations give excellent agreement with 3D simulations, if the smoothing length is in the range $0.6\leq\epsilon\leq0.7$. Thus, the smoothing length is set to $\epsilon=0.6$ in simulations presented in this section. Note that our additional study presented in Appendix\,\ref{apx:Smooth} confirmed that our main results are independent of the choice of the smoothing length being in the above-mentioned range.

\subsection{Migration and trapping of a $10\,\mathrm{M_\oplus}$ planetary core in 2D}

\begin{figure}
	\centering
	\includegraphics[width=\columnwidth]{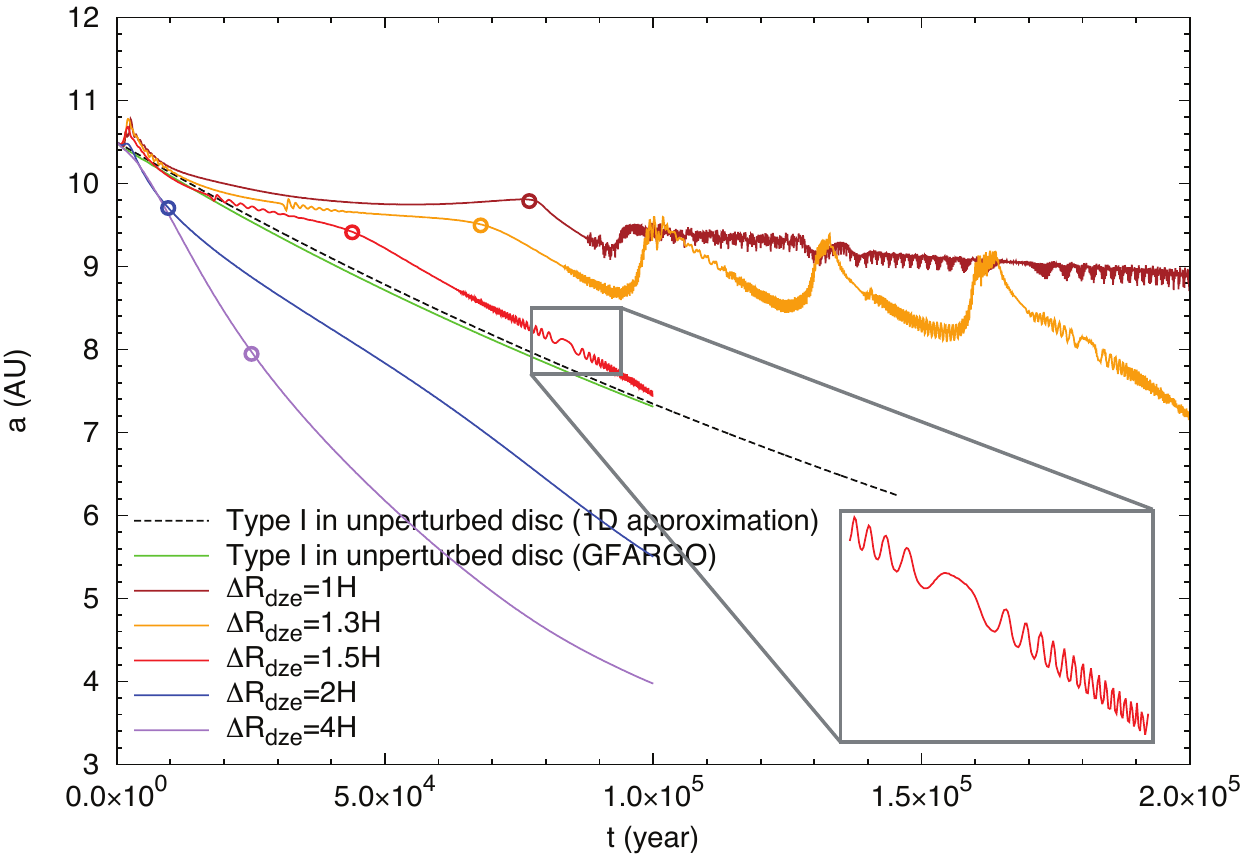}
	\caption{Migration of a $10\,\mathrm{M_\oplus}$ planetary core calculated by the 2D {\small GFARGO} code with different widths of the outer dead zone edge in the range $1H_\mathrm{dze}\leq\Delta R_\mathrm{dze}\leq4H_\mathrm{dze}$. Type I migration in an unperturbed disc is shown with green and black dashed curves calculated numerically and analytically, respectively. The late-time oscillation of the planetary semimajor axis observed in sharp dead zone edge models is shown zoomed-in the inset for model $\Delta R_\mathrm{dze}=2H_\mathrm{dze}$. The circles indicate the points where the planet passes through the density maximum.}
	\label{fig:mig-1}
\end{figure}

First, we compare the results of Type I migration in an unperturbed disc (where no viscosity reduction is applied) calculated by a 2D numerical simulation to the analytical prediction [by the solution of equations\,(\ref{eq:mig-analithic-1})-(\ref{eq:mig-analithic-3})] for a $10\,\mathrm{M_\oplus}$ planetary core. According to our $10^5$\,yr long calculations, the 2D numerical results are in excellent agreement with the analytical predictions (see Fig.\,\ref{fig:mig-1}, green and black dashed curves). The discrepancy in the planetary semimajor axis ($a$) for the 1D analytical approximation and the 2D numerical calculation is less than $\sim0.5\%$ at $t=10^5$\,yr.

We continued the above comparison of the 1D and 2D cases in the cases of perturbed disc models, too. We considered the migration of a $10\,\mathrm{M_\oplus}$ planetary core in discs with outer dead zone edges having different widths of the viscosity reduction being expressed with the local scale height as $1H_\mathrm{dze}\leq\Delta R_\mathrm{dze}\leq4H_\mathrm{dze}$. The radial distance of the dead zone edge is set to $R_\mathrm{dze}=12$\,au as in the 1D models. The planet is placed initially to $a_0=10.5$\,au, where the suspected planetary trap (i.e. the density maximum) appears at the onset of gas accumulation. Our results are shown in Fig.\,\ref{fig:mig-1}. 

After a short initial outward migration, due to a positive gradient in the density at $R=10.5$\,au, the migration is directed inwards, while the density maximum slowly drifts inwards. The rate of initial inward migration exceeds those in the Type I regime for all models. However, the migration history is significantly different for sharp ($\Delta R_\mathrm{dze}\leq1.5H_\mathrm{dze}$) and blunt ($\Delta R_\mathrm{dze}\geq2H_\mathrm{dze}$) dead zone edge models.

Trapping occurs exclusively in sharp dead zone edge models, while for blunt dead zone models the planet migrates inwards with a migration rate that exceeds that in the Type I regime. The trapping, if it happens, seems to be temporary and its duration is inversely proportional to the width of the viscosity reduction. For instance, it lasts about $40\times10^3$, $80\times10^3$ and $90\times10^3$\,yr for models with $\Delta R_\mathrm{dze}=1.5H_\mathrm{dze},\,1.3H_\mathrm{dze}$ and $1H_\mathrm{dze}$, respectively. After the planet is ejected from the trap, its migration speed increases and exceeds that in the Type I regime. 

However, the planetary core can also be retrapped as its migration turns outwards for models $\Delta R_\mathrm{dze}\leq1.3H_\mathrm{dze}$. The ejection and retrapping cycle is repeated two more times when $\Delta R_\mathrm{dze} = 1.3H_\mathrm{dze}$. After the retrapping, the planet finally gets ejected at about $170\times10^3$\,yr. We mention that using a slightly larger smoothing length ($\epsilon=0.7$), the retrapping appears only once, see Appendix\,\ref{apx:Smooth}. In the model $\Delta R_\mathrm{dze}=1H_\mathrm{dze}$, we observed only one ejection at about $t=90\times10^3$\,yr and a retrapping at about $t=95\times10^3$\,yr. Following the evolution of the planetary semimajor axis after the retrapping, we found that the planet remains in the trap up to $200\times10^3$\,yr.

A small-amplitude oscillation of the planetary semimajor axis can be detected temporarily from about $t\simeq15\times10^3$, $35\times10^3$ and $90\times10^3$\,yr for models $\Delta R_\mathrm{dze}=1.5H_\mathrm{dze}$, $1.3H_\mathrm{dze}$ and $1H_\mathrm{dze}$, respectively. After the ejection of the planet from the trap, a small-amplitude oscillation in the planetary semimajor axis reappears, but with higher frequency for models $\Delta R_\mathrm{dze}\leq1.5H_\mathrm{dze}$ (see the inset in Fig.\,\ref{fig:mig-1}).

\subsection{Evolution of the density jump and the position of the planetary core}

\begin{figure}
	\centering
	\includegraphics[width=\columnwidth]{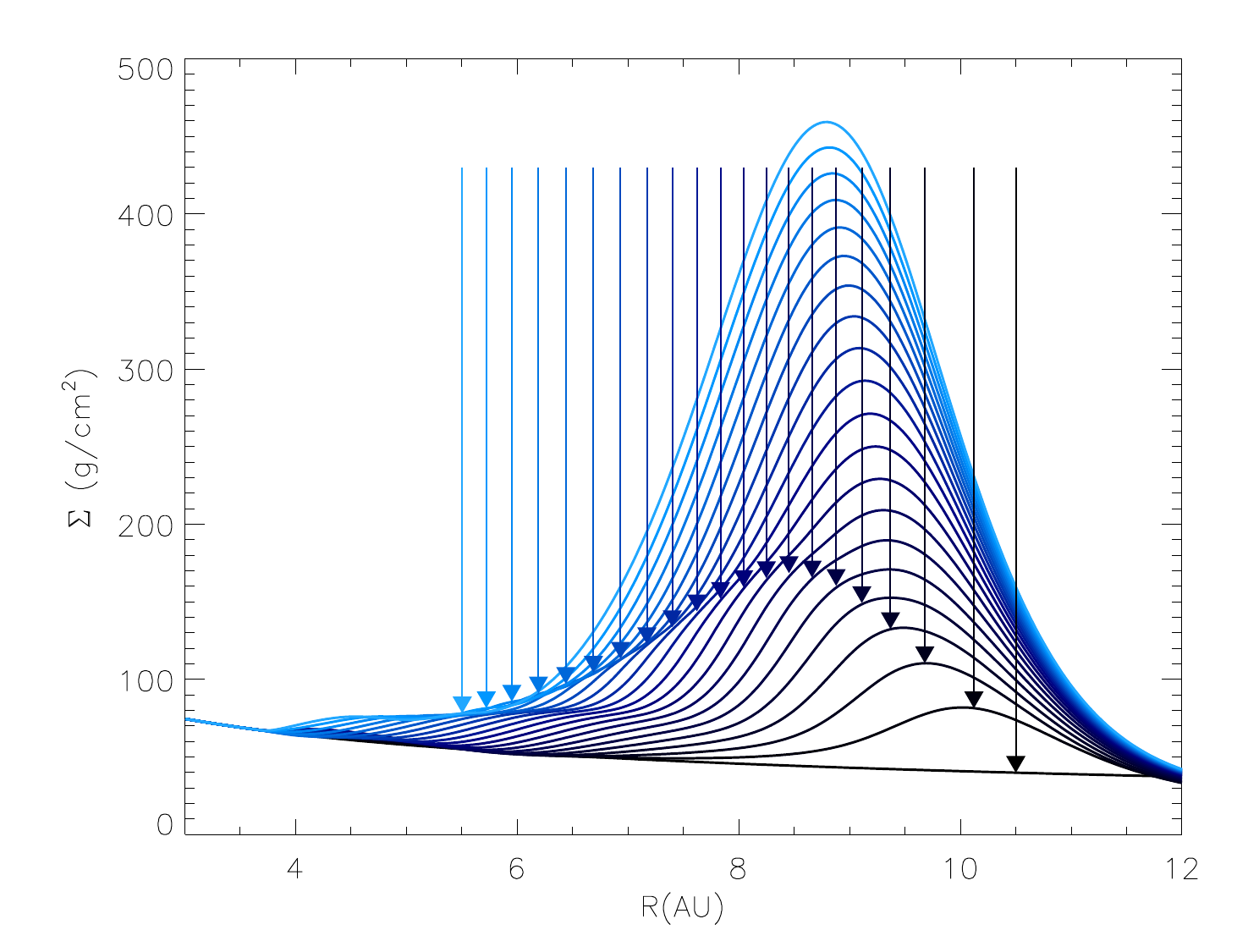}
	\includegraphics[width=\columnwidth]{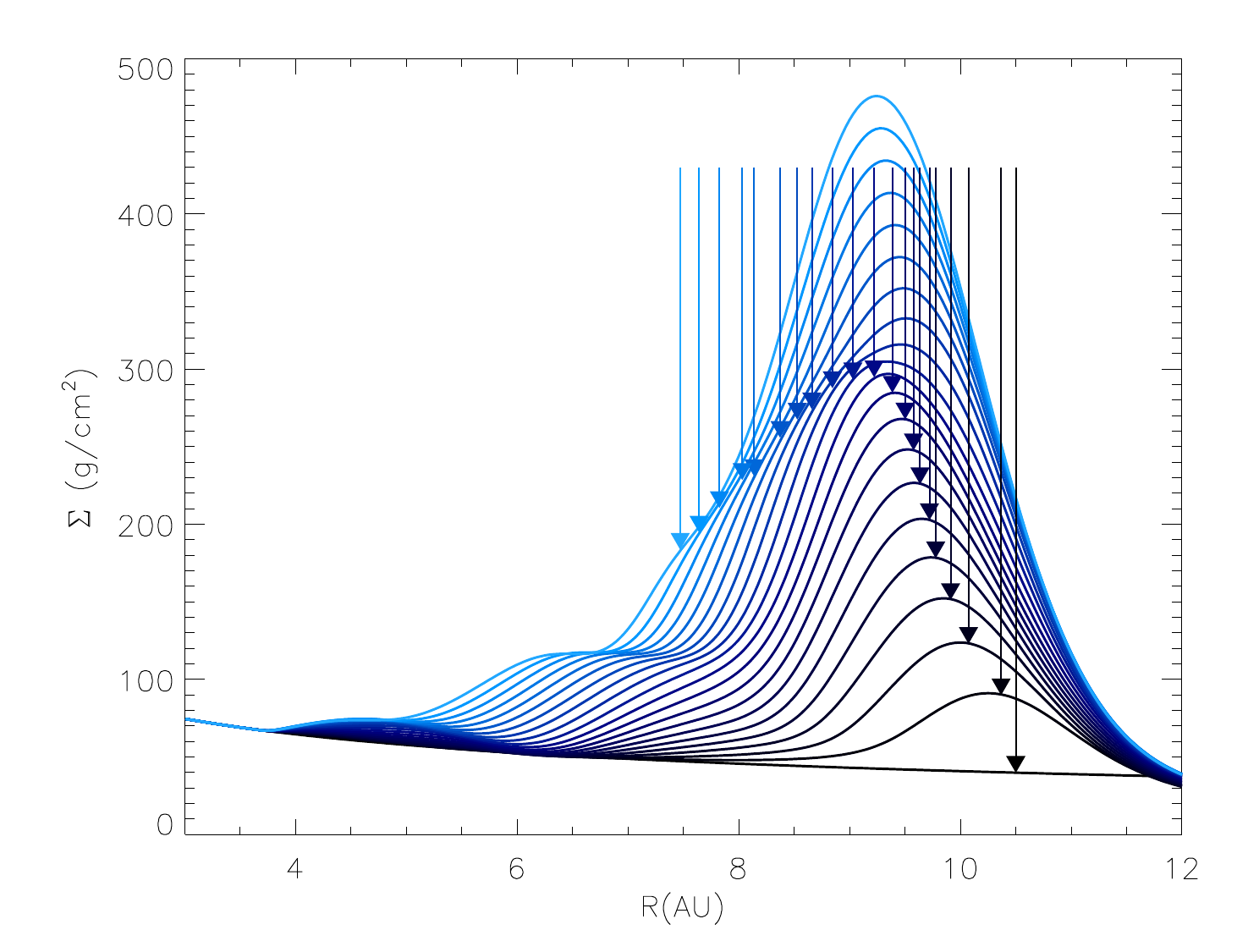}
	\caption{Azimuthally averaged surface mass density profile snapshots with a cadence of $5\times10^3$\,yr in models $\Delta R_\mathrm{dze}=2H_\mathrm{dze}$ (upper panel) and $\Delta R_\mathrm{dze}=1.5H_\mathrm{dze}$ (lower panel). The radial positions of the $10\,\mathrm{M_\oplus}$ planet are shown with arrows.} 
	\label{fig:profile1}
\end{figure}

Fig.\,\ref{fig:profile1} (upper panel) shows the evolution of the azimuthally averaged disc surface mass density profile and the radial position of the $10\,\mathrm{M_\oplus}$ planet for model $\Delta R_\mathrm{dze}=2H_\mathrm{dze}$.  The planet converges towards the density maximum with a significant migration rate, but contrary to the 1D simulation, it is not trapped. 

Surprisingly, in models where the width of the dead zone edge is decreased, the planetary core will be temporarily trapped. Fig.\,\ref{fig:profile1} (lower panel) shows such a case when $R_\mathrm{dze}=1.5H_\mathrm{dze}$. The planet converges towards the density maximum, but its migration is significantly slowed down. Similarly to the previous case (when no trapping happens), the planet reaches the density maximum and its migration is significantly slowed down. After some time, however, the planet leaves the density maximum, which in this case serves as a temporary trap.

We note that the radial density profiles in the 2D simulations slightly differ from those of in the 1D simulations. As one can see, the profiles at the left side of the density jump, where the radial gradient of the azimuthally averaged density is positive, show excess density compared to the 1D simulations. This can most likely be explained by the planetary backreaction on to the disc, i.e the $10\,\mathrm{M_\oplus}$ planetary core perturbs the surface density in its close vicinity.

\subsection{Vortex mode and trapping}
\label{sect:Fourier}
\begin{figure}
	\centering
	\includegraphics[width=\columnwidth]{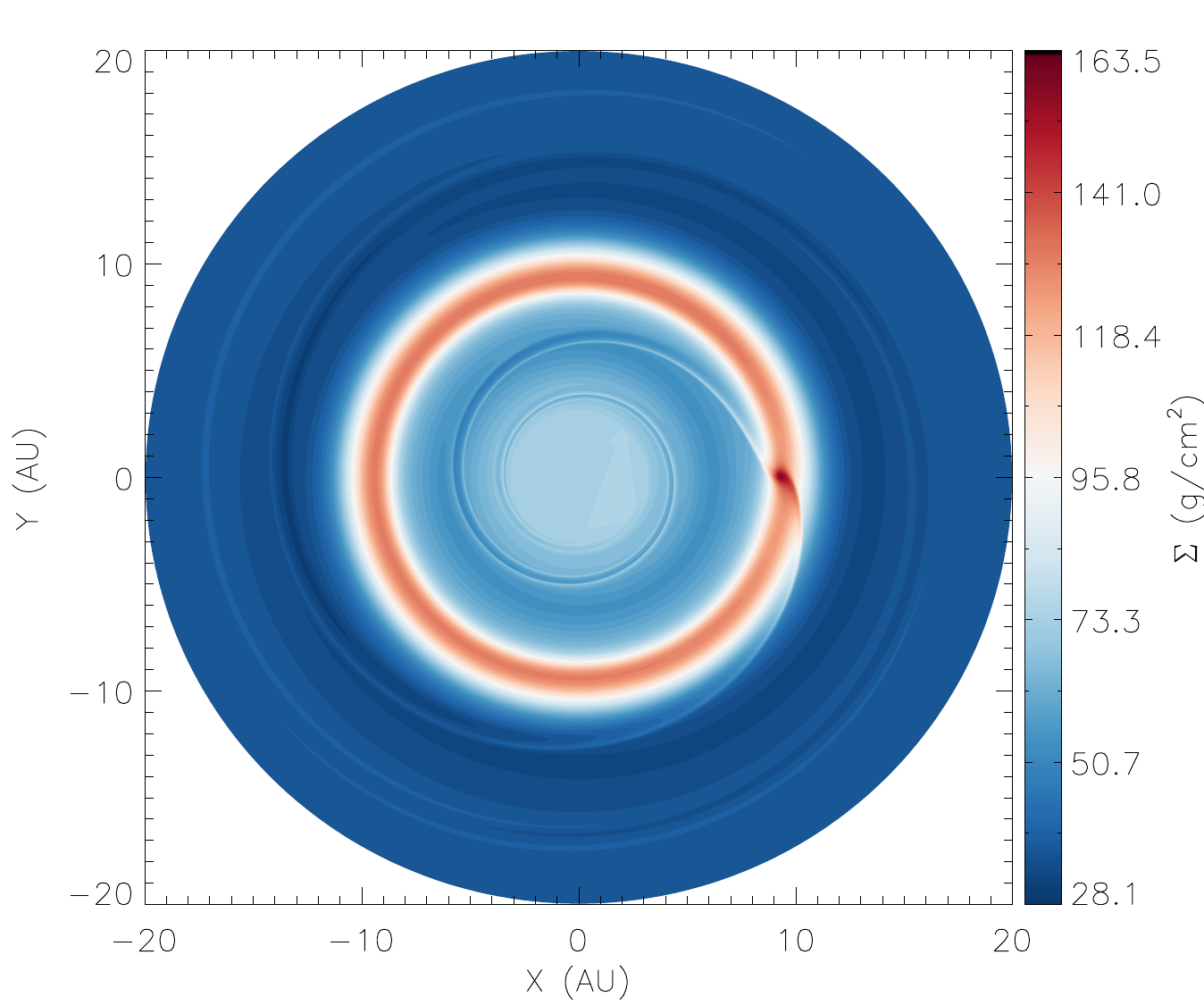}
	\includegraphics[width=\columnwidth]{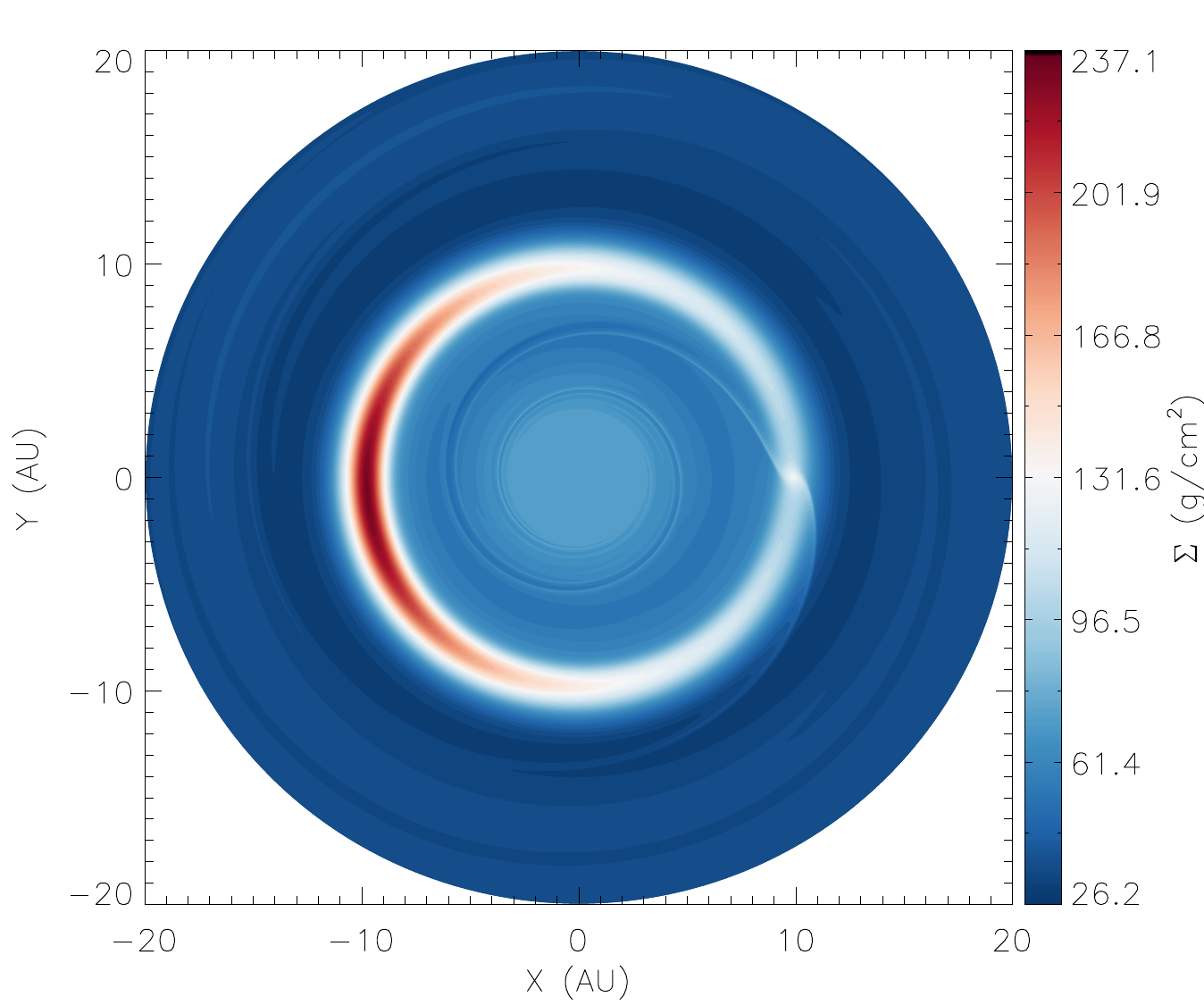}
	\caption{Snapshots of surface mass density distribution for models $\Delta R_\mathrm{dze}=2H_\mathrm{dze}$ (RWI stable, upper panel) and $\Delta R_\mathrm{dze}=1.5H_\mathrm{dze}$ (RWI unstable, lower panel) at $17.5\times10^3$\,yr.}
	\label{fig:gasdens}
\end{figure}

\begin{figure}
	\centering
	\includegraphics[width=\columnwidth]{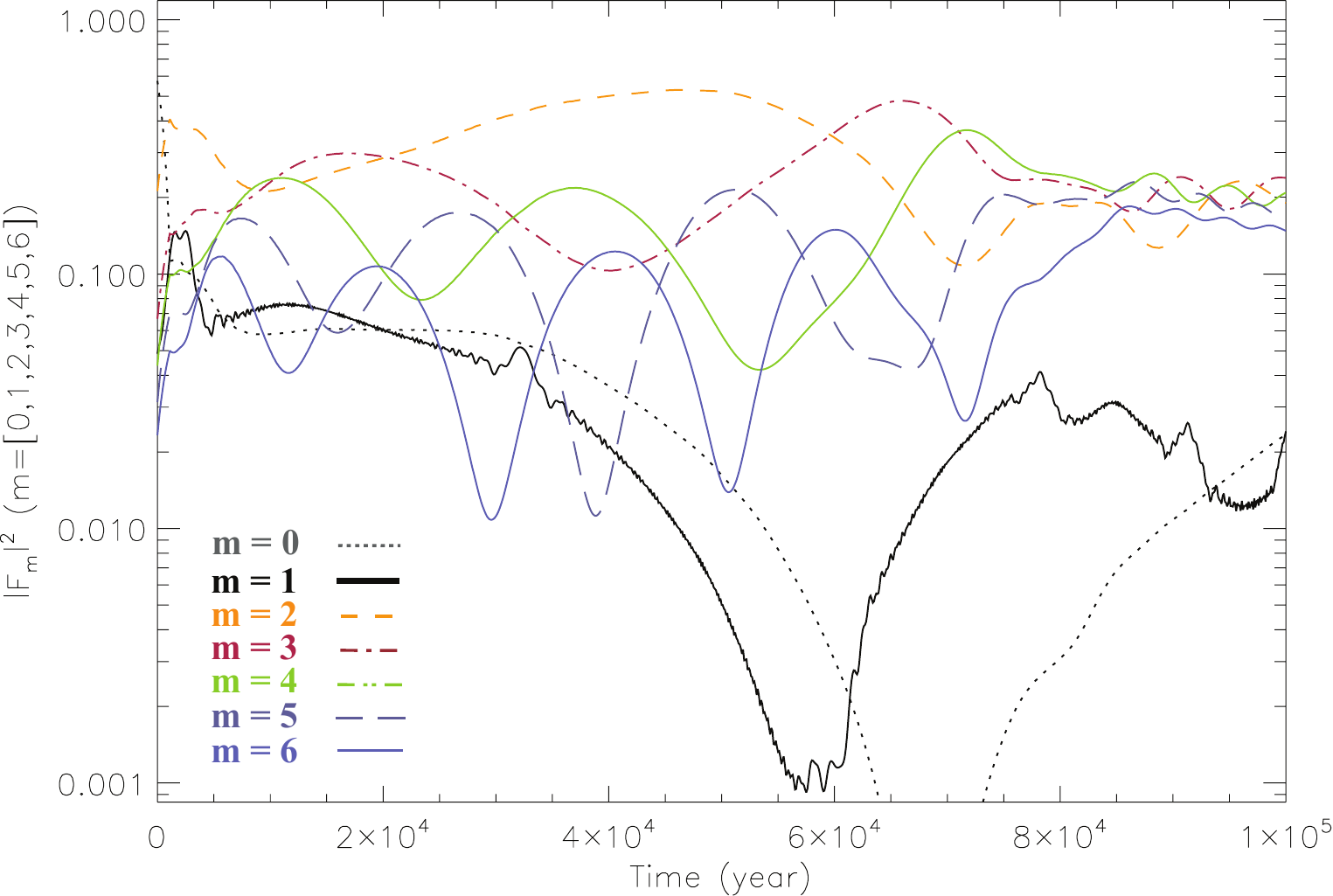}
	\includegraphics[width=\columnwidth]{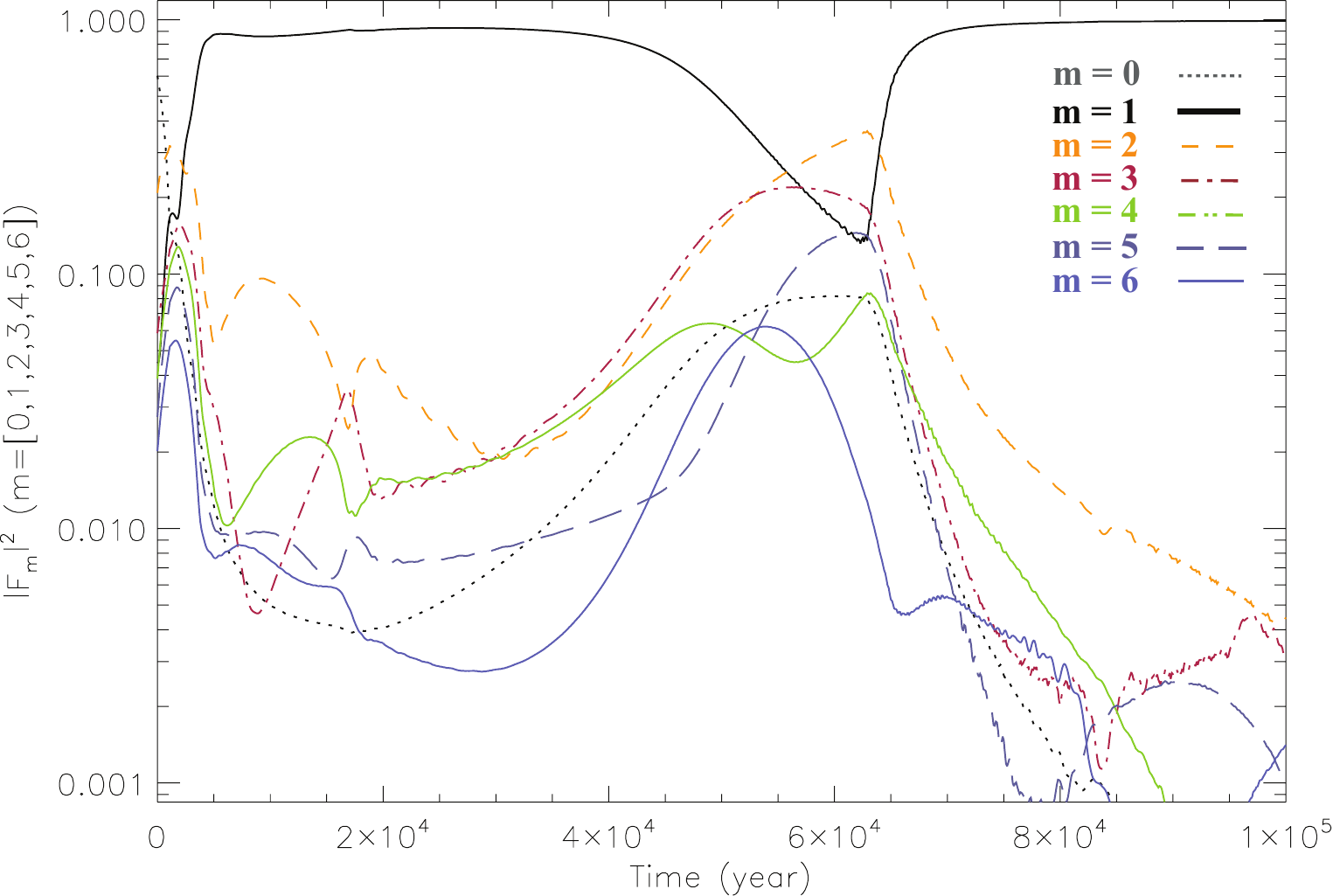}
	\includegraphics[width=\columnwidth]{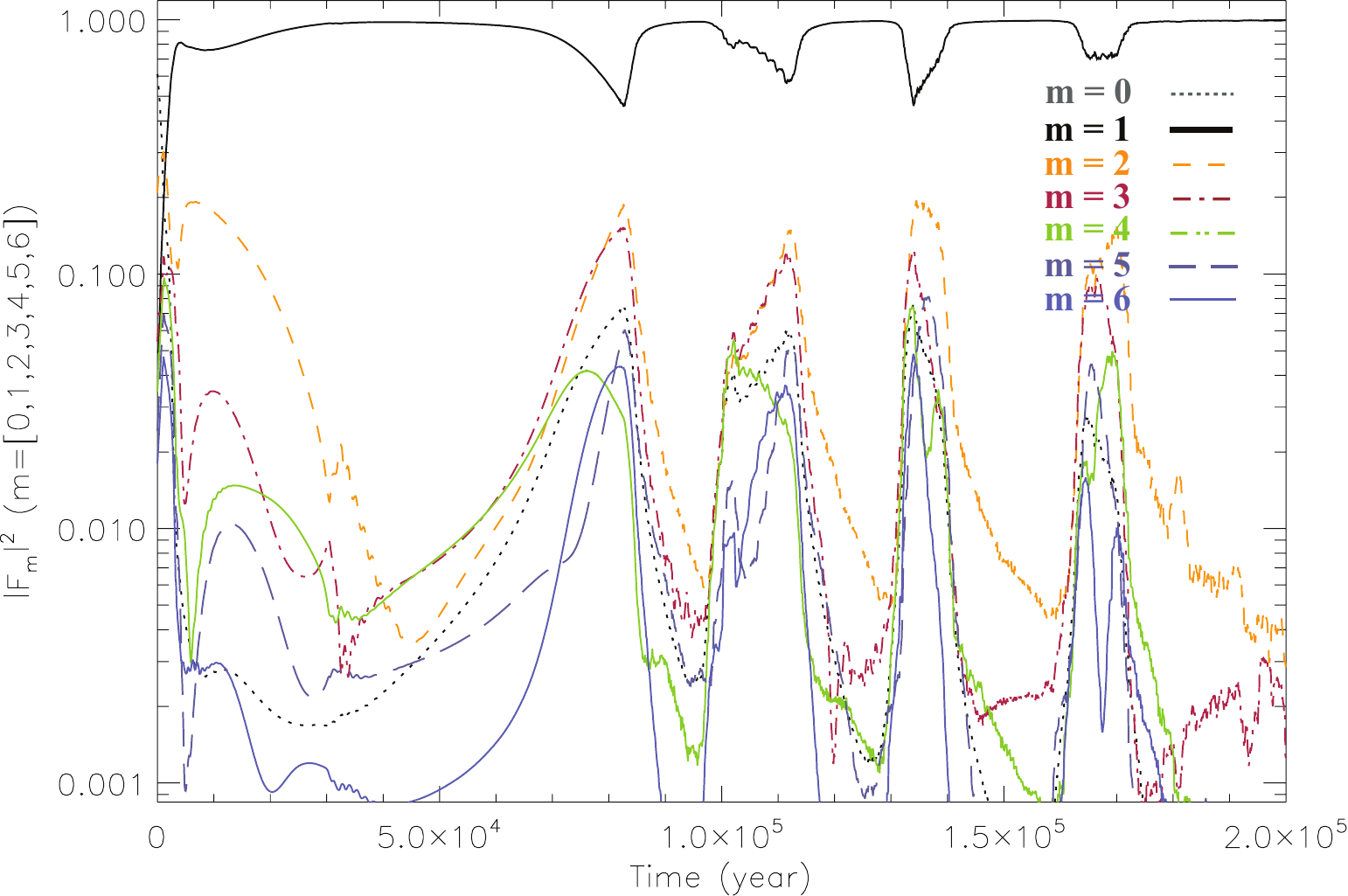}
	\includegraphics[width=\columnwidth]{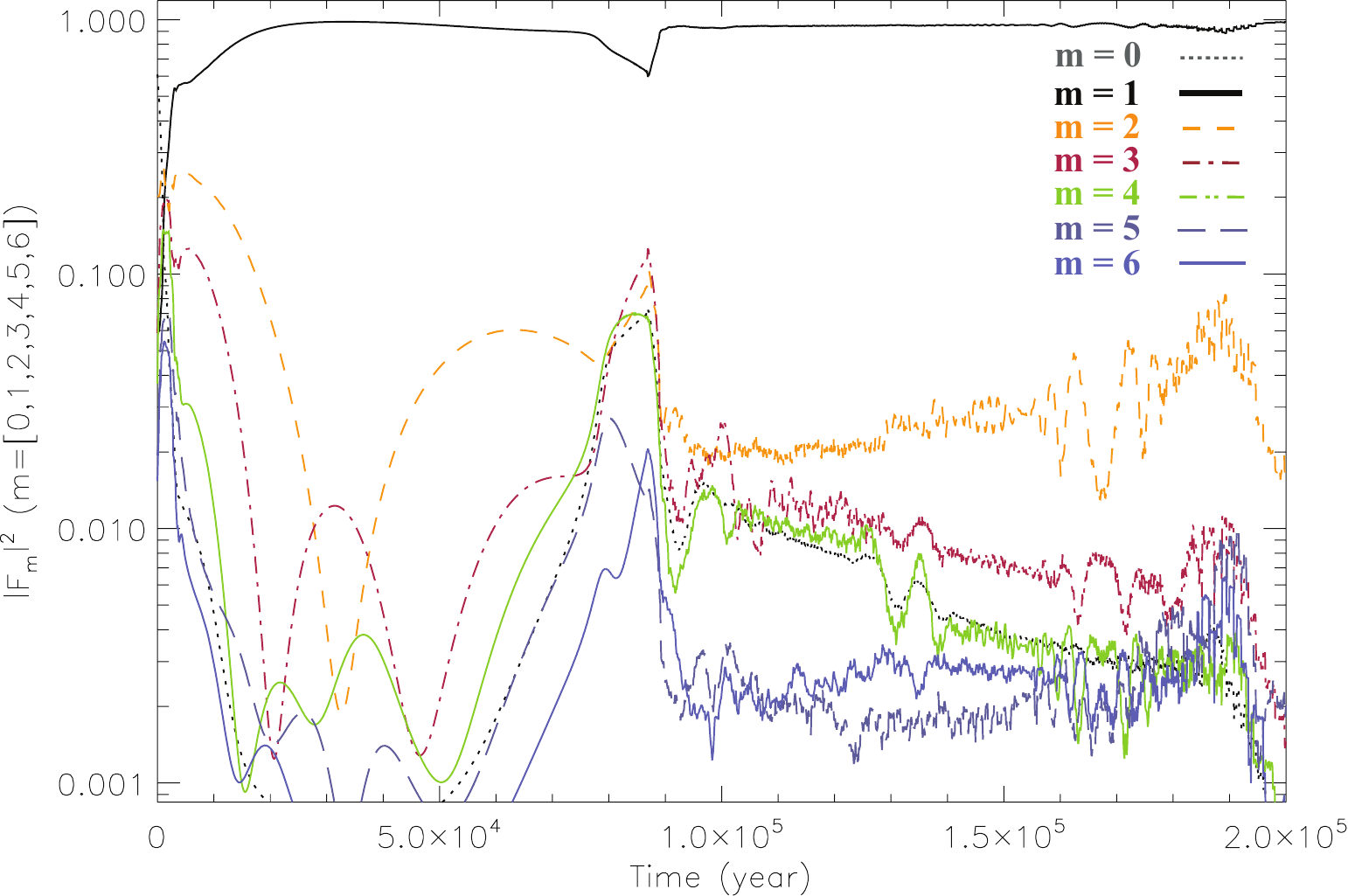}
	\caption{Azimuthal spectral power of the surface mass density distribution against time for models $\Delta R_\mathrm{dze}=2H_\mathrm{dze}$, $1.5H_\mathrm{dze}$, $1.3H_\mathrm{dze}$, and $1H_\mathrm{dze}$ (from top to down panels, respectively). The Fourier amplitudes are normalized by the total power measured in the $m=0-6$ modes.}
	\label{fig:fourier}
\end{figure}

If the viscosity reduction at the dead zone edge is sharp enough ($\Delta R_\mathrm{dze}<2H_\mathrm{dze}$), a long-lived, large-scale, horseshoe vortex forms via vortex coalescing \citep{Lyraetal2009, Regalyetal2012}. On the other hand, when $\Delta R_\mathrm{dze}\ge2H_\mathrm{dze}$, only a density jump forms that appears as a ring like axisymmetric perturbation in 2D simulations. Fig.\,\ref{fig:gasdens} presents the surface mass density distribution snapshots for models $\Delta R_\mathrm{dze}=2H_\mathrm{dze}$ (upper panel) and $1.5H_\mathrm{dze}$ (lower panel) at $t=17.5\times10^3$\,yr.

To investigate how the vortex formation affects the trapping of the $10\,\mathrm{M_\oplus}$ planetary core, we calculated the azimuthal spectral power of the surface mass density distribution as a function of time. First, a 2D Fourier transform of the density in the ring (period of $2\upi$ along the azimuthal direction at the radius where the vortices are) was calculated. Then the Fourier amplitudes of the $m=0-6$ modes are normalized by the total power in these seven modes.

The spectral power is shown in Fig.\,\ref{fig:fourier} for models $\Delta R_\mathrm{dze}=2H_\mathrm{dze}$, $1.5H_\mathrm{dze}$, $1.3H_\mathrm{dze}$ and $1H_\mathrm{dze}$. In the model where $\Delta R_\mathrm{dze}=2H_\mathrm{dze}$, the RWI is not excited, i.e. no large-scale vortex forms; therefore, the relative power of the $m=1$ mode to the $m=0$ and $m>1$ modes remains very small (Fig.\,\ref{fig:fourier}, upper panel). No trapping  of the planetary core occurs in this model. In contrast, the $m=1$ mode is at maximum relative to the $m=0$ and $m>1$ modes if RWI is excited, e.g. in models $\Delta R_\mathrm{dze}\leq1.5H_\mathrm{dze}$ (Fig.\,\ref{fig:fourier}, three lower panels). 

We observed that around $t=45\times10^3$\,yr in the model with $\Delta R_\mathrm{dze}=1.5H_\mathrm{dze}$, the relative power of the $m=1$ mode suddenly decreased, the $m=2$ mode increased and the planet was ejected from the trap. In the model with $\Delta R_\mathrm{dze}=1.3H_\mathrm{dze}$, several ejection and retrapping was observed at the same time, when the relative power in the $m=1$ mode declined. For model $\Delta R_\mathrm{dze}=1H_\mathrm{dze}$, when only one retrapping occurs, the decline in the relative power of the $m=1$ mode also appears at this time.

Thus, we can conclude that the large-scale horseshoe vortex is disintegrated, when the planet approaches the density maximum. After the planet leaves the density maximum, the large-scale vortex reappears, i.e. the power in the $m=1$ mode is at maximum again. Therefore, the retrapping of the planetary core might be connected to reappearance of the $m=1$ mode vortex.

\section{Discussion}

In this section by analysing carefully the torques exerted on the planet, we try to reveal the surprising discrepancy between the results of 1D and 2D simulations. We recall that according to the 1D results, the migrating planet reaches the zero torque position and stays there during the whole simulations. Due to the viscous evolution of the disc, the density jump slightly drifts inwards with time, also followed by the zero torque position. Thus, the planetary migration rate is greatly reduced, meaning that the planet is trapped in the density jump. In our 1D simulations, we found that the trapping efficiency of the $10\,\mathrm{M_\oplus}$ planetary core does not depend on the width of the viscosity reduction. 

Contrary to the 1D results, in our 2D hydrodynamic simulations, the planetary core has not been trapped in models with blunt viscosity reduction ($\Delta R_\mathrm{dze}\geq2H_\mathrm{dze}$). Surprisingly, the planetary core migrated through the density jump with an even higher migration rate than usual for Type I migration in unperturbed discs. For sharper viscosity reduction ($\Delta R_\mathrm{dze}<2H_\mathrm{dze}$), the planet has temporarily been trapped.

\subsection{Torques exerted on the planetary core}
\label{sect:torques}

\begin{figure}
	\centering
	\includegraphics[width=\columnwidth]{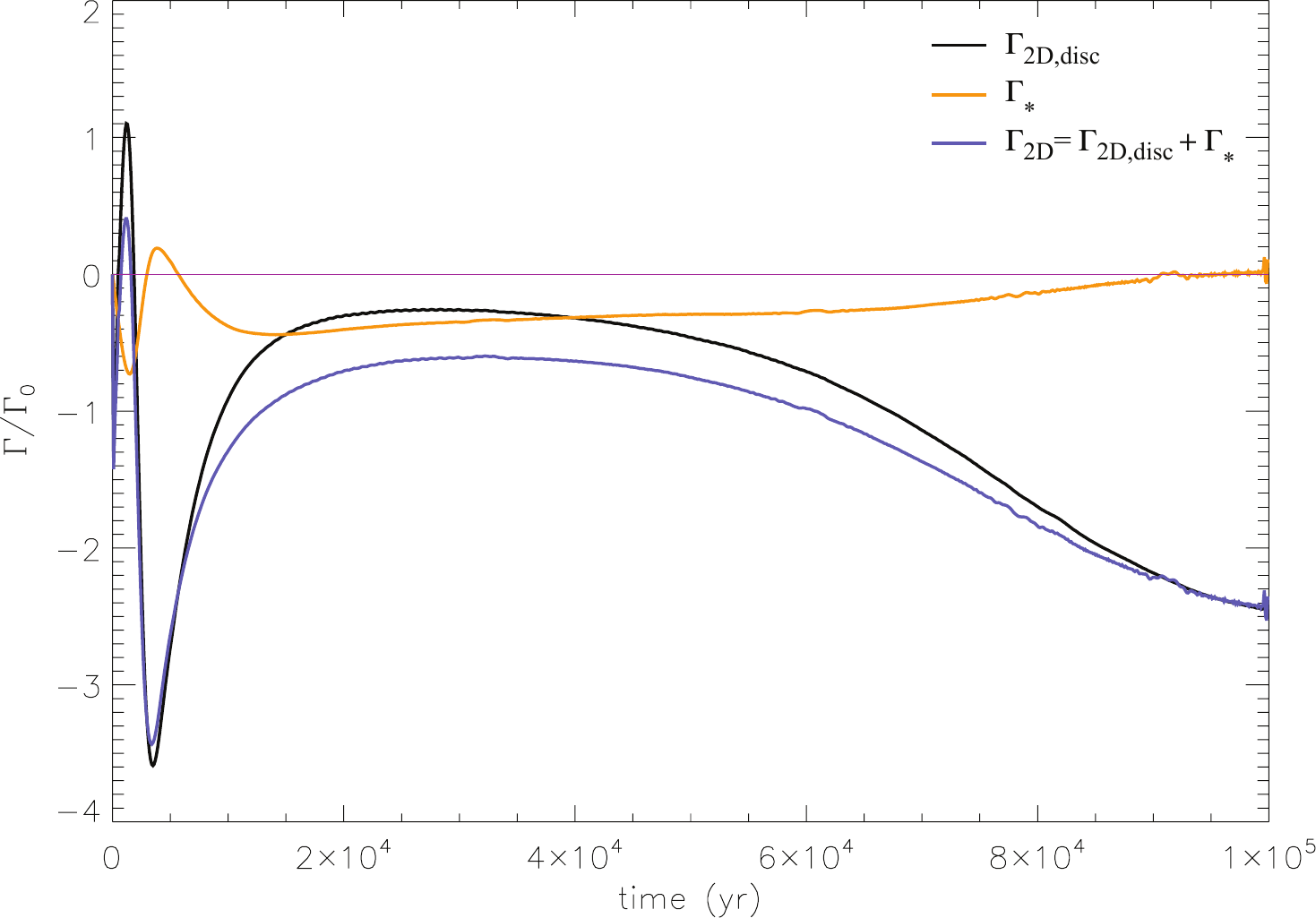}
	\includegraphics[width=\columnwidth]{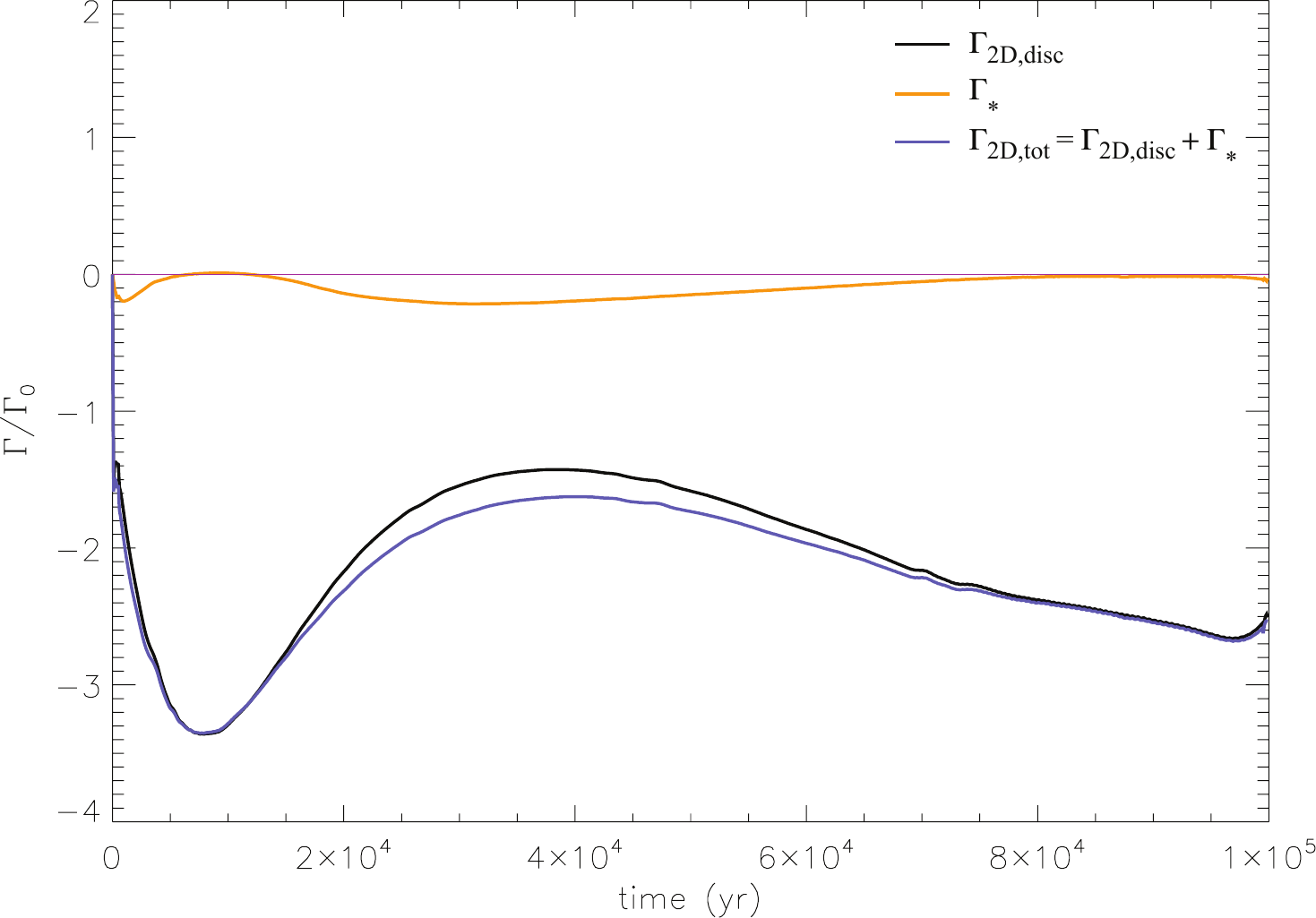}

	\caption{$100\times10^3$\, yr evolution of disc ($\Gamma_\mathrm{2D,disc}$), stellar ($\Gamma_\mathrm{*}$) and net ($\Gamma_\mathrm{2D,net}$) torques exerted on the planet in a model where no RWI is excited, i.e. for blunt dead zone edge models in which no trapping occurs. The upper and lower panels show torque evolution against time for models $\Delta R_\mathrm{dze}=2H_\mathrm{dze}$ and $4H_\mathrm{dze}$, respectively.}
	\label{fig:torque-evol-norwi}
\end{figure}

The migration of a planet is determined by the torque exerted by the disc on it. For an unperturbed disc (no viscosity reduction), the net torque is the sum of the inner and outer -- with respect to the radial position of the planet -- Lindblad torques (together called as the differential Lindblad torque) and the corotation torque. If the net torque is negative, the planet migrates towards the central star \citep{Ward1997}. In hydrodynamic simulations the total disc torque can be calculated numerically by summing the torques exerted on the planet by the elementary grid cells, i.e. 
\begin{equation}
	\Gamma_\mathrm{2D,disc}=\sum_{i,j=1}^{N_R,N_\phi}Fy_{i,j}x_\mathrm{p}-Fx_{i,j}y_\mathrm{p},
	\label{eq:torque2D}
\end{equation}
where $Fy_{i,j}$ and $Fx_{i,j}$ are the forces felt by the planet due to the gravitational pull of the material inside the grid cell at coordinates $x_{i,j}$ and $y_{i,j}$, while $x_\mathrm{p}$ and $y_\mathrm{p}$ are the planetary coordinates in the Cartesian system. Note that the centre of the coordinate system is fixed to the star. Since we use a frame corotating with the planet, $y_\mathrm{p}=0$ and $x_\mathrm{p}=a$, where $a$ is  the radial distance of the planet to the star. As a consequence, the second term of equation (\ref{eq:torque2D}) vanishes. Using equation (\ref{eq:torque2D}) the total disc torque can be given by
\begin{equation}
	\Gamma_\mathrm{2D,disc}=\sum_{i,j=1}^{N_R,N_\phi}\frac{A_{i,j}\Sigma_{i,j}\left(1-\exp{[d_{i,j}/R_\mathrm{H}]^2} \right)}{\left(d_{i,j}+\epsilon H(a)\right)^3} a,
	\label{eq:disc_torque}
\end{equation}
where $A_{i,j}$ and $\Sigma_{i,j}$ are the surface area and the surface mass density of the grid cell ${i,j}$, respectively, and $d_{i,j}$ is the distance of the centre of grid cell $(i,j)$ to the planet. The factor $(1-\exp{[d_{i,j}/R_\mathrm{H}]^2})$ in equation (\ref{eq:disc_torque}) appears due to the torque cut-off applied, while the expression $(d_{i,j}+\epsilon H(a))$ reflects the smoothing of the planetary potential. 

It is usually assumed that the barycentre of the system coincides with the star, meaning that the stellar torque felt by the planet vanishes. However, it is important to take into account the stellar torque ($\Gamma_\mathrm{*}$), if the barycentre of the star disc and planet system is not centred on the star. This can happen if the disc density distribution is non-axisymmetric due to the development of a large vortex. If the barycentre of the system is shifted away from the star, the stellar torque exerted on the planet is
\begin{equation}
	\Gamma_*=-y_\mathrm{bc} \frac{GM_*M_\mathrm{p}}{a^2},
\end{equation}
where $y_\mathrm{bc}$ is the vertical position of the barycentre, which can be given by
\begin{equation}
	y_\mathrm{bc}=\sum_{i,j=1}^{N_R,N_\phi} \frac{A_{i,j}\Sigma_{i,j} y_{i,j}}{M_*+M_\mathrm{p}+M_\mathrm{disc}}.
\end{equation}
Note that we use dimensionless units, thus $G=1$, and the unit of mass is the central stellar mass, thus $M_*=1$. Therefore, $M_\mathrm{p}=q$, where $q=M_\mathrm{p}/M_*=3\times10^{-5}$ for a $10\,\mathrm{M_\oplus}$ planetary core.

The evolution of the normalized torques [$\Gamma_\mathrm{2D,disc}/\Gamma_0$, $\Gamma_*/\Gamma_0$ and $(\Gamma_\mathrm{2D,disc}+\Gamma_\mathrm{*})/\Gamma_0$, where the normalization factor $\Gamma_0$ is given by equation (\ref{eq:mig-analithic-3})] is shown in Fig.\,\ref{fig:torque-evol-norwi} for the models with $\Delta R_\mathrm{dze}=2H_\mathrm{dze}$ and $4H_\mathrm{dze}$. The disc torque is negative throughout the simulations. It is noteworthy that the stellar torque does not vanish, and it is commensurable to the disc torque for the model with $\Delta R_\mathrm{dze}=2H_\mathrm{dze}$, while negligible for the model with $\Delta R_\mathrm{dze}=4H_\mathrm{dze}$. This means that the surface mass density distribution is not axisymmetric even in the RWI stable disc. The asymmetry is growing with decreasing dead zone edge width. The net torque ($\Gamma_\mathrm{2D,disc}+\Gamma_*$) felt by the planet is always negative; thus, the planet is not trapped in these models. As the magnitude of total torque is larger for model $\Delta R_\mathrm{dze}=4H_\mathrm{dze}$ than for $\Delta R_\mathrm{dze}=2H_\mathrm{dze}$, the migration rate is higher for a wider dead zone edge model, too.

\begin{figure}
	\centering
	\includegraphics[width=\columnwidth]{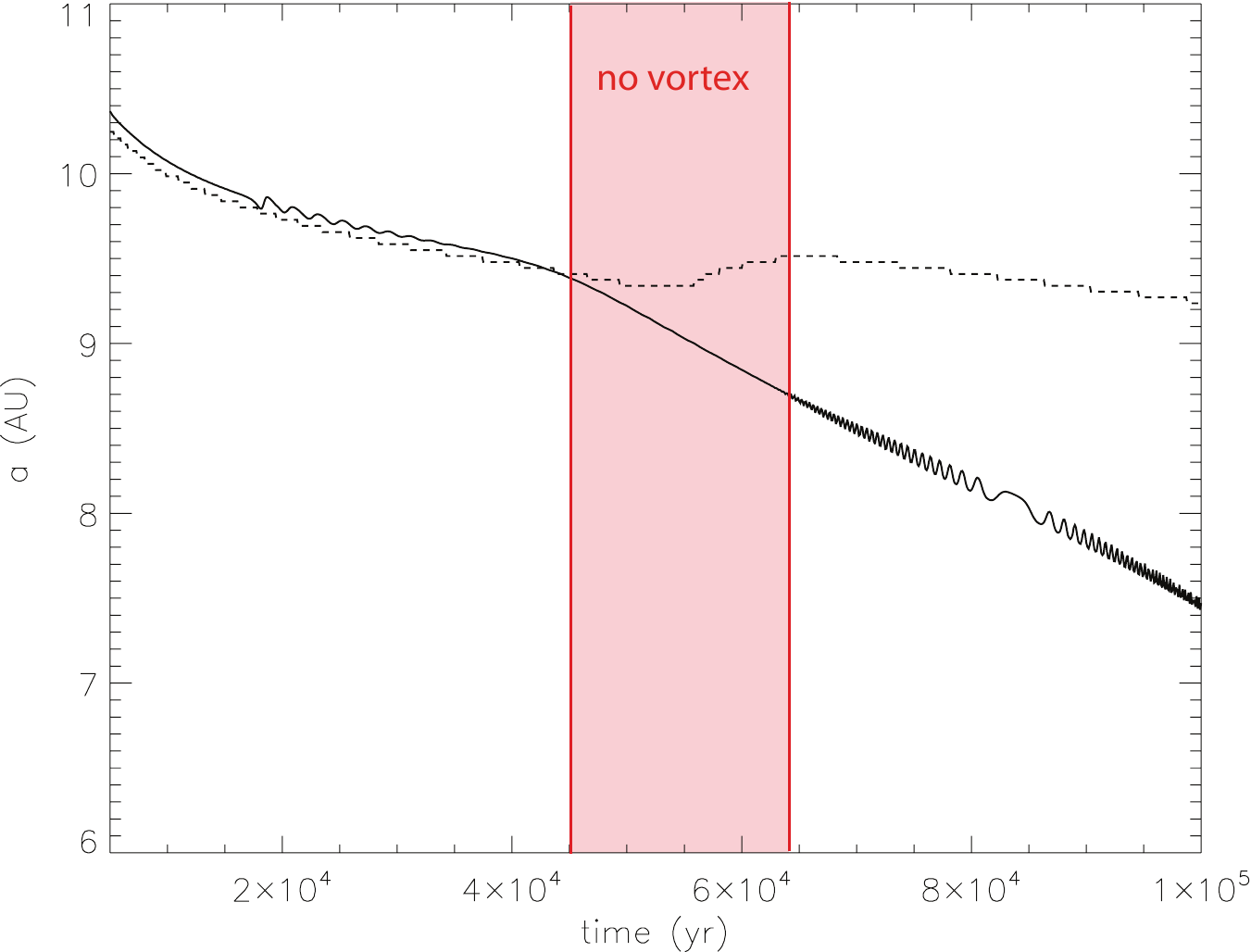}
	\caption{Planetary semimajor axis (solid black curve) and the radial position of the maximum of azimuthally averaged density profile (dotted black curve) against time in model $\Delta R_\mathrm{dze}=1.5H_\mathrm{dze}$. The shaded region shows the time interval, where the vortex is disintegrated.}
	\label{fig:dens-max-r}
\end{figure}

\begin{figure}
	\centering
	\includegraphics[width=\columnwidth]{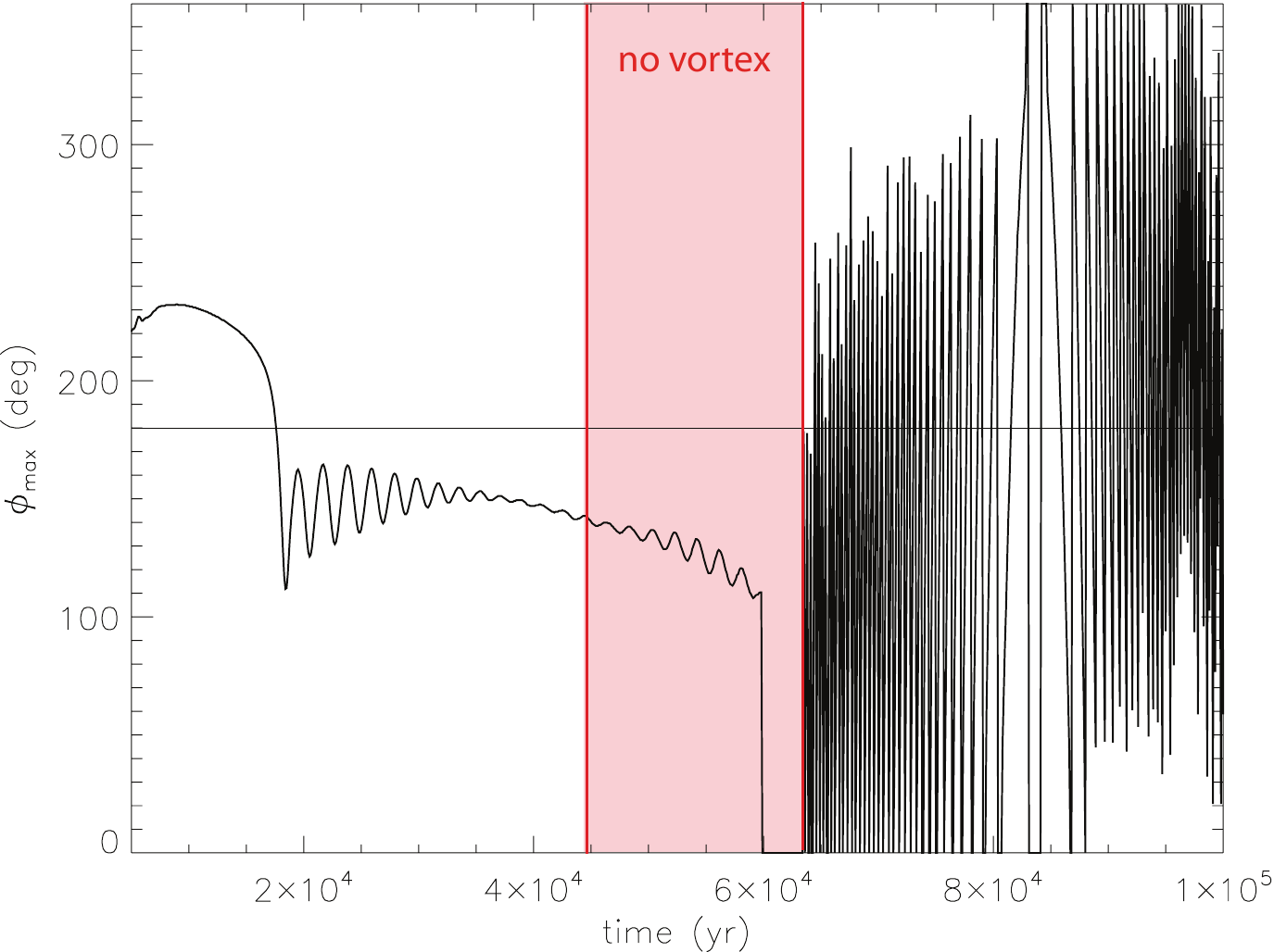}
	\caption{Azimuthal position of the maximum of radially averaged density profile against time in model $\Delta R_\mathrm{dze}=1.5H_\mathrm{dze}$. The shaded region shows the time interval, where the vortex is disintegrated.}
	\label{fig:dens-max-a}
\end{figure}

In the model with $\Delta R_\mathrm{dze}=1.5H_\mathrm{dze}$, in which the RWI is excited, the torque evolution and therefore the migration history of the planet are more complicated. Fig.\,\ref{fig:dens-max-r} shows the time evolution of the planetary semimajor axis and the radial position of the maximum of azimuthally averaged density profile (indicating the radial position of the density maximum). As one can see, the planet slowly migrates towards the density maximum, while the density maximum is also drifted inwards. Thus, initially, the planet orbits close to but beyond the density maximum until the ejection happens at $t\simeq45\times10^3$\,yr. When the planet goes through the density maximum, the $m=1$ mode vortex disintegrates (see Fig.\,\ref{fig:fourier}), and the planetary migration begins to accelerate.

The azimuthal position of the maximum of the radially averaged density profile ($\phi_\mathrm{max}$) -- where the radial averaging is confined to a certain radial extension centred on the maximum of azimuthally averaged density profile -- can refer to the vortex centre position. Fig.\,\ref{fig:dens-max-a} shows how the vortex azimuthal position changes for the model with $\Delta R_\mathrm{dze}=1.5H_\mathrm{dze}$. The vortex centre is at the rear of planet ($180^\circ<\phi_\mathrm{max}<360^\circ$) during the first $18\times10^3$\,yr of the simulation. However, as the planet migrates towards the vortex orbital radius, the vortex is shifted to the front of planet ($0^\circ<\phi_\mathrm{max}<180^\circ$). Note that $\phi_\mathrm{max}$ starts to oscillate around $\phi_\mathrm{max}\simeq150^\circ$, i.e. the vortex librates with decreasing amplitude from $t=18\times10^3$\,yr up to the ejection. At later times, after $60\times10^3$\,yr, when the planet orbits at a smaller distance than the re-formed vortex itself, the relative position of the planet and the vortex is rapidly changing due to the significant difference in the azimuthal velocity of the vortex and the planet.

\begin{figure}
	\centering
	\includegraphics[width=\columnwidth]{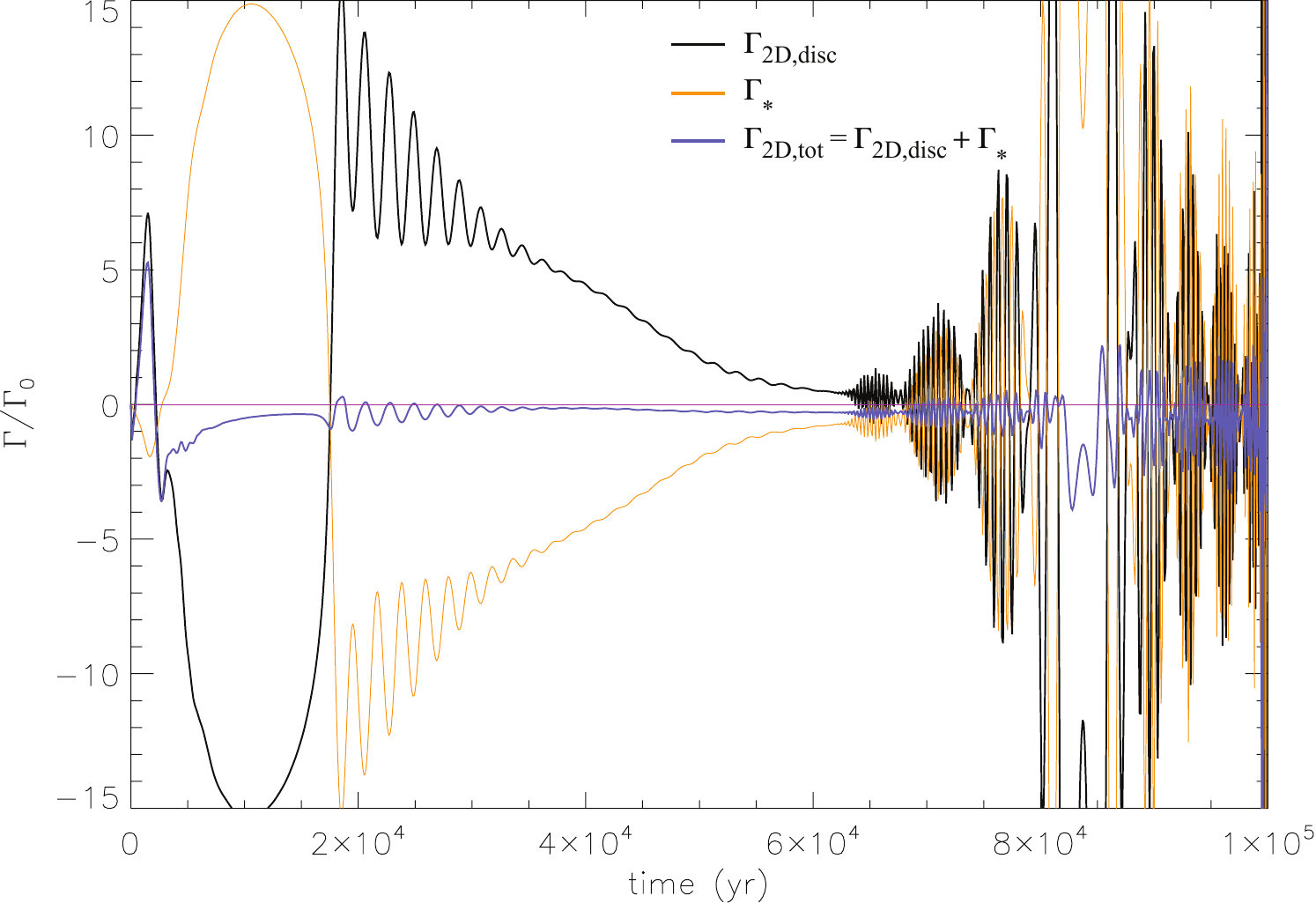}
	\caption{$100\times10^3$\,yr evolution of disc ($\Gamma_\mathrm{2D,disc}$), stellar ($\Gamma_\mathrm{*}$) and net ($\Gamma_\mathrm{2D,net}$) torques exerted on the planet in model $\Delta R_\mathrm{dze}=1.5H_\mathrm{dze}$, where RWI is excited. In this thin dead zone edge model, temporary trapping of the planet occurs.}
	\label{fig:torque-evol-rwi}
\end{figure}

Fig.\,\ref{fig:torque-evol-rwi} displays the evolution of the disc and stellar torques exerted on the planet in model $\Delta R_\mathrm{dze}=1.5H_\mathrm{dze}$. When the vortex is at the rear of the planet ($t<17\times10^3$\,yr), the disc torque is negative, while it is positive if the vortex is at the front of the planet ($t>18\times10^3$\,yr). Note that when $\phi_\mathrm{max}=180^\circ$ (at $t=18\times10^3$\,yr), the disc torque is still negative, while slightly later (at $t=20\times10^3$\,yr) it vanishes and becomes positive. As the vortex centre librates around the azimuthal position of $\phi_\mathrm{max}\simeq150^\circ$, the disc torque will show oscillations in the time interval $20\times10^3\leq t \leq 35\times10^3\,\mathrm{yr}$. The disc torque will again oscillate from about $t=60\times10^3$\,yr, right after the ejection of the planet, but with much higher amplitude than in the previous case.

Let us emphasize that the stellar torque plays a significant role in the migration of the planetary core since the disc density distribution is highly non-axisymmetric, and the barycentre of the system is significantly shifted away from the star. As one can see in the upper panel of Fig.\,\ref{fig:torque-evol-rwi}, the magnitudes of stellar and disc torques are comparable, being about one order of magnitude larger in amplitude than in the RWI stable case (see Fig.\,\ref{fig:torque-evol-norwi} for comparison). The stellar and disc torques are always opposite in sign, resulting in a small magnitude but net negative torque throughout the whole simulation (Fig.\,\ref{fig:torque-evol-rwi}). At $t\simeq45\times10^3$\,yr, the magnitude of the net negative torque is slightly increasing, which eventually causes the ejection of the planet from the trap.

The disc torque becomes highly oscillating at $t>60\times 10^3$\,yr, when the planet gets inside of the density maximum. Since the barycentre of the system shifts as the vortex azimuthal position ($\phi_\mathrm{max}$) changes (Fig.\,\ref{fig:dens-max-a}), the stellar torque is also highly oscillating. The high-frequency oscillation of the torques is due to that the angular velocity of the planet and the vortex differs significantly; therefore, their mutual distance changes rapidly. The disc torque is positive, if the vortex is at the front of the planet, while negative if the vortex is at the rear of the planet. The opposite is true for the stellar torque. As a result, the net torque felt by the planet is negative on average causing fast inward migration of the planet. To explain the curious retrapping phenomenon of the planetary core observed in models with $\Delta R_\mathrm{dze}\leq1.3H_\mathrm{dze}$ requires a more detailed torque analysis, which will be presented in the next section.

\subsection{Comparison of the 1D and 2D torques}

\begin{figure}
	\centering
	\includegraphics[width=\columnwidth]{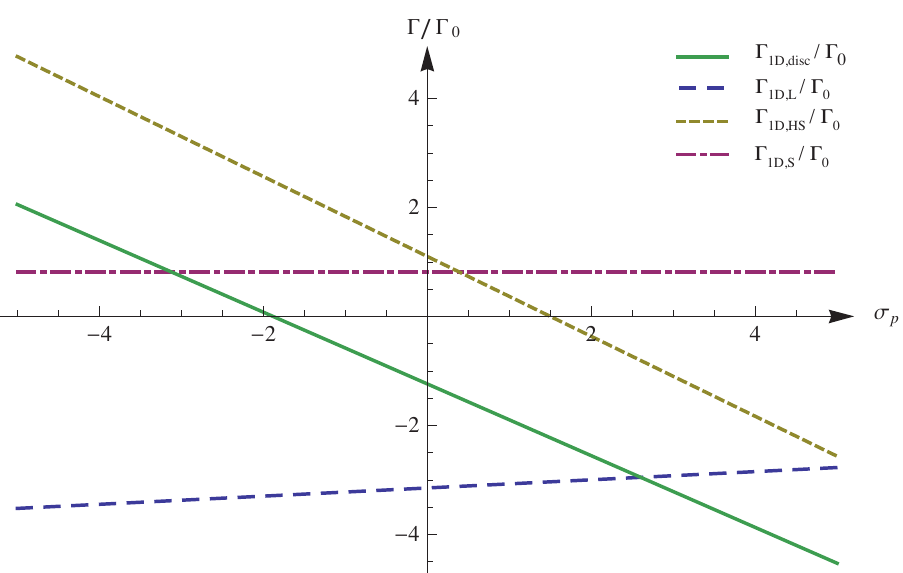}
	\caption{Normalized torque terms (differential Lindblad, entropy- and vortensity-related  horseshoe drags, and total disc torque) that are given by equations (\ref{eq:1D,L}) and (\ref{eq:1D,C}) as a function of $\sigma_\mathrm{p}$.}
	\label{fig:1dtorque}
\end{figure}

According to \citet{Paardekooperetal2010}, the 1D approximation of the differential Lindblad torque can be given by
\begin{equation}
	\frac{\Gamma_\mathrm{1D,L}}{\Gamma_0}=-(2.5+1.7\beta_\mathrm{p}-0.1\sigma_\mathrm{p})\left(\frac{0.4}{\epsilon}\right)^{0.71},
	\label{eq:1D,L}
\end{equation}
while for a locally isothermal case the unsaturated corotation torque is given by
\begin{eqnarray}
	\frac{\Gamma_\mathrm{1D,C}}{\Gamma_0}=&-&1.4 \beta_\mathrm{p} \left(\frac{0.4}{\epsilon}\right)^{1.26}  + 2.2\beta_\mathrm{p}\left(\frac{0.4}{\epsilon}\right)^{0.71}\nonumber\\
 	&+& 1.1 \left(\frac{3}{2} - \sigma_\mathrm{p}\right) \frac{0.4}{\epsilon}.
	\label{eq:1D,C}
\end{eqnarray}
The expression $(\Gamma_\mathrm{1D,L}+\Gamma_\mathrm{1D,C})/\Gamma_0$ is equivalent to $\Gamma/\Gamma_0$ given by equation (\ref{eq:mig-analithic-2}). The first two terms of equation (\ref{eq:1D,C}) are together the entropy-related corotation torque ($\Gamma_\mathrm{1D,S}$) caused by the radial change in the temperature of the gas. The last term of equation (\ref{eq:1D,C}) is the vortensity-related horseshoe drag ($\Gamma_\mathrm{1D,HS}$) caused by the angular momentum exchange between the planet and the gas that experienced U–turn near the planet \citep{Ward1997}. Fig.\,\,\ref{fig:1dtorque} displays the normalized differential Lindblad torque, entropy-related corotation torque, vortensity-related horseshoe drag and net disc torque $(\Gamma_\mathrm{1D,L}+\Gamma_\mathrm{1D,S}+\Gamma_\mathrm{1D,HS})/\Gamma_0$ given by equations (\ref{eq:1D,L}) and (\ref{eq:1D,C}) as a function of the density slope at the radial position of the planet $\sigma_\mathrm{p}$. The value of $\sigma_\mathrm{p}$ is found to be in the range $-5\leq\sigma_\mathrm{p}\leq5$ for all models presented in this work. The differential Lindblad torque weakly, while the vortensity-related horseshoe drag strongly, depends on the slope of the density profile $\sigma_\mathrm{p}$. The entropy-related corotation torque is obviously independent of the density slope. Thus, it is plausible to assume that the 1D approximation of the differential Lindblad torque is adequate for 2D simulations.

We note that equation (\ref{eq:1D,L}) was derived under the condition that $\sigma_\mathrm{p}$ is constant over a radial range on either side of the planet, which is not true in models with viscosity transitions. Thus, in models where the slope of the density profile significantly changes its value within the radial range of several $H(R)$, the applicability of the above 1D approximation of torques is questionable as stated by \citet{DAngeloLubow2010}.

The differential Lindblad and corotation torques cannot be separated geometrically in 2D simulations since the density perturbation within the horseshoe region triggers evanescent waves that go beyond the boundaries of the horseshoe region contributing to the corotation torque \citep{CasoliMasset2009,MassetCasoli2009}. However, accepting the validity of the 1D approximation of differential Lindblad torque ($\Gamma_\mathrm{2D,L}=\Gamma_\mathrm{1D,L}$ ), the value of 2D corotation torque can be inferred by subtracting the Lindblad contribution given by the 1D approximation from the net disc torque calculated by equation (\ref{eq:disc_torque}), i.e. $\Gamma_\mathrm{2D,C}=\Gamma_\mathrm{2D,disc}-\Gamma_\mathrm{2D,L}$. To calculate the 1D torques based on equations (\ref{eq:1D,L}) and (\ref{eq:1D,C}) using the results of 2D simulations, the value of density slope is required at the radial position of planet, which can be approximated by the radial gradient of the azimuthally averaged 2D density distribution.

\subsubsection{Models without viscosity reduction}

\begin{figure}
	\centering
	\includegraphics[width=\columnwidth]{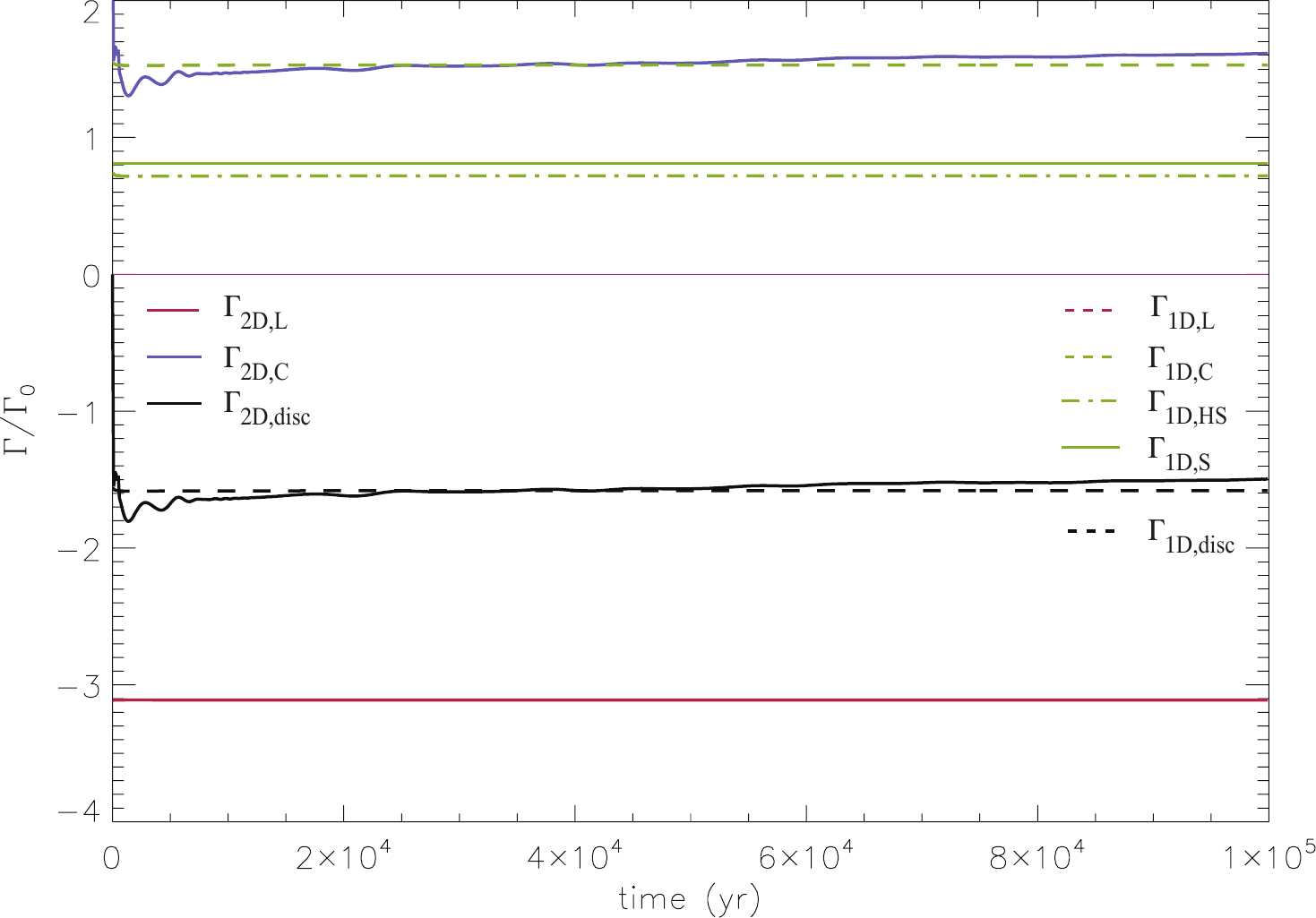}
	\caption{$100\times10^3$\,yr evolution of the approximated 1D differential Lindblad ($\Gamma_\mathrm{1D,L}$) and corotation ($\Gamma_\mathrm{1D,C}$) torques calculated from the 2D density distribution compared to the appropriate 2D torques ($\Gamma_\mathrm{2D,L}$ and $\Gamma_\mathrm{2D,C}$) in the unperturbed disc model, where no viscosity transition is applied. The evolution of 1D and 2D net disc torques ($\Gamma_\mathrm{1D,L}+\Gamma_\mathrm{1D,C}$ and $\Gamma_\mathrm{2D,L}+\Gamma_\mathrm{2D,C}$) is also displayed. The torques are normalized by $\Gamma_0$.}
	\label{fig:1D-out2}
\end{figure}

First, we analyse models, where no viscosity reduction has been applied, and the migration rates of the $10\,\mathrm{M_\oplus}$ planetary core in the 1D and 2D simulations are in excellent agreement. Fig.\,\ref{fig:1D-out2} shows the $100\times10^3$\,yr evolution approximated 1D and 2D differential Lindblad and corotation torques for the unperturbed disc model. Not only the total disc torques ($\Gamma_\mathrm{2D,disc}$ and $\Gamma_\mathrm{1D,disc}$), but $\Gamma_\mathrm{1D,C}$ and $\Gamma_\mathrm{2D,C}$ are also in good agreement. Comparing the 1D and 2D corotation torques, the difference is less than $3\%$ on average at the end of the $100\times10^3$\,yr long simulation. Thus, we can confirm that the 1D approximation of the corotation torque given by equation (\ref{eq:1D,C}) is quite satisfactory for the unperturbed disc models.

In models that involve viscosity reduction, however, the planetary core migrates in a regime with lower viscosity, and the viscosity in the dead zone is $\alpha=10^{-4}$. Analysing the results of a simulation without viscosity reduction (unperturbed disc), but using $\alpha=10^{-4}$, we found that the migration is faster in this low-viscosity model compared to the $\alpha=10^{-2}$ unperturbed case.

\begin{figure}
	\centering
	\includegraphics[width=\columnwidth]{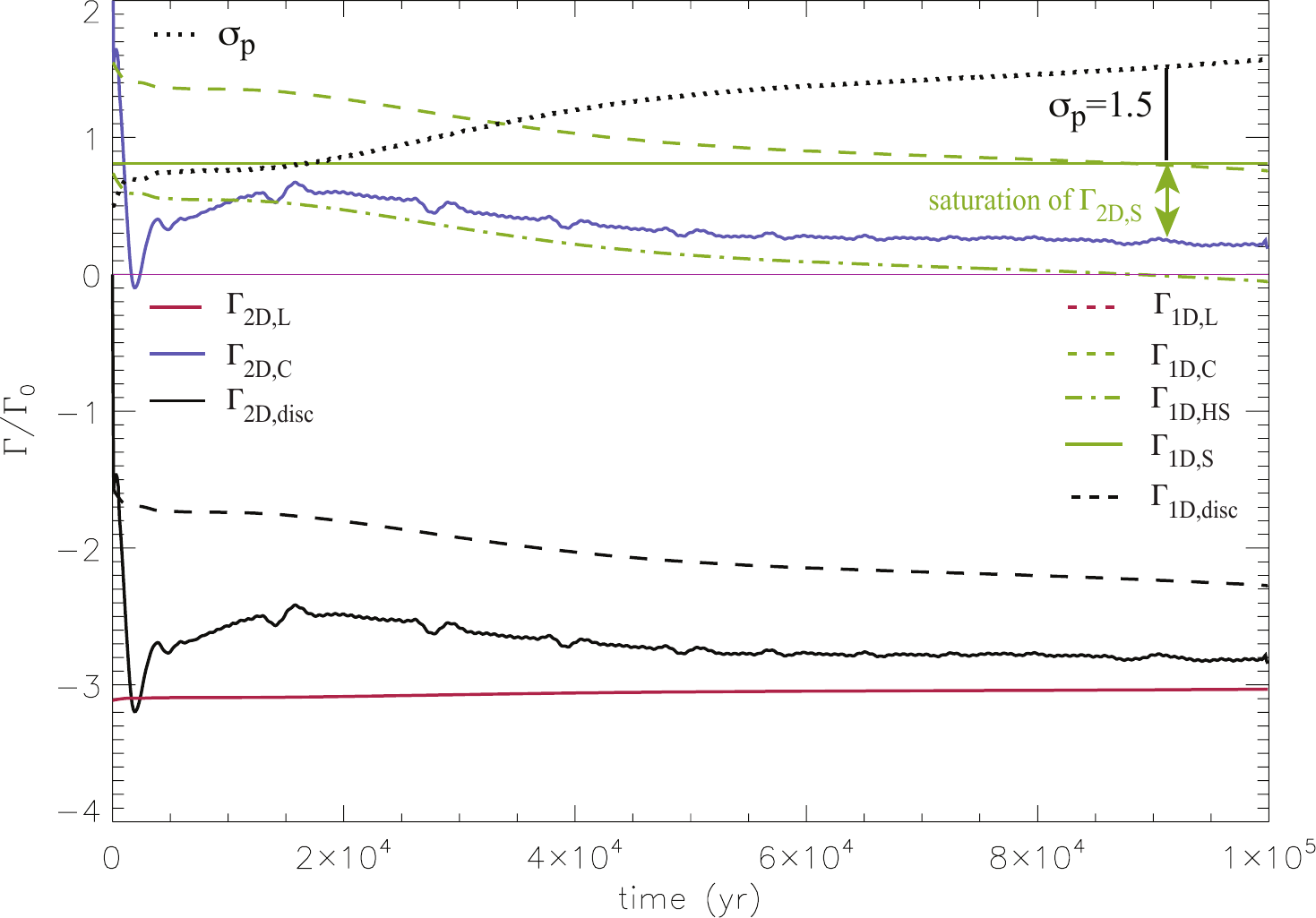}
	\caption{$100\times10^3$\,yr evolution of the approximated 1D differential Lindblad ($\Gamma_\mathrm{1D,L}$), entropy-related corotation ($\Gamma_\mathrm{1D,S}$), vortensity-related horseshoe drag ($\Gamma_\mathrm{1D,HS}$) and the total corotation torque ($\Gamma_\mathrm{1D,C}=\Gamma_\mathrm{1D,S}+\Gamma_\mathrm{1D,HS}$) calculated from the 2D density distribution compared to the appropriate 2D torques ($\Gamma_\mathrm{2D,L}$ and $\Gamma_\mathrm{2D,C}$) in a model where no viscosity reduction is applied and the global viscosity is set to $\alpha=10^{-4}$. The evolution of 1D and 2D net disc torques ($\Gamma_\mathrm{1D,L}+\Gamma_\mathrm{1D,C}$ and $\Gamma_\mathrm{2D,L}+\Gamma_\mathrm{2D,C}$) together with the evolution of the local density slope $\sigma_\mathrm{p}$ are also displayed. The torques are normalized by $\Gamma_0$.}
	\label{fig:1D-out2e}
\end{figure}

Plotting the evolution of the 1D and 2D torques for the low-viscosity model (Fig.\,\ref{fig:1D-out2e}), one can see that the 2D total disc torque is larger than in the corresponding 1D approximation. Moreover, both the 1D and 2D corotation torques are declining, while the differential Lindblad torque remains constant with the same magnitude as in the high-viscosity ($\alpha=10^{-2}$) case. It is also noteworthy that the 1D approximation widely overestimates the magnitude of the corotation torque compared to that of the 2D case. Since the 1D entropy-related corotation torque ($\Gamma_\mathrm{1D,S}$) is larger than the 2D total corotation torque ($\Gamma_\mathrm{2D,C}$), the discrepancy between the 1D and 2D corotation torques can be explained either by that the 1D estimation of vortensiy-related horseshoe drag is completely wrong (it should be negative to lower the corotation torque below the entropy-related corotation torque) or that the entropy-related corotation torque saturates in 2D simulations. 

As $\beta_\mathrm{p}$ is constant in our locally isothermal approximation, the decline of corotation torque must be attributed to the change in the vortensity-related horseshoe drag or the slow saturation of the entropy-related corotation torque. Fig.\,\ref{fig:1D-out2e} also shows the evolution of the local density slope at the planetary radius ($\sigma_\mathrm{p}$). Since $\sigma_\mathrm{p}$ continuously increases, the horseshoe drag declines, and that can explain the decline of total corotation torque. According to equation (\ref{eq:1D,C}), the horseshoe drag should vanish if $\sigma_\mathrm{p}=1.5$. Thus, for this case (shown by an arrow in Fig.\,\ref{fig:1D-out2e}), the 2D corotation torque equals the entropy-related corotation torque, which is much smaller than its 1D estimation. Therefore, we conclude that the entropy-related corotation torque partially saturates due to the low viscosity.

\begin{figure}
	\centering
	\includegraphics[width=\columnwidth]{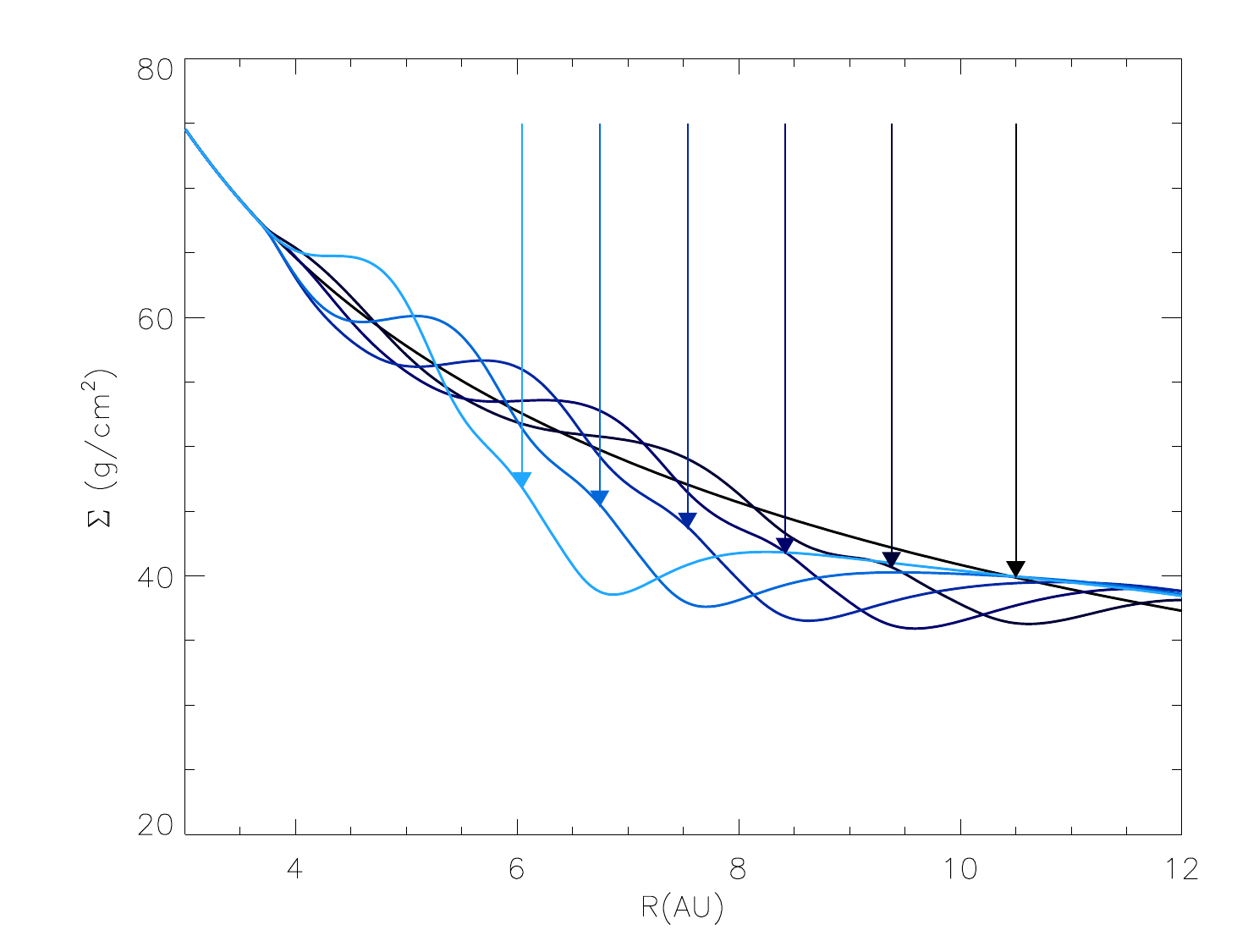}
	\caption{Azimuthally averaged surface mass density profile snapshots with a cadence of $20\times10^3$\,yr in a model where no viscosity reduction is applied and the global viscosity is set to $\alpha=10^{-4}$. The radial positions of the $10\,\mathrm{M_\oplus}$ planet are shown with arrows.} 
	\label{fig:profile}
\end{figure}

\begin{figure}
	\centering
	\includegraphics[width=\columnwidth]{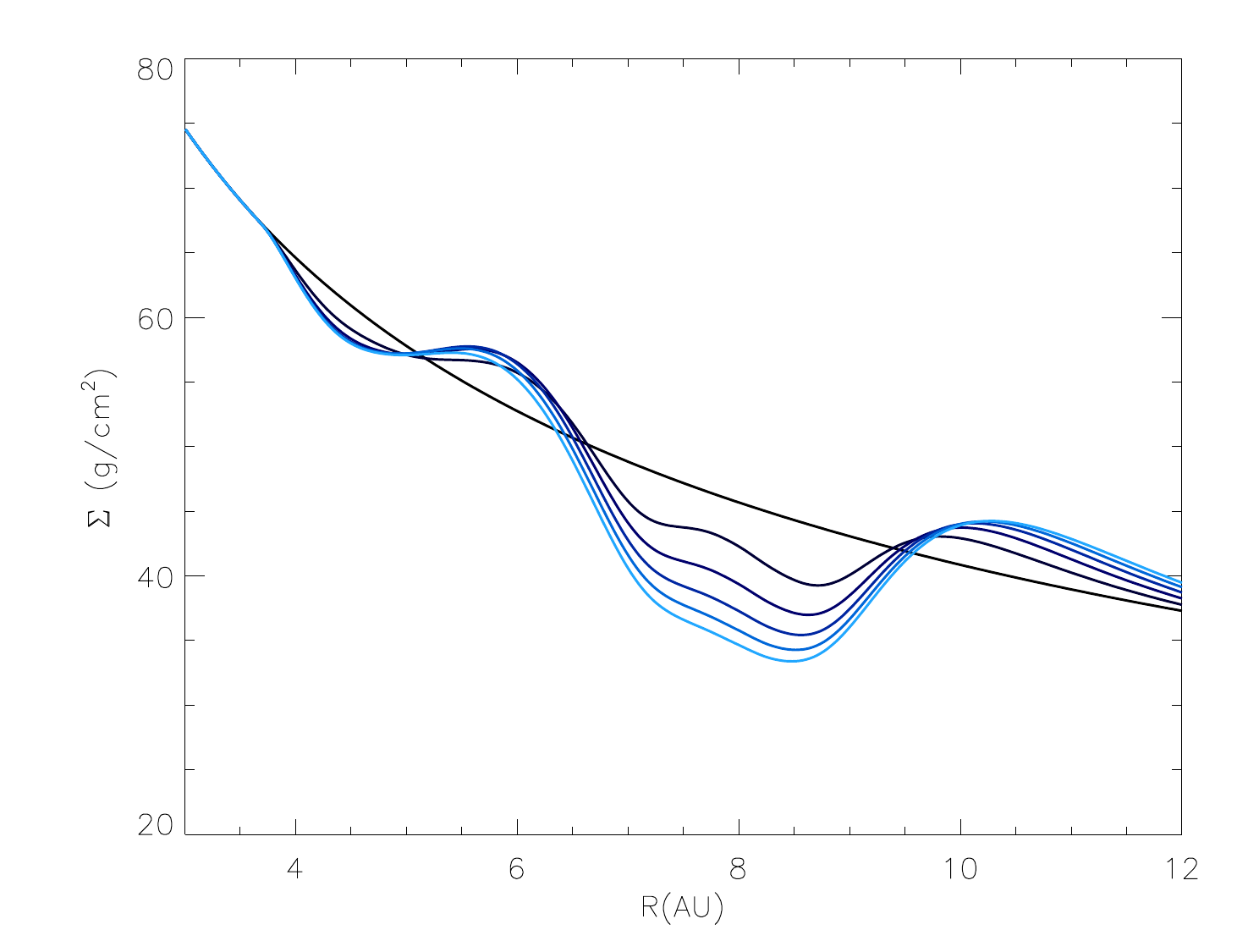}
	\caption{Azimuthally averaged surface mass density profile snapshots with a cadence of $20\times10^3$\,yr in a model where no viscosity reduction is applied and global viscosity is set to $\alpha=10^{-4}$. The planetary mass is $10\,\mathrm{M_\oplus}$ and its orbit is fixed at $R=8\,\mathrm{au}$.}
	\label{fig:plpos-out2h}
\end{figure}

Plotting the azimuthally averaged surface mass density profiles with the cadence of $20\times10^3$\,yr for this particular model (Fig.\,\ref{fig:profile}), one can see that the disc's density distribution is significantly perturbed near the planet. If the angular momentum is not carried away by the spiral waves, which may occur through viscous dissipation, the material inside (outside) the planet loses (gains) angular momentum and recedes from the planet. Consequently, the material appears to be ‘pushed’ away from the location of the planet and a small gap begins to open in the disc. According to \citet{LinPapaloizou1993a}, the viscous criterion for gap formation is
\begin{equation}
	q>\frac{40\nu(a)}{a^2\Omega(a)}.
\end{equation}
Using the $\alpha$ prescription the kinematic viscosity is $\nu(R)=\alpha c_\mathrm{s}^2(R)/\Omega(R)$; in our flat disc approximation the sound speed is $c_\mathrm{s}(R)=\Omega(R)H(R)$. Therefore, a planet opens a gap, if the viscosity is low enough, i.e. for the case in which
\begin{equation}
	\alpha<\frac{q}{h^2 40}.
	\label{eq:viscgapcrit}
\end{equation}
As a consequence, the $10\,\mathrm{M_\oplus}$ planetary core is able to open a gap, if $\alpha<3\times10^{-4}$, which condition is satisfied in our models that include dead zone. We run an additional simulation with constant $\alpha=10^{-4}$ global viscosity, in which the $10\,\mathrm{M_\oplus}$ planet orbiting at $R=8$\,au is not allowed to migrate. According to this simulation, the planet indeed opens a partial gap (see Fig.\,\ref{fig:plpos-out2h}). 

\begin{figure}
	\centering
	\includegraphics[width=\columnwidth]{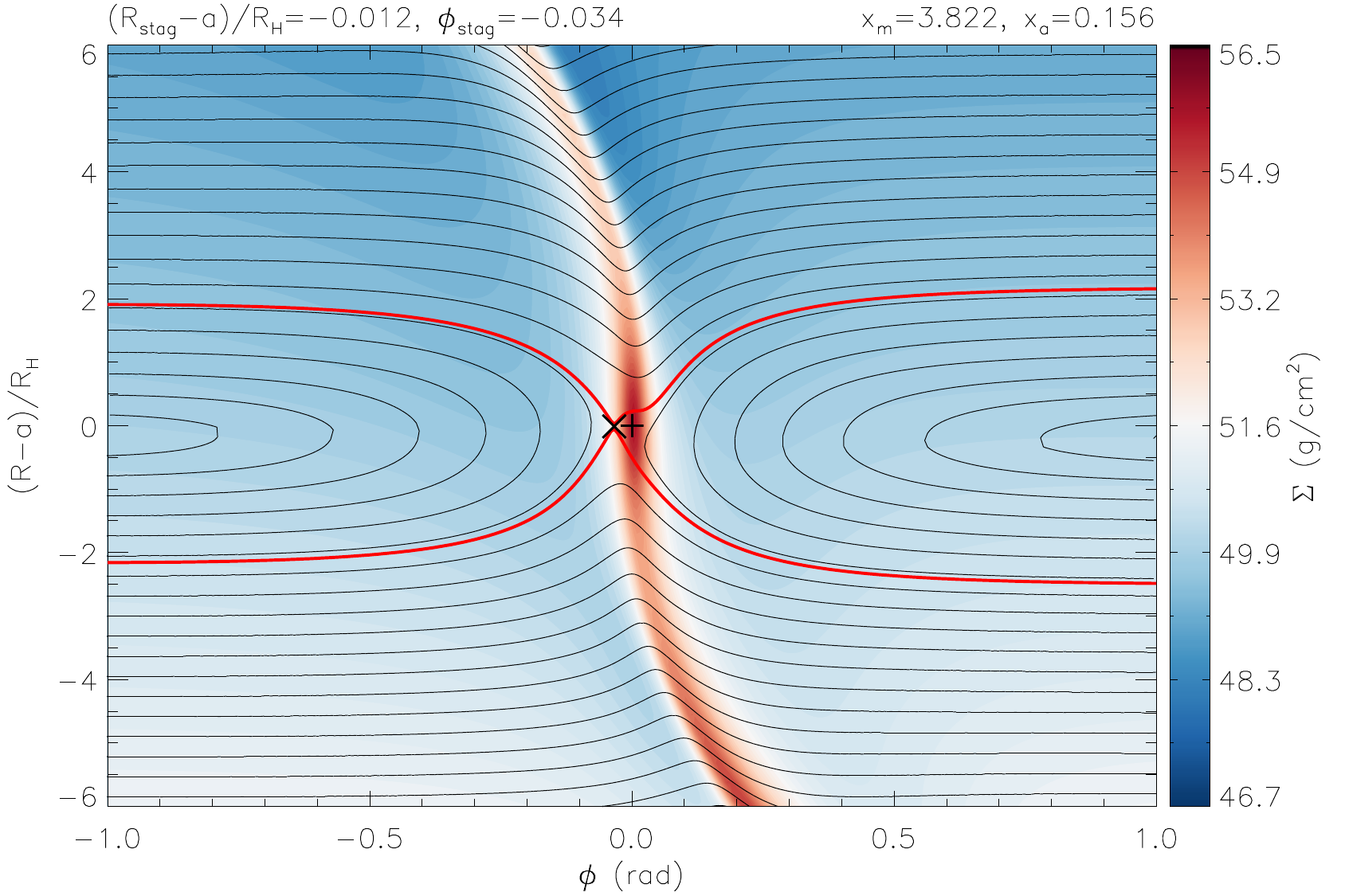}
	\includegraphics[width=\columnwidth]{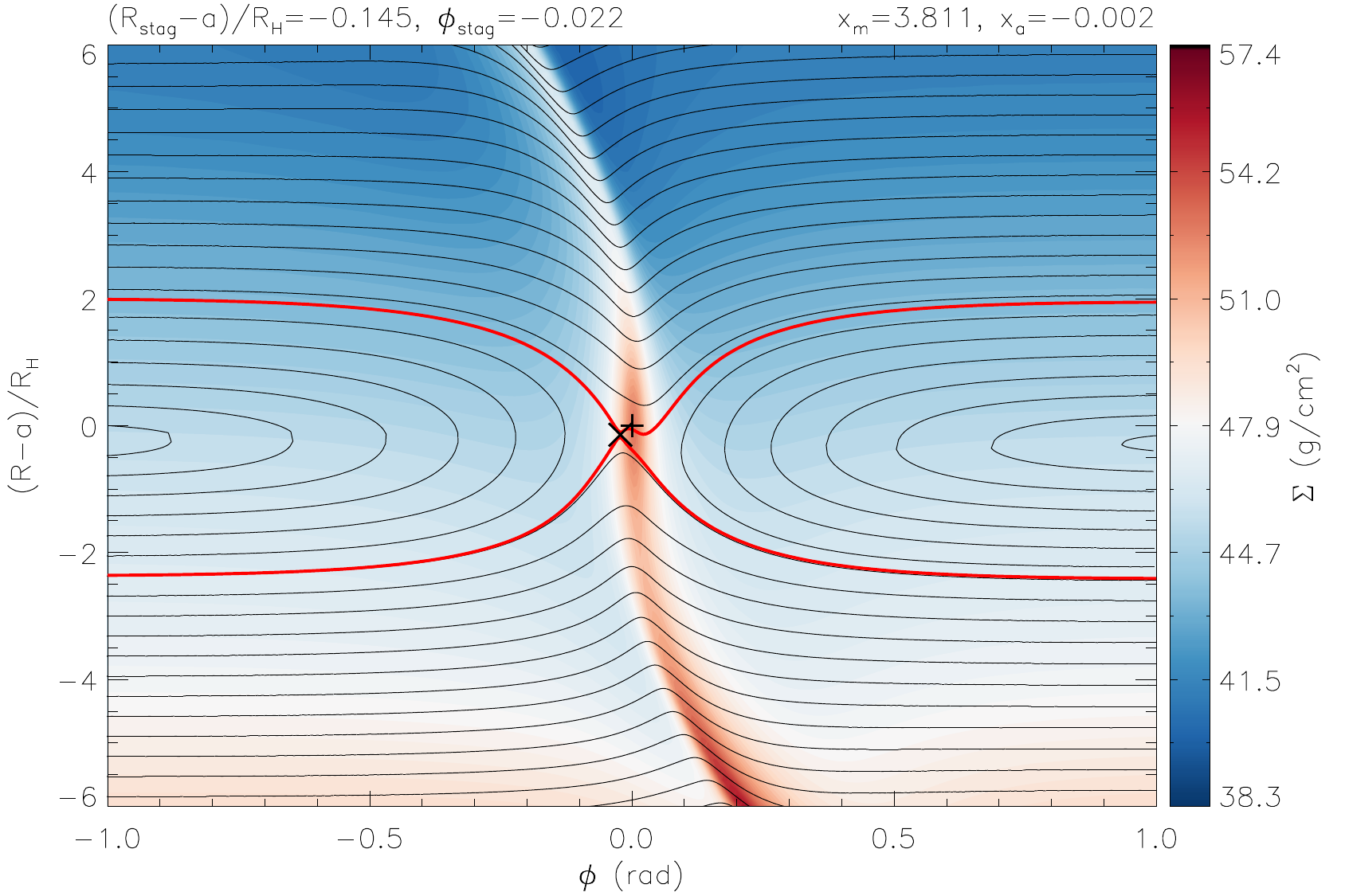}

	\caption{Streamlines (black curves) and the horseshoe separatrix (red curves) in the co-orbital region of  planet overplotted by the gas density distribution for models with constant viscosity of $\alpha=10^{-2}$ (upper panel) and $\alpha=10^{-4}$ (lower panel). The position of planet and the stagnation point are denoted by $+$ and $\times$, respectively.}
	\label{fig:streamout2-2e}
\end{figure}

Although the corotation (therefore the vortensity-related horseshoe drag and the entropy-related corotation torque) cannot be separated geometrically from the differential Lindbald torque in 2D simulations, some qualitative conclusions can be drawn by analysing the horseshoe dynamics. Fig.\,\ref{fig:streamout2-2e} shows the gas streamlines in the horseshoe region of planet overplotted by the surface mass density distribution for models $\alpha=10^{-2}$ (upper panel) and $\alpha=10^{-4}$ (lower panel) at $t=100\times10^3$ and $65\times10^3$\,yr, when the planet orbits at the same radial distance $a=7.3$\,au. The vortensity-related horseshoe drag emerges due to the front--rear asymmetry (with respect to planetary azimuthal position) of the horseshoe region defined by
\begin{equation}
	x_\mathrm{a}=\frac{1}{2}\frac{(x_s^F-x_s^R)}{x_s^F+x_s^R},
	\label{eq:asym}
\end{equation}
where $x_s^F$ and $x_s^R$ are the front and rear horseshoe width measured at azimuth $\phi=\pm1$, respectively. The mean horseshoe width is $x_\mathrm{m}\simeq0.68$\,au for both models; however, the horseshoe asymmetry is nearly vanishing for the low-viscosity models: $x_\mathrm{a}=0.15$ and $x_\mathrm{a}=0.002$ for models $\alpha=10^{-2}$ and $\alpha=10^{-4}$, respectively. Consequently, the vortensity-related horseshoe drag, and therefore the corotation torque, is smaller for the low-viscosity model, which results in more negative net disc torque, i.e. faster inward migration of the planetary core.

Summarizing the above findings, the $10\,\mathrm{M_\oplus}$ planet partially opens a gap around its orbit and the local density slope continuously increases resulting in a decrease of the horseshoe drag for models in which the planet orbits in a low viscosity $\alpha=10^{-4}$ region. As a result, the migration of  the $10M_\uplus$ planetary core is faster in models that include a dead zone than in the conventional Type I regime. Since the gap opening is only partial, the migration is neither in the Type II nor in the Type III regime, as it would require significant co-orbital mass deficit, which is absent because the planetary core has sub-Saturnian mass \citep{MassetPapaloizou2003}. As a conclusion, the 1D analytical estimation of the corotation torque given by equation (\ref{eq:1D,C}) is inadequate for a $10\,\mathrm{M_\oplus}$ migrating planet in a low-viscosity $\alpha=10^{-4}$ regime. The reason of this is that either the entropy-related corotation torque almost saturates due to low viscosity or the the back reaction of the planet on to the disc is neglected.

\subsubsection{Models with a blunt dead zone edge}

\begin{figure}
	\centering
	\includegraphics[width=\columnwidth]{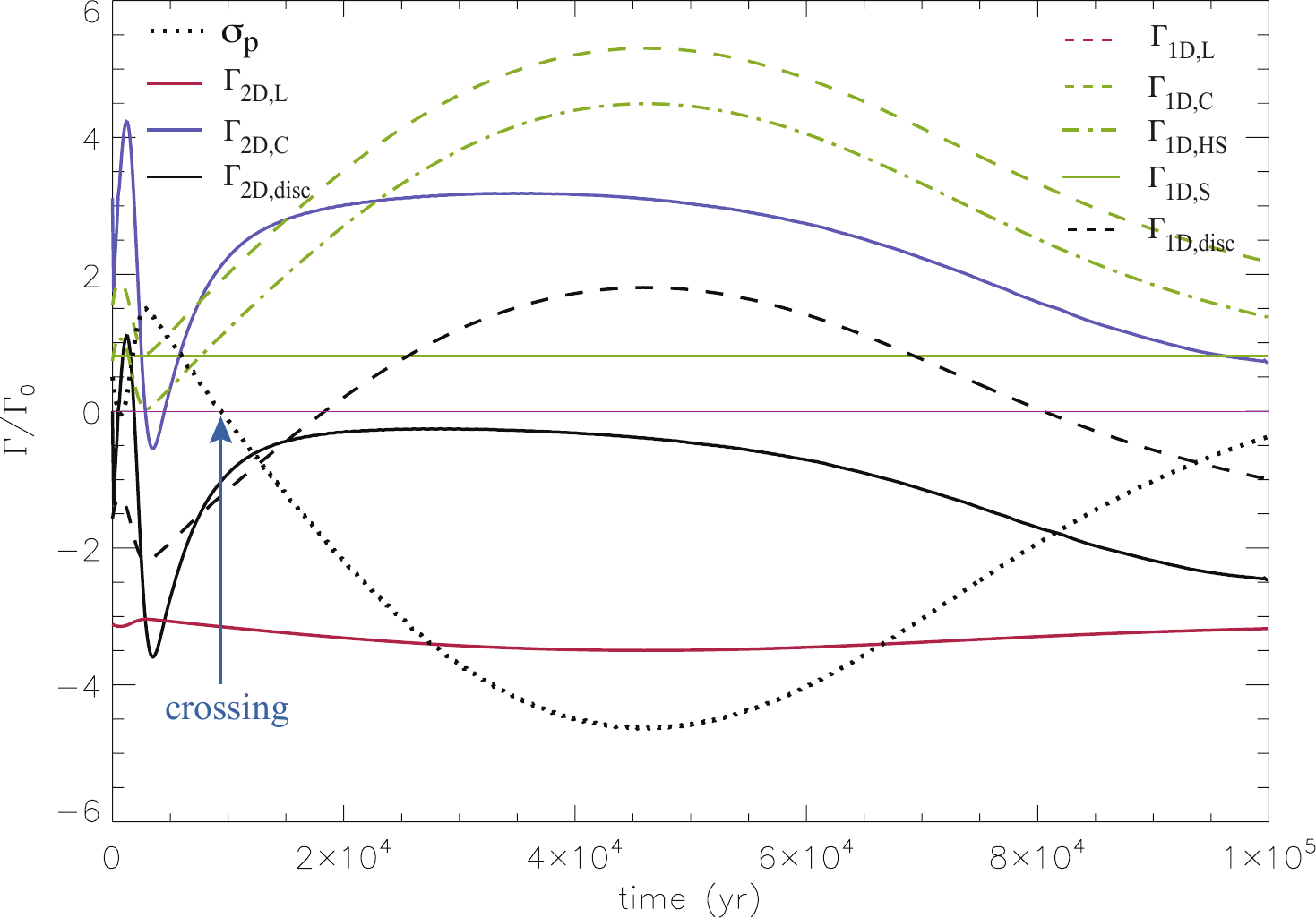}
	\includegraphics[width=\columnwidth]{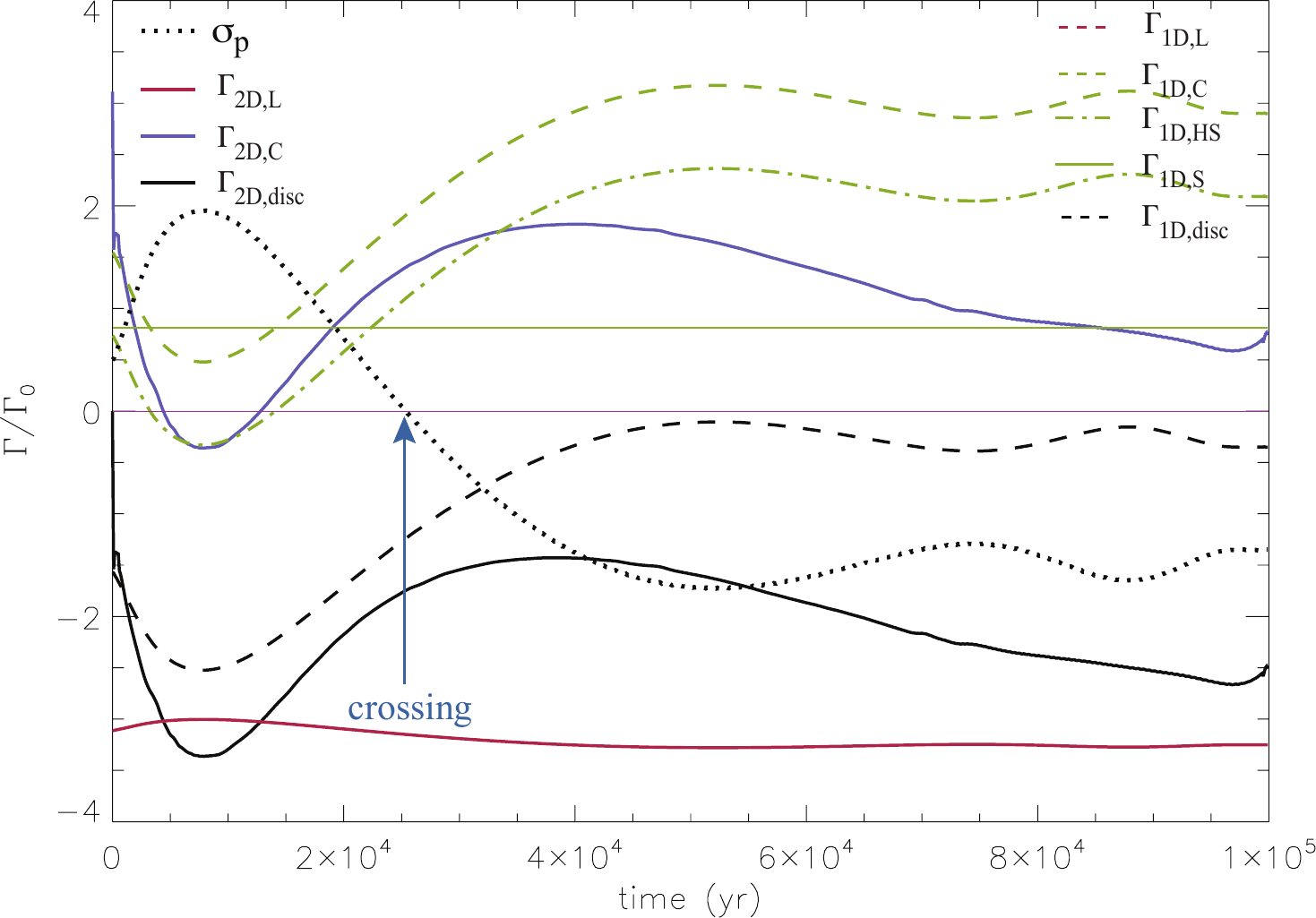}
	\caption{$100\times10^3$\,yr evolution of the approximated 1D differential Lindblad ($\Gamma_\mathrm{1D,L}$), the vortensity-related horseshoe drag ($\Gamma_\mathrm{1D,HS}$) and entropy-related corotation ($\Gamma_\mathrm{1D,S}$) torques calculated from the 2D density distribution compared to the appropriate 2D torques ($\Gamma_\mathrm{2D,L}$ and $\Gamma_\mathrm{2D,C}$) for models $\Delta R_\mathrm{dze}=2H_\mathrm{dze}$ (upper panel) and $\Delta R_\mathrm{dze}=4H_\mathrm{dze}$ (lower panel). The evolution of net disc torques ($\Gamma_\mathrm{1D,L}+\Gamma_\mathrm{1D,C}$ and $\Gamma_\mathrm{2D,L}+\Gamma_\mathrm{2D,C}$) together with the evolution of the local density slope $\sigma_\mathrm{p}$ are also displayed. The torques are normalized by $\Gamma_0$.}
	\label{fig:1D-out5a6}
\end{figure}

Now let us investigate the blunt viscosity reduction models, where contrary to the 1D simulations, no trapping happened in our 2D simulations. Fig.\,\ref{fig:1D-out5a6} (lower panel) shows the evolution of the approximated 1D and 2D torques for the model with $\Delta R_\mathrm{dze}=4H_\mathrm{dze}$, where the fastest inward migration of the $10\,\mathrm{M_\oplus}$ planetary core has been observed. It can be seen that the magnitude of net disc torque is significantly overestimated by the 1D approximation. From the point, where the planet passes through the density maximum, the discrepancy between 1D and 2D net disc torques increases. It is also noticeable that the magnitude of the 2D corotation torque is well below that of the 1D approximation after the crossing point. The differential Lindblad torque slightly changes its value after the planet crosses the density jump; however, its contribution to the change in net disc torque is negligible. Since the total disc torque given by the 1D approximation nearly vanishes after $\sim50\times10^3$\,yr, according to the 1D torque approximation, the planet should be trapped at a radius around $\sim7$\,au.

The same analysis as described above can be done for the model with $\Delta R_\mathrm{dze}=2H_\mathrm{dze}$, see Fig.\,\ref{fig:1D-out5a6} (upper panel). Here a similar over- and underestimation of the net disc torque and corotation torques can be observed after the planet passes through the density maximum at $t\simeq10\times10^3$\,yr. The error of the corotation torque in the 1D approximation is much higher than in the previous case; moreover, even the sign of net disc torque is not given correctly by the 1D approximation in the time interval $20\times10^3\,\mathrm{yr}\leq t\leq80\times10^3\,\mathrm{yr}$. 

\begin{figure}
	\centering
	\includegraphics[width=\columnwidth]{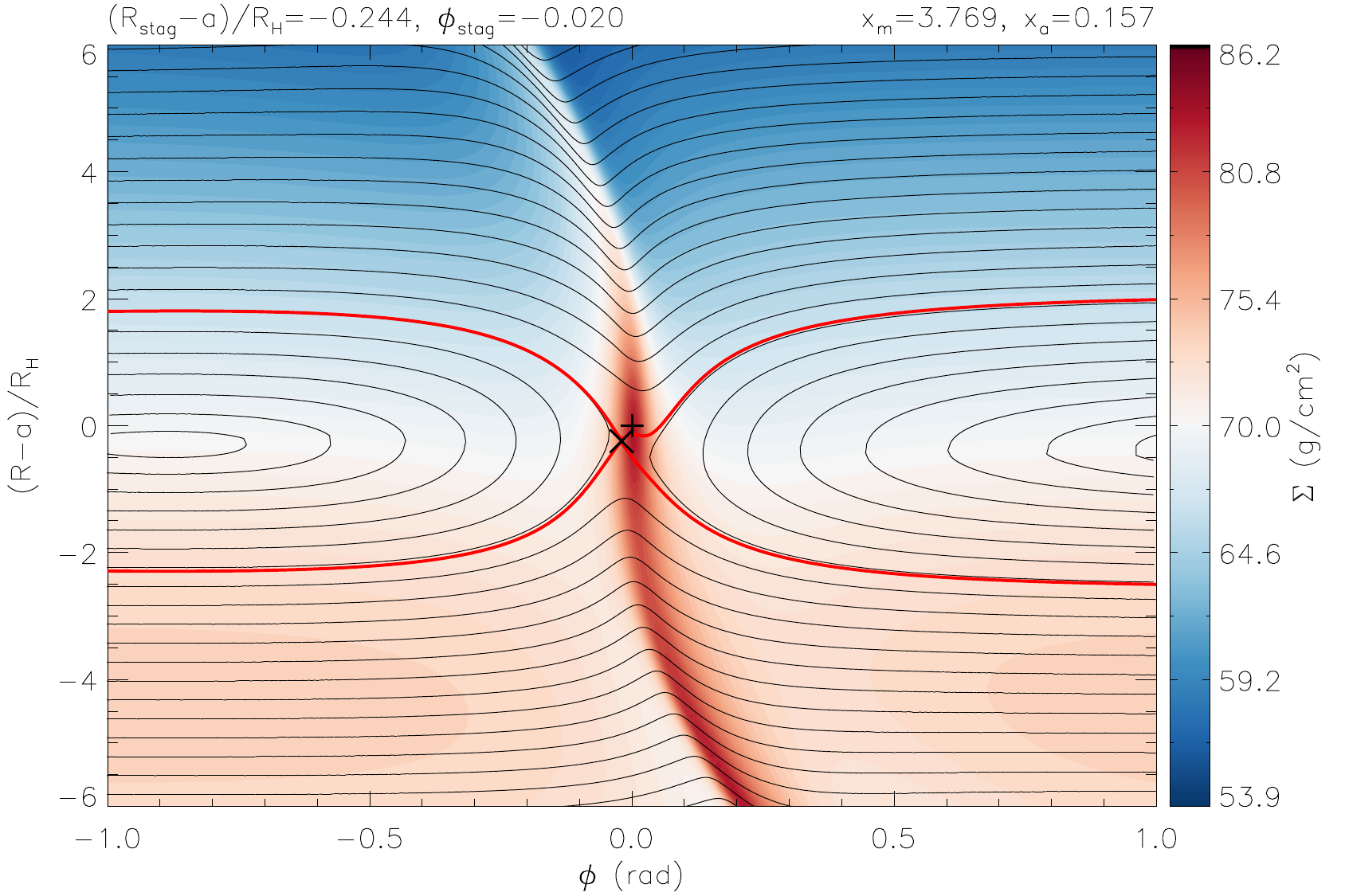}
	\includegraphics[width=\columnwidth]{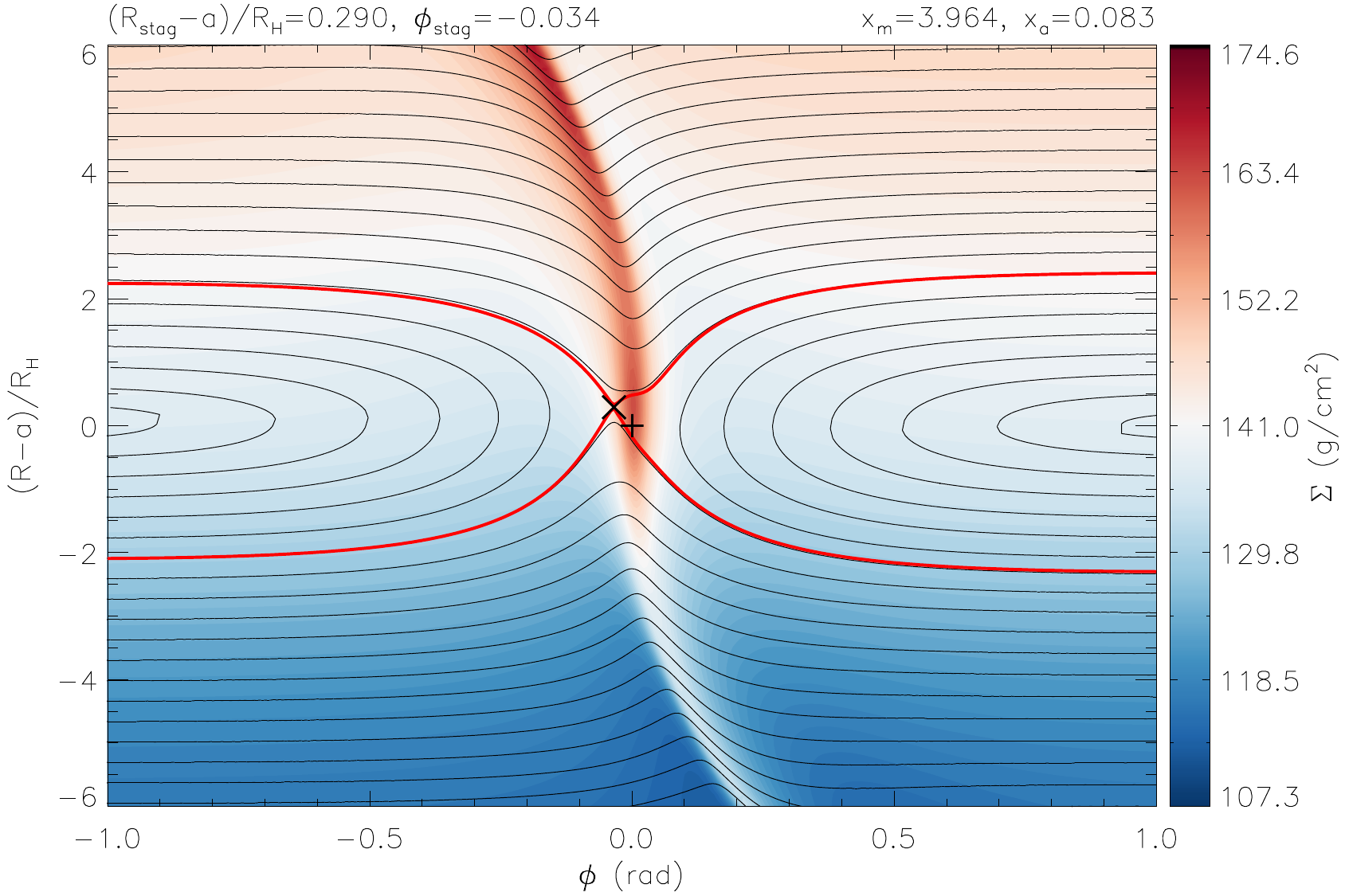}

	\caption{Streamlines (black curves) and the horseshoe separatrix (red curves) in the co-orbital region of planet overplotted by the gas density distribution for model $\Delta R_\mathrm{dze}=4H_\mathrm{dze}$ at $t=10\times10^3$\,yr (upper panel) and $t=40\times10^3$\,yr (lower panel). The position of planet and the stagnation point are denoted by $+$ and $\times$, respectively.}
	\label{fig:streamout6}
\end{figure}

Fig.\,\ref{fig:streamout6} shows the horseshoe region of the planet for the model with $\Delta R_\mathrm{dze}=4H_\mathrm{dze}$ at $t=10\times10^3$ and $40\times10^3$\,yr, when the planet is beyond and inside the density maximum, respectively. The  stagnation point of the horseshoe separatrices is beyond (inside) the planetary orbit before (after) the crossing of density maximum. Moreover, the width of the horseshoe region decreases after the planet crossed the density maximum: the front--rear asymmetry of the horseshoe region is $x_\mathrm{a}=0.092$ and $0.082$ before and after the crossing, respectively. Since the horseshoe asymmetry is continuously decreasing as the planet passes the density maximum, the magnitude of the corotation torque also decreases. As a consequence, the planet feels an increasing total negative net torque (in absolute value), and its inward migration accelerates, thus, no trapping occurs. The same decline in horseshoe asymmetry, but with smaller average asymmetry can be measured in the model where $\Delta R_\mathrm{dze}=2H_\mathrm{dze}$ ($x_\mathrm{a}=0.03$ and $0.013$ before and after the crossing of density maximum, respectively).

To summarize the above findings, we found that the 1D approximation overestimates the magnitude of the positive corotation torque (especially the vortensity-related horseshoe drag) for blunt viscosity reduction models. According to the streamline analysis, this can be attributed to the difference in horseshoe dynamics compared to the unperturbed models. Namely, the horseshoe asymmetry is continuously decreasing after the planet crosses the density maximum in the blunt dead zone edge models. Taking into account that the differential Lindblad torque remains nearly constant, and the vortensity-related horseshoe drag decreases after the crossing, the planet is not trapped, rather it migrates inwards with a higher migration rate than in the conventional Type I regime.

\subsubsection{Models with a sharp dead zone edge}
\label{sect:discussion-sharp}

\begin{figure}
	\centering
	\includegraphics[width=\columnwidth]{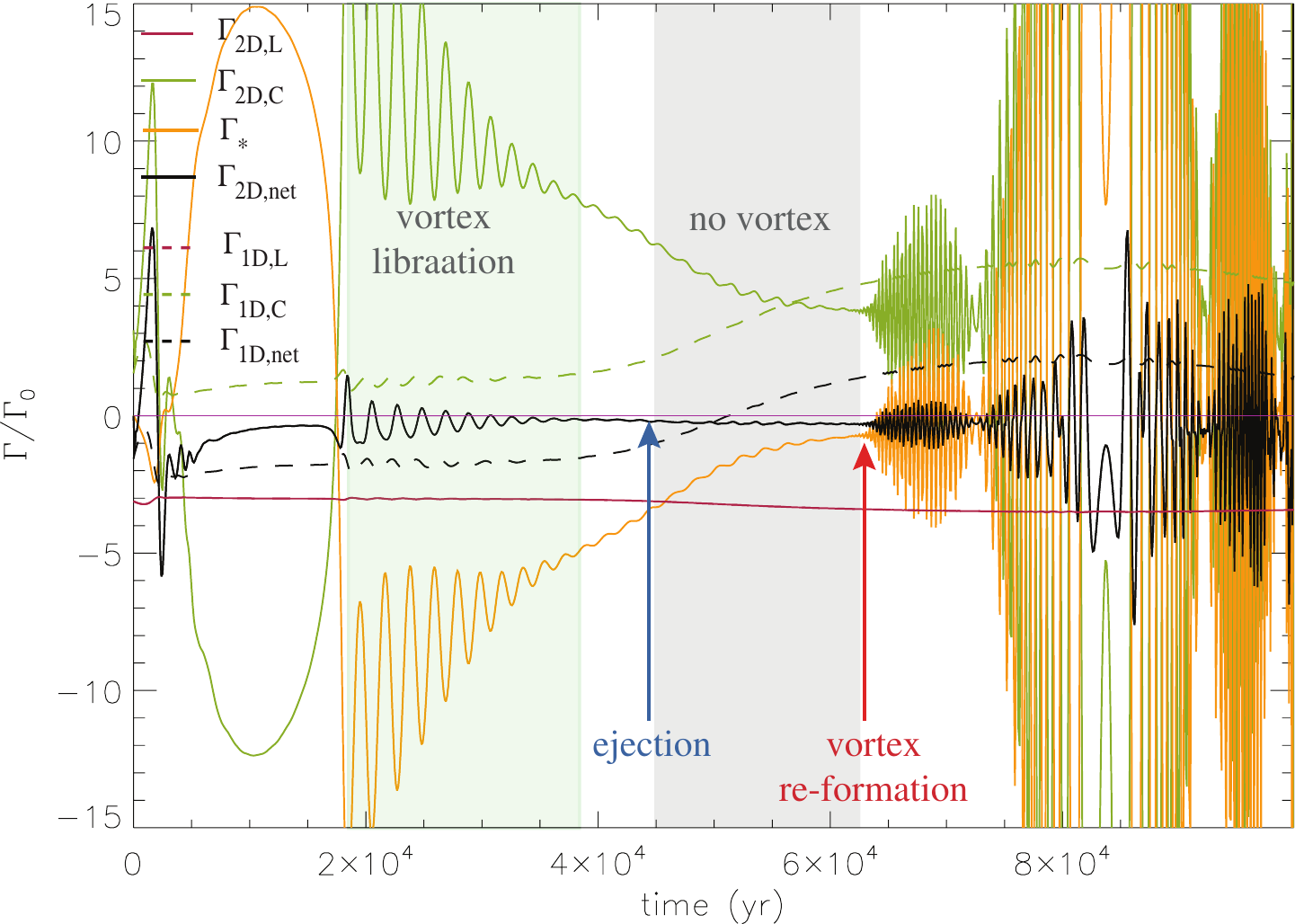}

	\caption{Approximated 1D differential Lindblad ($\Gamma_\mathrm{1D,L}$) and corotation ($\Gamma_\mathrm{1D,C}$) torques calculated by the 2D density distribution compared to the corresponding 2D torques ($\Gamma_\mathrm{2D,L}$ and $\Gamma_\mathrm{2D,C}$) for model $\Delta R_\mathrm{dze}=1.5H_\mathrm{dze}$. The stellar ($\Gamma_*$) and net ($\Gamma_\mathrm{2D,net}=\Gamma_\mathrm{2D,L}+\Gamma_\mathrm{2D,C}+\Gamma_\mathrm{*}$) torques are also shown. The torques are normalized by $\Gamma_0$.}
	\label{fig:t1d-out7b}
\end{figure}

In what follows, we will analyse sharp viscosity reduction models, in which (at least) a temporary trapping mechanism happens. In models with $\Delta R_\mathrm{dze}\leq1.5H_\mathrm{dze}$, a significant stellar torque ($\Gamma_*$) is observed due to the barycentre shift caused by the non-axisymmetric density distribution; thus, it must also be included in the torque analysis. Fig.\,\ref{fig:t1d-out7b} shows the evolution of the approximated 1D and 2D torques as well as the stellar torque for model $\Delta R_\mathrm{dze}=1.5H_\mathrm{dze}$. As one can see, the 1D approximation of the corotation torque is even worse when compared to the blunt viscosity reduction models. Before $t\simeq17\times10^3$\,yr, $\Gamma_\mathrm{1D,C}$ and $\Gamma_\mathrm{2D,C}$ differ in sign. Before (after) the ejection at $t\simeq45\times10^3$\,yr, $\Gamma_\mathrm{1D,C}$ is underestimated (overestimated) compared to $\Gamma_\mathrm{2D,C}$. $\Gamma_\mathrm{2D,C}$ and $\Gamma_\mathrm{*}$ are always opposite in sign. Moreover, $\Gamma_\mathrm{2D,C}$ becomes oscillating together with $\Gamma_\mathrm{*}$ from $t\simeq18\times10^3$\,yr. Since $\Gamma_\mathrm{2D,L}$ remains nearly constant, we assume that both the corotation and stellar torques play the crucial role in the migration history and the temporary trapping.

\begin{figure}
	\centering
	\includegraphics[width=\columnwidth]{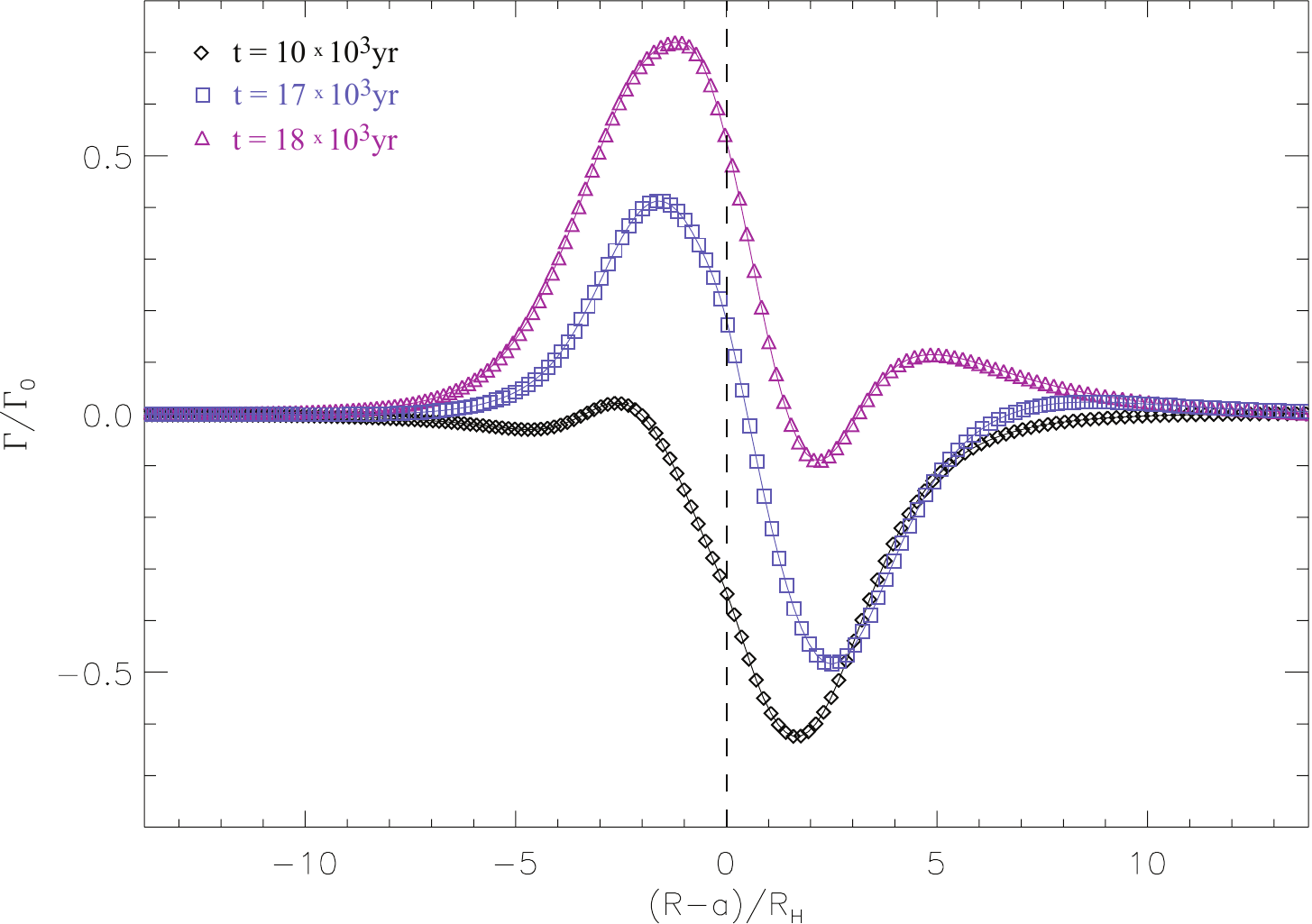}
	\caption{Radial torque density profile in model $\Delta R_\mathrm{dze}=1.5H_\mathrm{dze}$ for three different phases of simulation, where $\phi_\mathrm{max}<180^\circ$ ($t=10\times10^3$\,yr), $\phi_\mathrm{max}=180^\circ$ ($t=18\times10^3$\,yr), and $\phi_\mathrm{max}>180^\circ$ ($t=60\times10^3$\,yr).}
	\label{fig:tp-out7b}
\end{figure}

Fig.\,\ref{fig:tp-out7b} shows the normalized radial torque density profile of the disc (no stellar torque included) calculated as
\begin{equation}
	\frac{d\Gamma_\mathrm{disc}}{dR}/\Gamma_0=\sum_{j=1}^{N_\phi}\frac{A_{i,j}\Sigma_{i,j}\left(1-\exp{[d_{i,j}/R_\mathrm{H}]^2} \right)}{\left(d_{i,j}+\epsilon H(a)\right)^3}(y_{i,j}-a) a,
	\label{eq:disc_torque_profile}
\end{equation}
for model $\Delta R_\mathrm{dze}=1.5H_\mathrm{dze}$ at three different phases before the ejection: when the corotation torque reaches its smallest value (at $t=10\times10^3$\,yr), its largest value (at $t=18\times10^3$\,yr) and when it vanishes (at $t=17\times10^3$\,yr). The disc torque density profile vanishes beyond $(R-a)/R_\mathrm{H}=\pm10$. The inner(outer) disc torque is weak(strong) at $t=10\times10^3$\,yr, when the vortex is at the rear of the planet. At this time the corotation torque has a large negative magnitude, while the stellar torque has a large positive magnitude. However, at $t=18\times10^3$\,yr, when the vortex is at the front of the planet, the inner(outer) disc torque is strong(weak). At this time the corotation torque has a large positive magnitude, while the stellar torque has a large negative magnitude. The torque density profile shows nearly symmetric inner and outer disc contribution, when the vortex and the planet are in the opposite side of the disc (the azimuth of the vortex centre is $\phi=180^\circ$) at $t=17\times10^3$\,yr. At this time the  stellar torque temporarily vanishes, while the corotation torque has a small but positive value.

\begin{figure}
	\centering
	\includegraphics[width=\columnwidth]{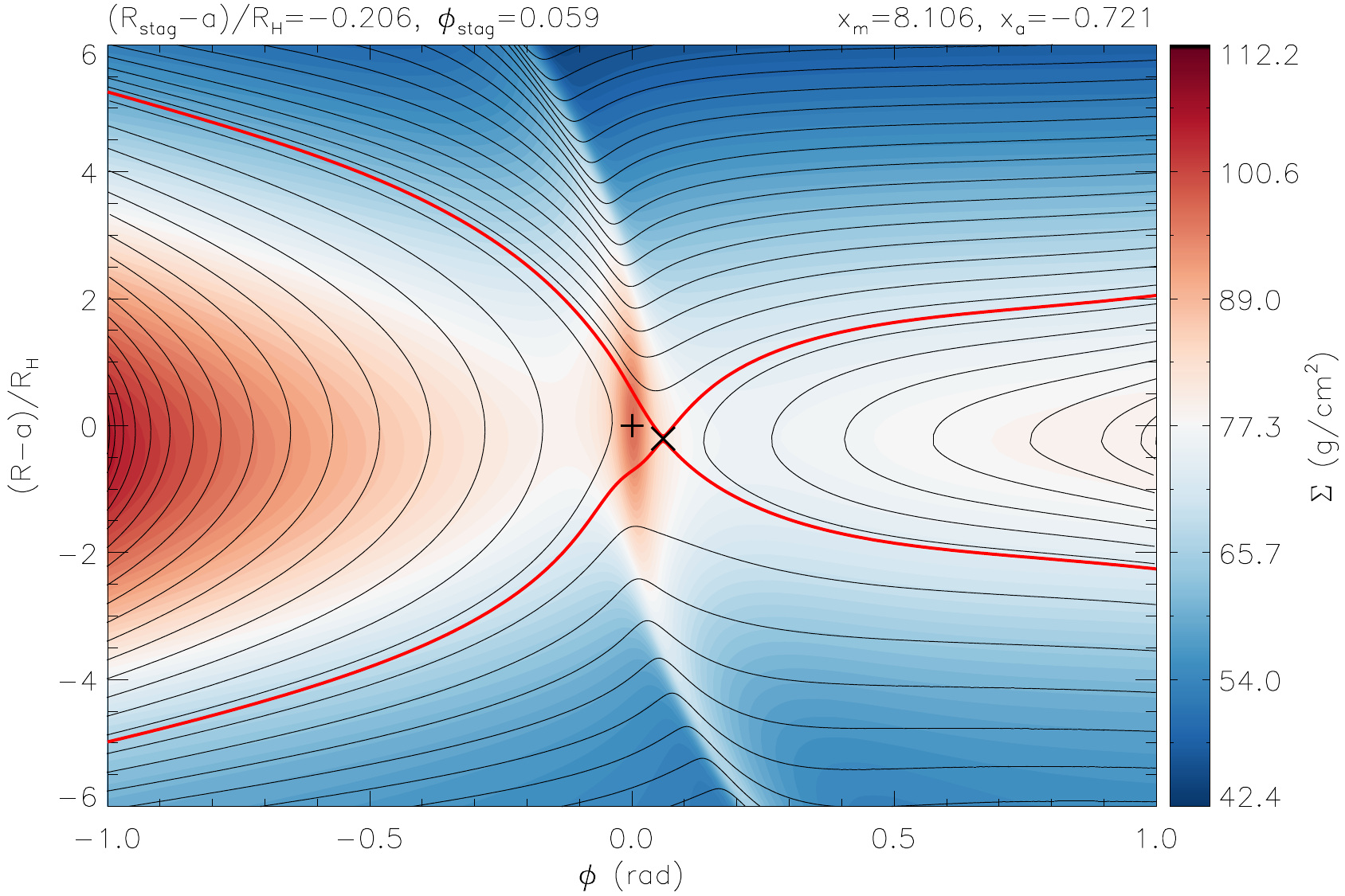}
	\includegraphics[width=\columnwidth]{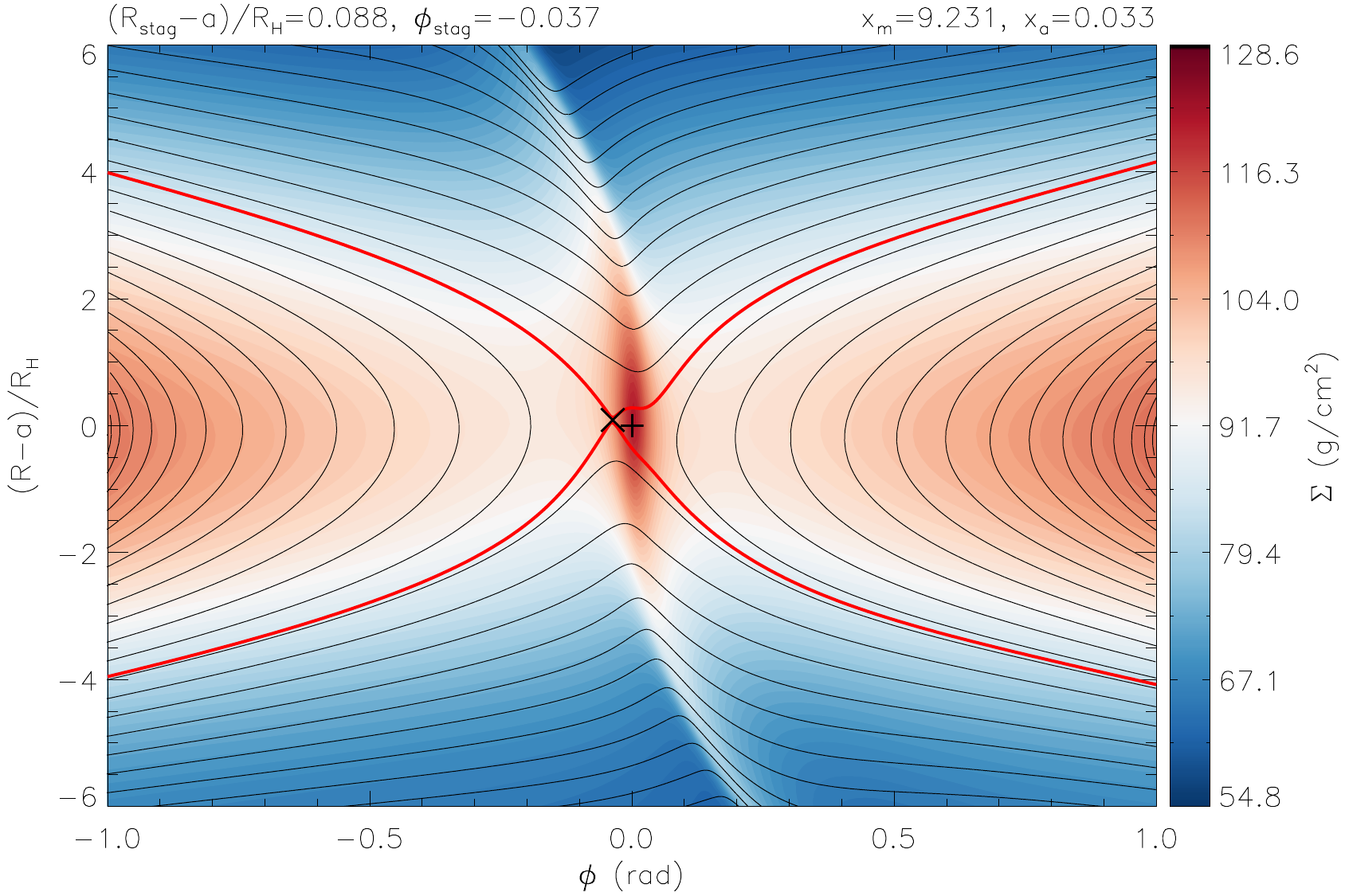}
	\includegraphics[width=\columnwidth]{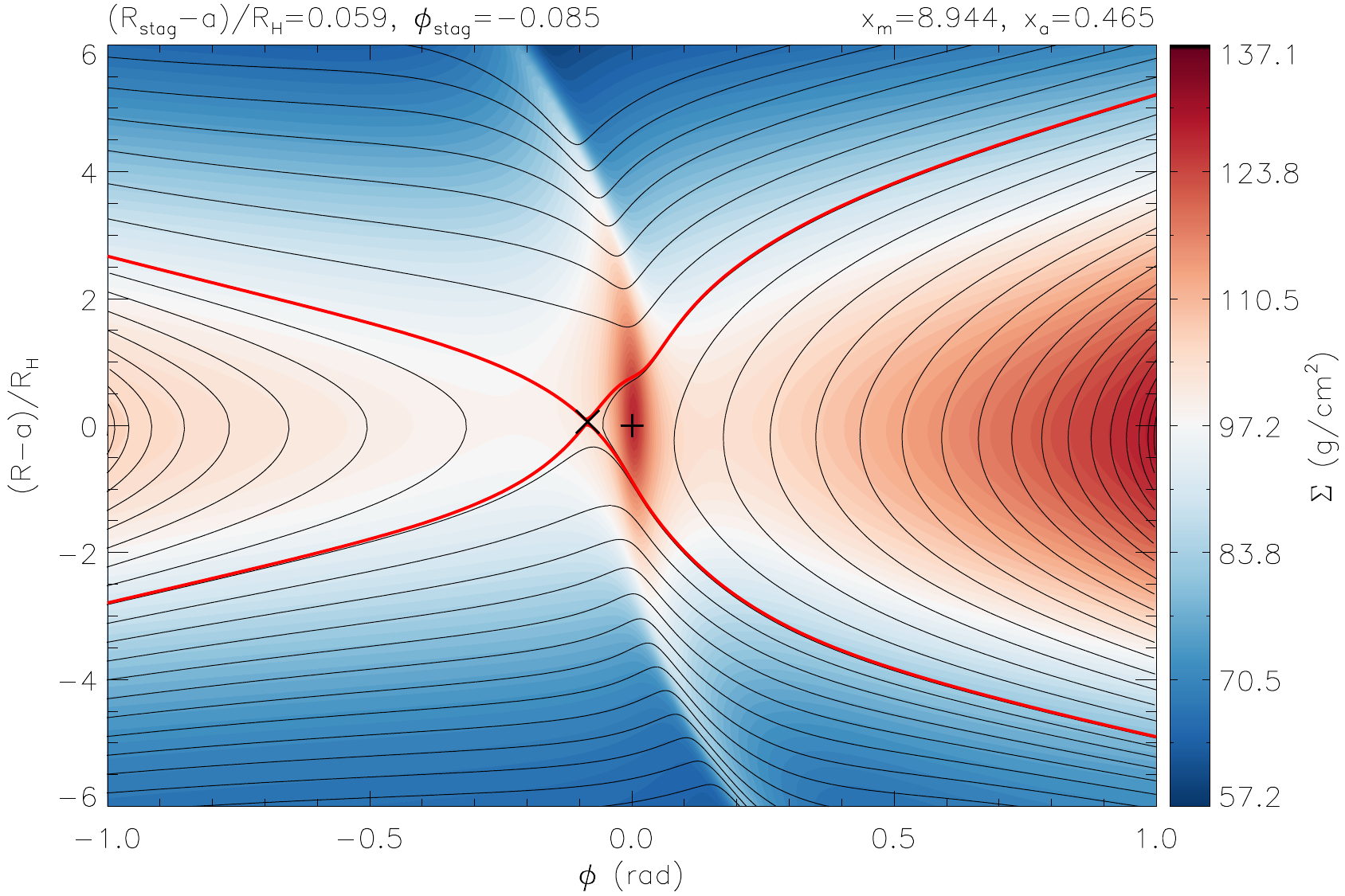}
	\caption{Streamlines (black curves) and the horseshoe separatrix (red curves) in the co-orbital region of  planet overplotted by the gas density distribution for model $\Delta R_\mathrm{dze}=1.5H_\mathrm{dze}$ at three different phases, before the ejection of planet: at $t=10\times10^3$\,yr (upper panel), at $t=17\times10^3$\,yr (middle panel) and at $t=18\times10^3$\,yr (lower panel). The position of planet and the stagnation point are denoted by $+$ and $\times$, respectively.}
	\label{fig:streamout7b}
\end{figure}

Fig.\,\ref{fig:streamout7b} shows the streamlines around the planet at the above-mentioned three phases. As one can see, the horseshoe region is extremely large ($x_\mathrm{m}\simeq1.6\textrm{--}1.7$\,au) compared to the previous (blunt viscosity reduction or no viscosity reduction) cases. Moreover, the front-rear asymmetry of the horseshoe region is $x_\mathrm{a}=-0.72$ and $0.46$ before and after the corotation torque changes its sign and nearly vanishes and $x_\mathrm{a}=0.03$ when the vortex is at azimuth $\phi=180^\circ$. Since the vortex centre is far from the planet ($|\phi|>\upi/2$, see Fig.\,\ref{fig:dens-max-a}), the material accumulated inside does not cause considerable excess torque. Note that the radially averaged disc torque profile vanishes in the azimuthal range $\upi/2\geq\phi\geq-\upi/2$. Therefore, the change in the sign of the corotation torque can be associated with the change in the sign of the front--rear horseshoe asymmetry.

\begin{figure}
	\centering
	\includegraphics[width=\columnwidth]{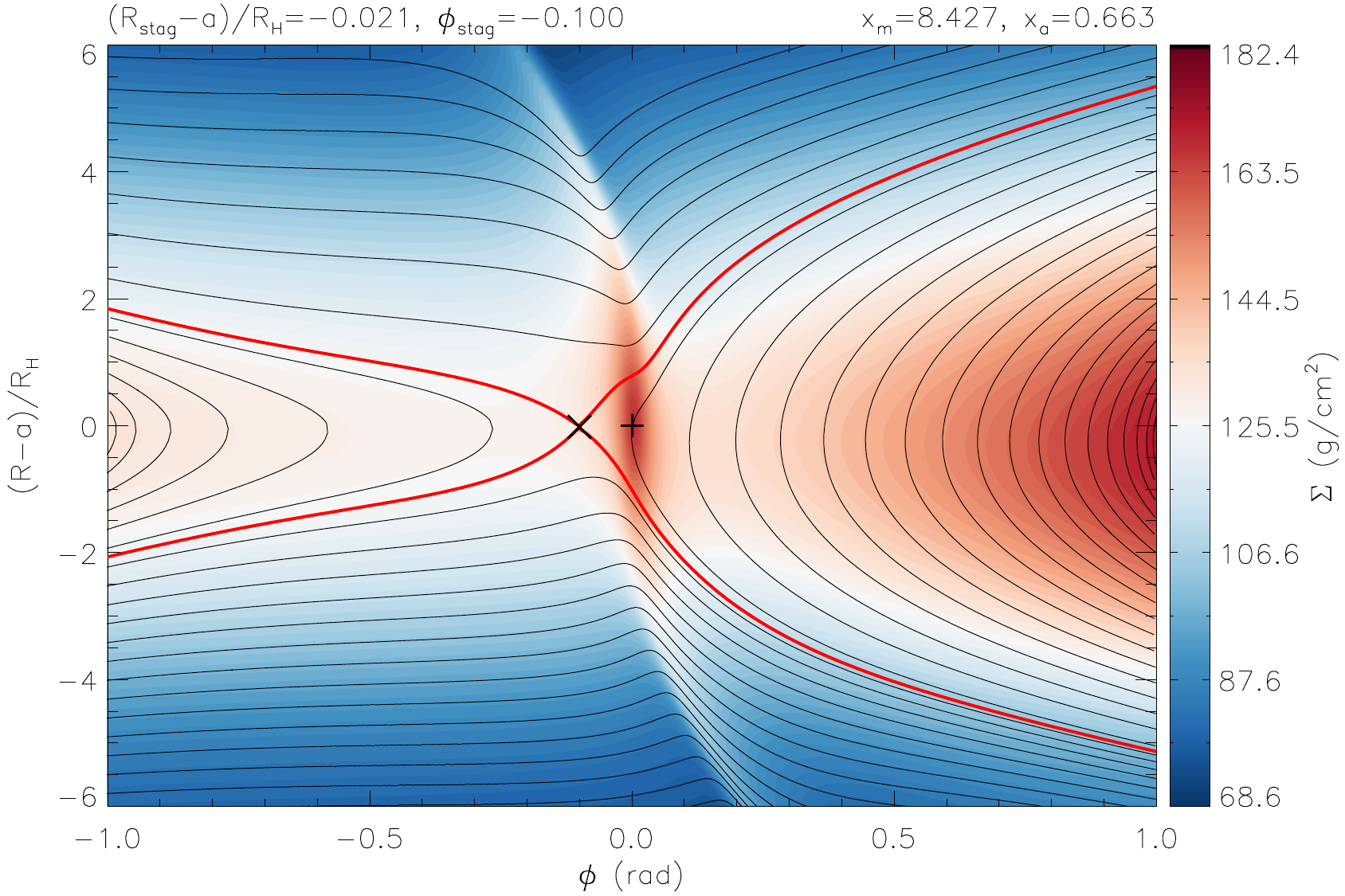}
	\includegraphics[width=\columnwidth]{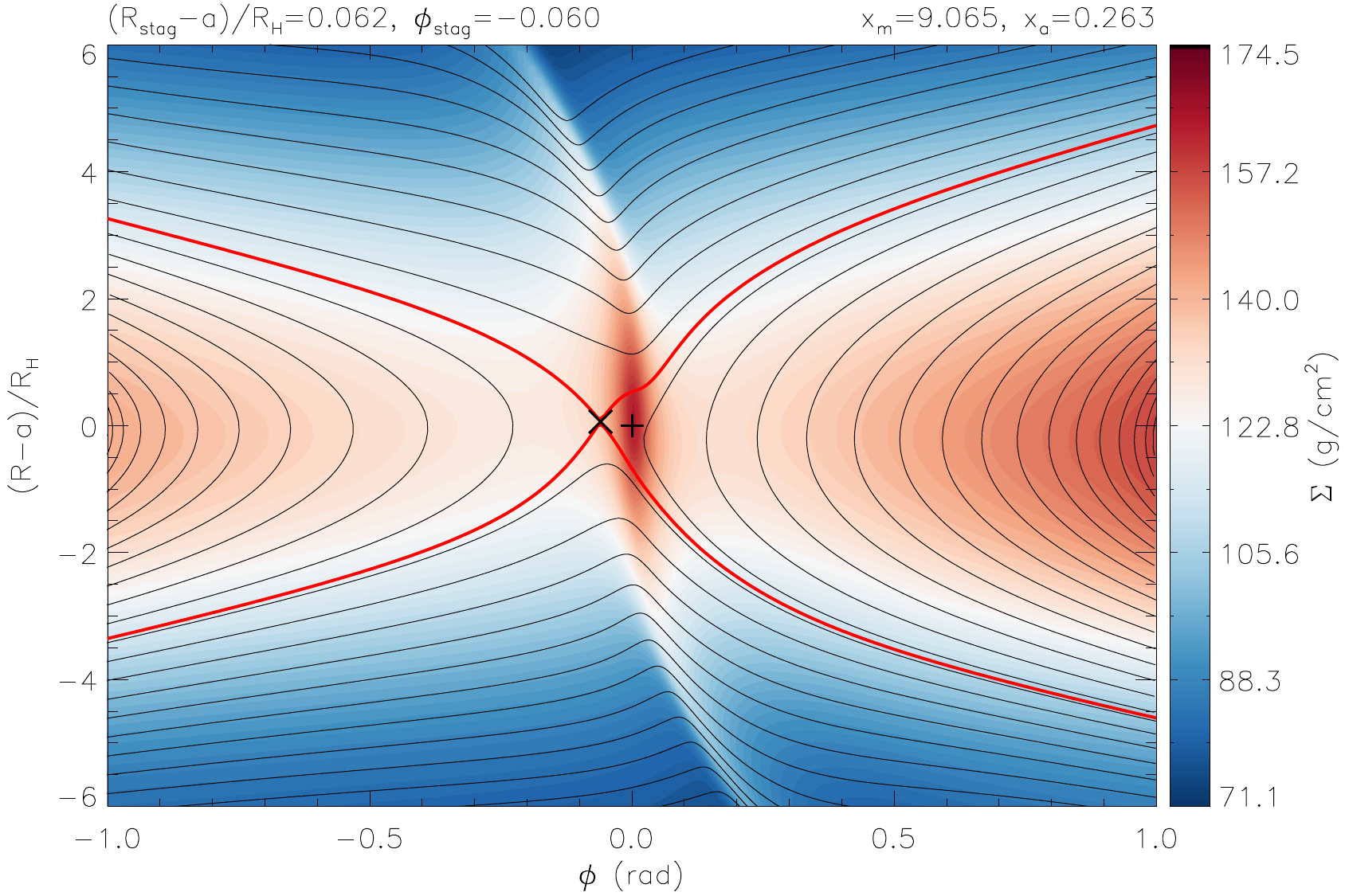}
	\caption{Streamlines (black curves) and the horseshoe separatrix (red curves) in the co-orbital region of  planet overplotted by the gas density distribution for model $\Delta R_\mathrm{dze}=1.5H_\mathrm{dze}$ at two different phases, during the first occurrence of the oscillation of planetary semimajor axis: at $t=25\times10^3$\,yr (upper panel) and at $t=25\times10^3$\,yr (lower panel). The position of planet and the stagnation point are denoted by $+$ and $\times$, respectively.}
	\label{fig:streamout7b-osc}
\end{figure}

At later times, the vortex librates around $\phi=150^\circ$ (see Fig.\,\ref{fig:dens-max-a}) that causes an oscillation of the stellar torque. Although the sign of the front--rear horseshoe asymmetry defined by equation (\ref{eq:asym}) is positive during the vortex libration, the magnitude of asymmetry is oscillating, shown in Fig.\,\ref{fig:streamout7b-osc} at $t=25\times10^3$ and $26\times10^3$\,yr. Thus, one can conclude that the oscillation of the corotation torque emerges due to the periodic oscillation of the horseshoe asymmetry by the vortex as the vortex centre librates around $\phi=150^\circ$. The net torque felt by the planet nearly vanishes on average during the vortex libration; thus, the planet is trapped. However, a small oscillation in its semimajor axis can be observed. 

At the ejection of the planetary core (at $t\simeq45\times10^3$\,yr), both the magnitudes of the corotation and stellar torques begin to decrease. This can be explained by the fact that the vortex dissolves (see Fig.\,\ref{fig:fourier}) causing an azimuthally symmetric gas distribution with a nearly vanishing stellar torque. In the meantime, the front--rear asymmetry of the horseshoe region is also decreasing ($x_\mathrm{a}=0.16$ at $t=60\times10^3$\,yr) causing a smaller magnitude corotation torque. Note that the 1D approximation of the corotation torque is increasing in magnitude at this time and the net torque would be positive. At the same time, the magnitude of the negative differential Lindbald torque slightly increases. As a result, the magnitude of the negative net torque felt by the planet slightly increases, which results in the ejection of the planet from the trap.

As the vortex is reformed after the ejection of planet within $20\times10^3$\,yr, the 2D corotation and the stellar torque become oscillating again, but with higher frequency and amplitude than in the previous oscillation stage. The high-frequency oscillation of the stellar torque is caused by the periodical shifts of the barycentre of the system, as the orbital speed of the vortex centre and planet significantly departs due to the difference in their orbital distance. The oscillation of the corotation torque is caused by oscillatory front--rear horseshoe asymmetry. As a result, the net torque acting on the planet is highly oscillating, but it is negative on average. Therefore, the planet migrates towards the star, with small oscillations in its semimajor axis (see the inset in Fig.\,\ref{fig:mig-1}), but with higher migration rate than  in the Type I regime.

\begin{figure}
	\centering
	\includegraphics[width=\columnwidth]{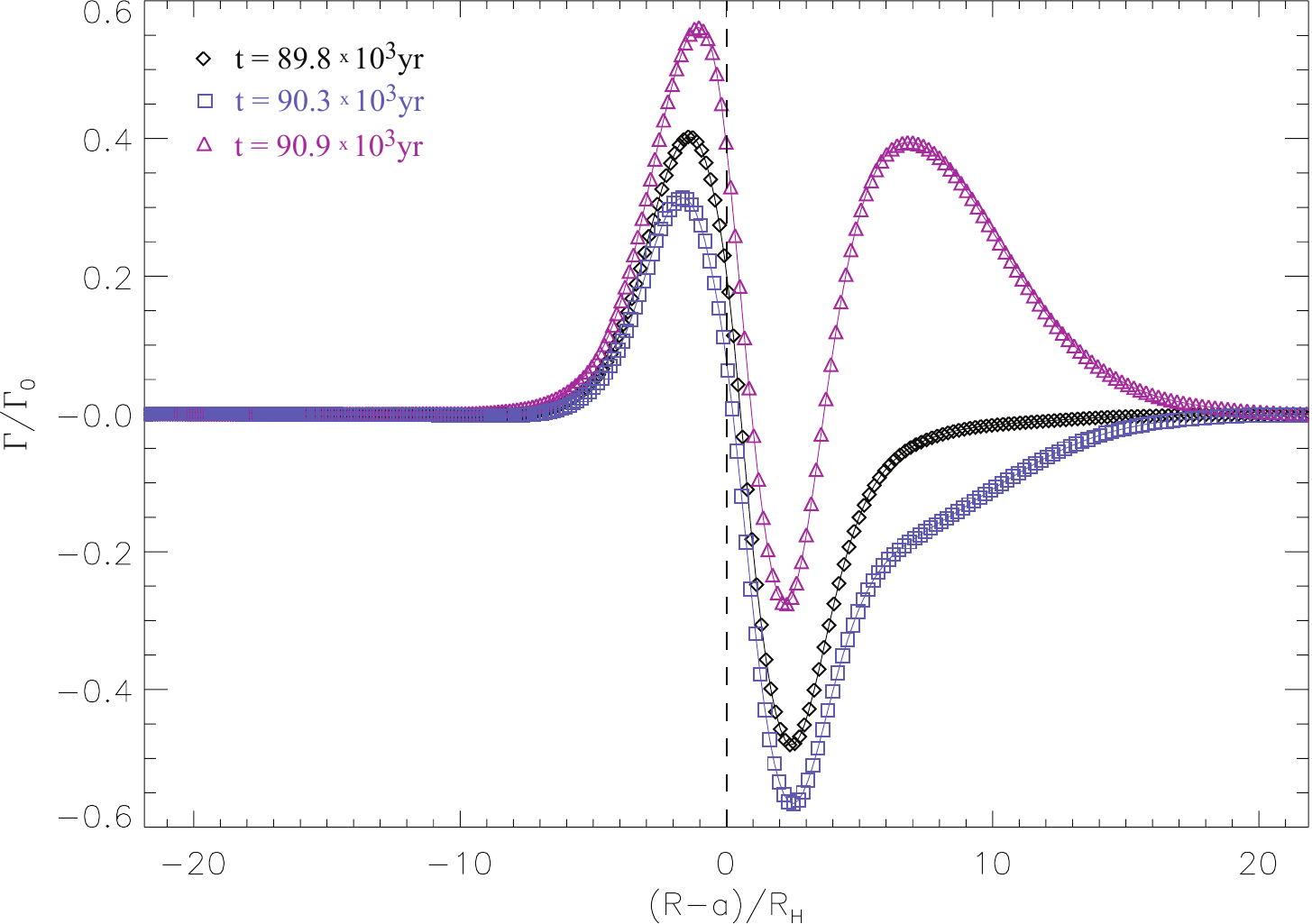}
	\caption{Radial torque density profile in model $\Delta R_\mathrm{dze}=1.3H_\mathrm{dze}$ for three different phase of simulation, where the vortex situated near the planet. $\phi_\mathrm{max}>0^\circ$ ($t=89.8\times10^3$\,yr), $\phi_\mathrm{max}=0^\circ$ ($t=90.3\times10^3$\,yr), and $\phi_\mathrm{max}<0^\circ$ ($t=90.9\times10^3$\,yr).}
	\label{fig:tpout8b}
\end{figure}

\begin{figure}
	\centering
	\includegraphics[width=\columnwidth]{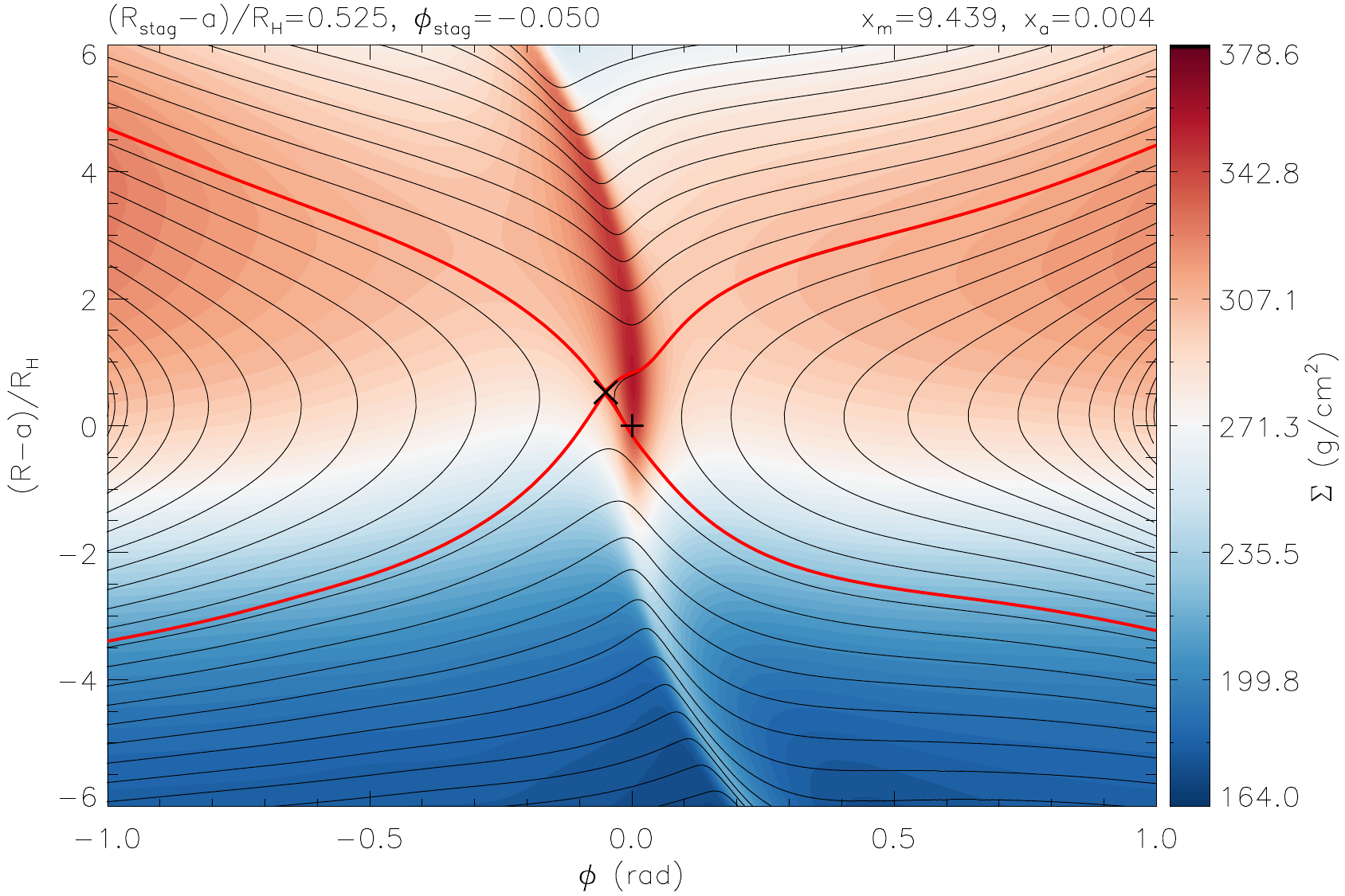}
	\includegraphics[width=\columnwidth]{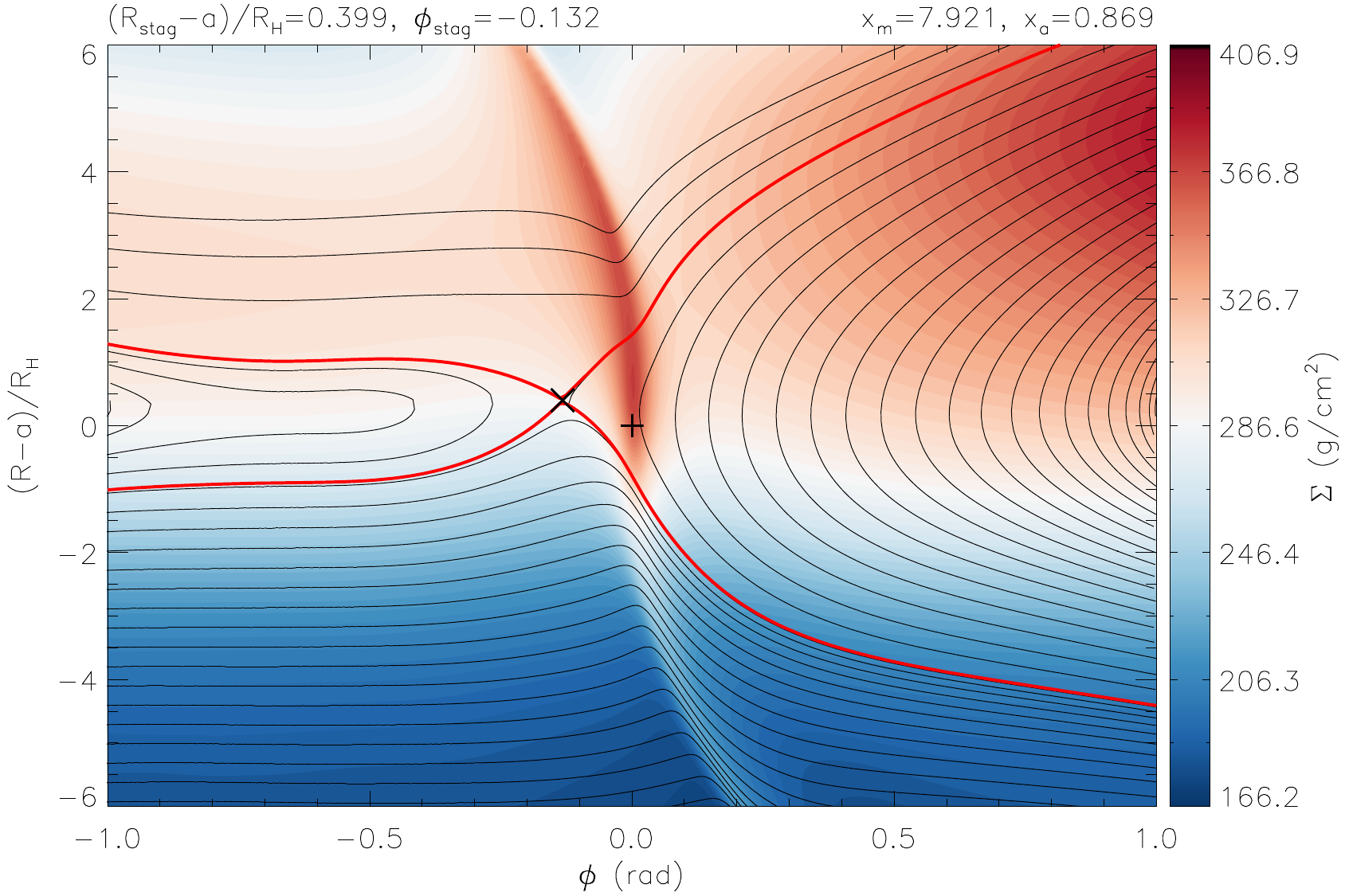}
	\includegraphics[width=\columnwidth]{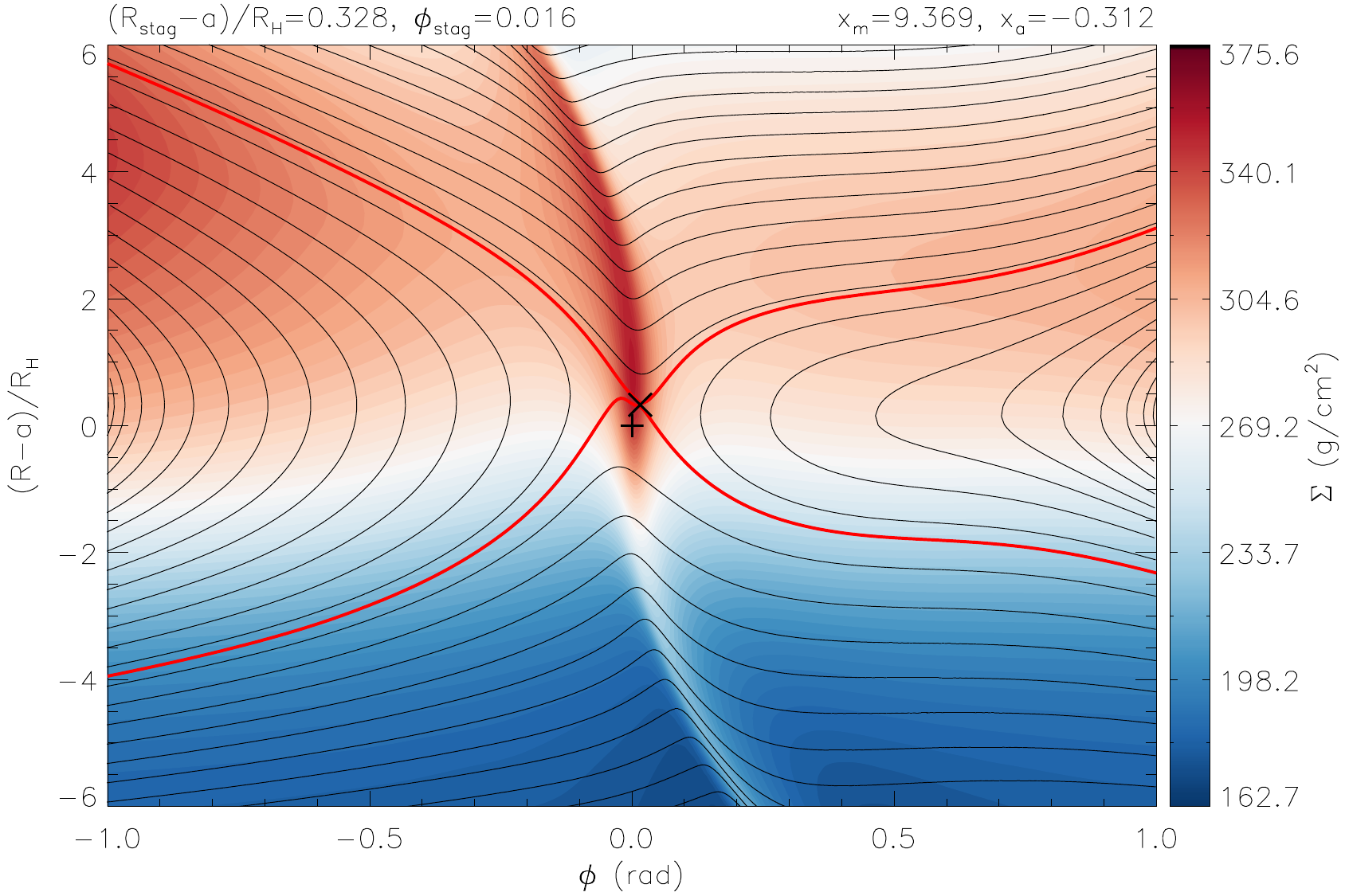}
	\caption{Streamlines (black curves) and the horseshoe separatrix (red curves) in the co-orbital region of  planet overplotted by the gas density distribution for model $\Delta R_\mathrm{dze}=1.3H_\mathrm{dze}$ at three different phases, during the re-capturing of the planet: at $t=89.8\times10^3$\,yr (upper panel), at $t=90.3\times10^3$\,yr (middle panel), and at $t=90.9\times10^3$\,yr (lower panel). The position of planet and the stagnation point are denoted by $+$ and $\times$, respectively.}
	\label{fig:streamout8b-osc}
\end{figure}

For models with sharper viscosity reduction ($\Delta R_\mathrm{dze}=1.3H_\mathrm{dze}$), we observed several ejection and re-trapping of the planetary core, before its final ejection. After the first ejection, the vortex reappears within $\sim10\times10^3$\,yr, much faster than in the previous case (see Fig.\,\ref{fig:fourier}). Due to the effect of vortex, the planet feels a high-amplitude oscillatory torque emerging from the outer disc, see  Fig.\,\ref{fig:tpout8b} showing the radial torque density profile exerted on the planet during its first recapturing at three consecutive phases. By analysing the horseshoe region at this time (see Fig.\,\ref{fig:streamout8b-osc}), one can find that at $t=89.8\times10^3$\,yr the vortex centre is at $\phi\simeq180^\circ$, causing a nearly vanishing front--rear horseshoe asymmetry ($x_\mathrm{a}=0.004$, upper panel), and therefore the radial torque density profile is nearly symmetric. At $t=90.3\times10^3$ and $90.9\times10^3$\,yr, the vortex centre is at the front and rear of the planet, respectively, causing high positive ($x_\mathrm{a}=0.87$, middle panel) and negative ($x_\mathrm{a}=-0.31$, lower panel) front--rear horseshoe asymmetry, which can explain the positive and negative excess in the radial torque density profile. 

\begin{figure}
	\centering
	\includegraphics[width=\columnwidth]{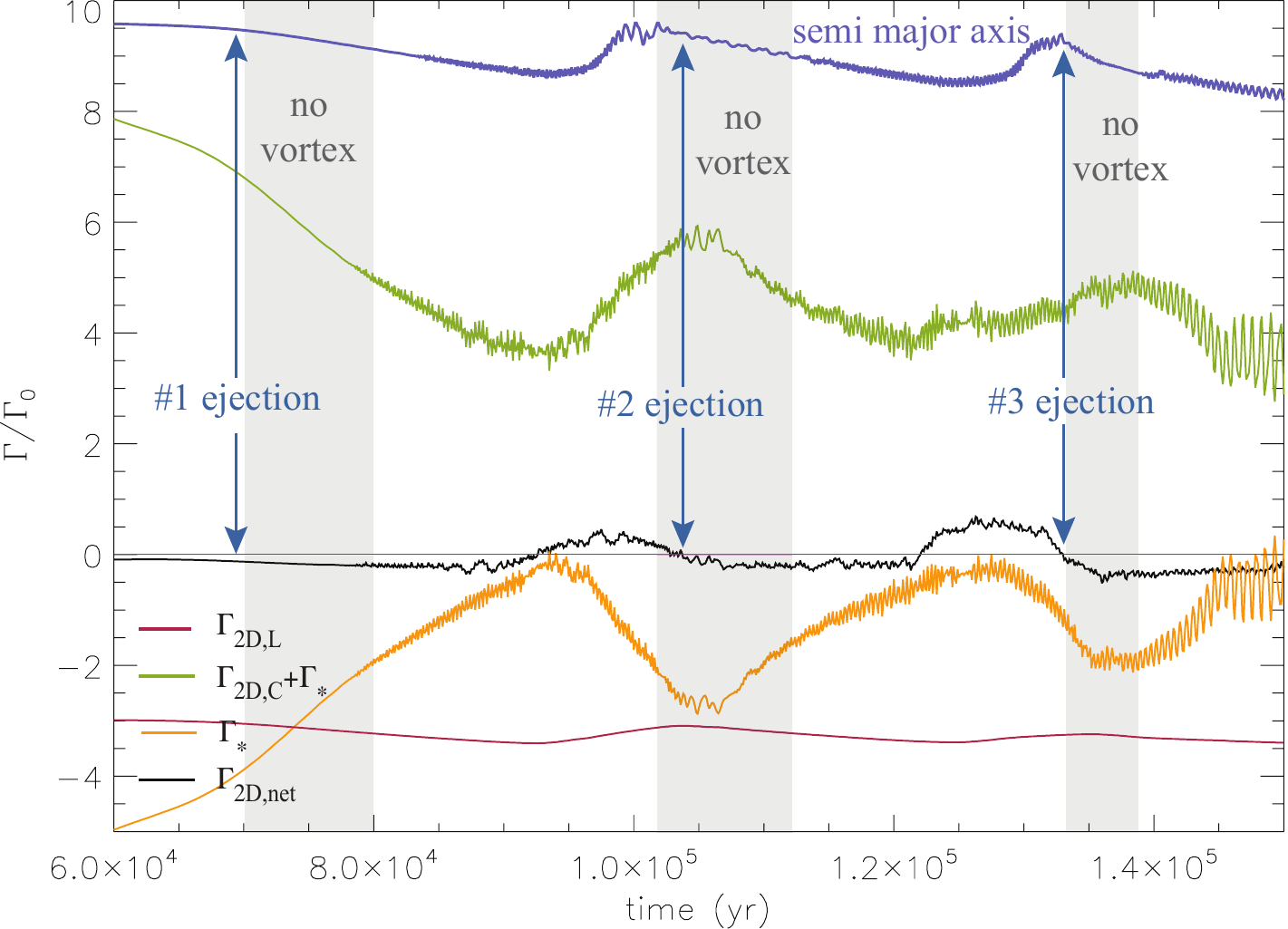}
	\caption{2D corotation ($\Gamma_\mathrm{2D,C}$), differential Lindblad ($\Gamma_\mathrm{2D,L}$), stellar ($\Gamma_*$) and net ($\Gamma_\mathrm{2D,net}=\Gamma_\mathrm{2D,C}+\Gamma_\mathrm{2D,L}+\Gamma_\mathrm{*}$) torques felt by the planet for model $\Delta R_\mathrm{dze}=1.3H_\mathrm{dze}$. The torques are normalized by $\Gamma_0$. The change in the semimajor axis of planet is also shown. The vortex is disintegrated inside the grey regions.}
	\label{fig:t1dout8b}
\end{figure}

Fig.\,\ref{fig:t1dout8b} shows the evolution of the differential Lindblad, corotation, stellar and net torques acting on the planet during the recapturing--ejecting phase. We note that the corotation, stellar and net torques are smoothed over a 100\,yr long interval; thus, the curves represent the average values around the torques which are oscillating. As the redeveloped vortex becomes more prominent, the magnitude of the corotation and stellar torques begins to increase at $t\simeq90\times10^3$\,yr. The magnitude of the negative differential Lindblad torque slightly decreases in the meantime. Since the horseshoe asymmetry is positive on average, the planet feels a high-magnitude positive corotation torque on average. As a consequence, the net torque becomes positive on average, and the planet begins to migrate outwards with small oscillations in its semimajor axis due to the oscillatory behaviour of stellar and corotation torques. As the orbital distance of the planetary core coincides with the vortex centre radius, the vortex is again dissolved at $t\simeq102\times10^3$\,yr. The magnitude of both the corotation and stellar torques begins to decrease on average, causing the second ejection of the planet. At this time, the magnitude of the differential Lindblad torque is slightly increasing. The recapturing--ejection phenomenon is repeated several times, while the planet eventually leaves the density maximum and migrates inwards with higher migration rate than in the Type I regime.

For even sharper viscosity reduction ($\Delta R_\mathrm{dze}\leq1H_\mathrm{dze}$), although the planet is ejected from the trap at $t=80\times10^{3}$\,yr due to the vortex disintegration (see Fig.\,\ref{fig:fourier}), it remains near the trap. In this particular model after the first redevelopment of the vortex, the vortex becomes so strong that the $10\,\mathrm{M_\oplus}$ planetary core cannot destroy it anymore. Moreover, as the magnitude of the stellar torque oscillates around a zero average value, the net torque felt by the planet although oscillating is nearly vanishing. As a result, the planet is slowly drifted inwards with the density maximum as the disc viscously evolves. Since this inward drift is much slower than in the conventional Type I migration, we can state that the planet is trapped.

Summarizing the above results, we found that the stellar torque caused by the strong azimuthal asymmetry in the disc density distribution due to large-scale vortex plays a significant role in the migration of the $10\,\mathrm{M_\oplus}$ planetary core for sharp dead zone edge models ($\Delta R_\mathrm{dze}\leq1.5H_\mathrm{dze}$). The stellar torque always counteracts the corotation torque. The planet is temporarily trapped due to a high-magnitude corotation torque caused by an asymmetric horseshoe region as long as the asymmetry is maintained by the vortex. For a medium sharp dead zone edge model ($\Delta R_\mathrm{dze}=1.3H_\mathrm{dze}$), the planet is ejected from the trap as the vortex is disintegrated by the planetary perturbation and the positive corotation torque declines, as well as the negative stellar torque on average. Being the planet departed from the density maximum, the vortex redevelops. If the vortex redevelopment is fast enough, the planet cannot migrate far from the density maximum, and it is recaptured in the trap. However, after several recapturing events, the planet is eventually ejected from the trap. For even sharper viscosity reduction ($\Delta R_\mathrm{dze}=1H_\mathrm{dze}$), the planet stays in the trap up to $t=2\times10^5$\,yr. This can be explained by the fact that the corotation torque remains positive, while the stellar torque is small on average. We conclude that the discrepancy between 1D and 2D models for sharp viscosity reduction is mainly due to the highly non-axisymmetric disc, which is completely neglected in 1D simulations.

\subsection{Limitations and future prospects}

Here we mention some limitations of our numerical simulations. The most important one could be that we ran simulations in 2D models assuming that the disc is thin; thus, the vertical motions can be neglected and have no severe effect on RWI. According to a recent investigation presented in \citet{Lin2012arXiv}, RWI is predominantly a two-dimensional instability. It is concluded that in the locally isothermal approximation, the vortex centre remains in approximate vertical hydrostatic equilibrium; thus, the two-dimensional approximation seems to be physically plausible. 

In our approach [$\alpha$-type viscosity prescription of \citet{ShakuraSunyaev1973} with a power-law profile of temperature, $T(R)\sim R^{-1}$], the kinematic viscosity ($\nu$) that drives the viscous evolution of the disc has a radial profile of $\nu(R)\sim R^{0.5}$. Therefore, the density profile of $\Sigma(R)\sim R^{-0.5}$ serves an equilibrium solution in an unperturbed disc in which no viscosity reduction is applied. However, in models that include dead zone, the density jump or Rossby vortices will not reach an equilibrium state within several hundred years of simulation. Rather, the density jump evolves such that it strengthens and shifts towards the star with time. Therefore, it is worth investigating the situation in which the planet is inserted to the disc at different phases of the viscous evolution of the density jump to reveal the dependence of trapping mechanism on the evolutionary state of the density jump. Indeed, this investigation will be presented in the second part of this work. Note, however, that if the kinematic viscosity has a profile such that $\nu(R)\sigma(R)\sim R^0$, the evolution of an evolved density jump can be freezed and the migration or trapping of low-mass planet near a steady-state density jump or vortex can be explored. This investigation will be presented in a separate paper of S. Ataiee et al. (in preparation).

Our simulations are done by applying the isothermal equation of state; thus, we assume that the thermal diffusion is infinitely efficient, i.e. the generated heat is radiated away perfectly. In a more realistic disc model, the temperature should change in the density jump depending on the cooling time, which certainly has an impact on the viscous evolution of the disc, and on the vortex itself. It is known that the inclusion of finite thermal diffusion and taking into account the radiative effects, planets that are smaller than $50\,\mathrm{M_\oplus}$ migrate outwards \citep{PaardekooperMellema2006,KleyCrida2008}. This effect is found to be more severe in 3D simulations \citep{Kleyetal2009}. Thus, it is worth investigating the temporary trapping of a  giant-planet core with the inclusion of more realistic thermodynamic effects.

The effect of the photoevaporation driven by the central star on the viscous evolution of the density jump was neglected in our simulations. Due to the photoevaporation, the disc material and the total disc mass are decreasing with time. This could have some consequences on the erosion of the density jump, as well as the migration and trapping of the planet. Nevertheless, the time-scale of photoevaporation is of the order of $10^6$\,yr for discs hosted by stars in the mass range of $M_*\leq3\,\mathrm{M_\odot}$ \citep{Gortietal2009}. Since the temporary trapping of giant-planet cores is modelled on the time-scale of several $100\times10^3$\,yr, neglecting photoevaporation processes might have severe effects only for $M_*>3\,\mathrm{M_\odot}$ star hosted discs.

Although the effects of several disc parameters such as the power-law index of density profile, the disc mass, disc temperature (equivalent to the disc aspect ratio), the magnitude of global viscosity and the depth of viscosity reduction have not been investigated in the present work, they do certainly influence the evolution of the density jump formed at the viscosity transition. If the RWI excitation mechanism or the vortex evolution is considerably affected by the above physical parameters, and therefore the stellar torque too, the migration history, and thus the recapturing phenomenon, might depend on the physical details of the model. Thus, it is worth investigating the migration and possible trapping of giant-planet cores in disc models where the above physical parameters are systematically changed. This study will be presented in a forthcoming work.

We also found that the vortex disintegrates as the $10\,\mathrm{M_\oplus}$ planetary core reaches the radial position of the vortex. The disintegration is presumably caused by the fact that the flow inside the vortex is perturbed by the planetary potential, if the vortex and the horseshoe region overlap. Although we did not investigate the effect of the planetary mass on the vortex disintegration, we suspect that the smaller the planetary mass, the later is the disintegration. This would mean longer temporary trapping for smaller planets as well as later ejection. Thus, it is also worth investigating how the longevity of temporary trapping depends on the planetary mass.

\section{Conclusion}

In this paper, we investigated the migration of a $10\,\mathrm{M_\oplus}$ giant-planet core in discs having viscosity transition at the outer edge of the dead zone. Only locally isothermal disc having initial surface mass density with power-law index of $-0.5$ was investigated assuming $0.01\,\mathrm{M_\odot}$ for the disc. The $\alpha$ prescription was used for the disc viscosity. The viscosity reduction near the outer dead zone edge (set to 12\,au) was modelled by a smooth decrease in $\alpha$ from $\alpha=10^{-2}$ to $10^{-4}$ within a radial distance $1H_\mathrm{dze}\leq\Delta R_\mathrm{dze}\leq4H_\mathrm{dze}$, called as the width of dead zone edge, where $H_\mathrm{dze}$ is the local disc pressure scale height at the dead zone edge. We assumed a flat disc model with the conventional aspect ratio of $h=0.05$.

The density jump forms at the dead zone edge, where the planet is subject to trapping according to the 1D approximation of Type I migration. The trapping of the planet does not depend on the width of the viscosity reduction in the 1D models. However, our 2D hydrodynamic simulations revealed that the trapping of the planet occurs only if certain conditions are fulfilled. 

We also revealed that the stellar torque exerted on the planet due to the shift of the barycentre, caused by the highly non-axisymmetric density distribution of the disc, plays a significant role in the planetary migration. The magnitude of the stellar torque exerted on the planet is significant in sharp dead zone edge models, and it is always opposite in sign compared to the corotation torque.

For blunt viscosity transitions ($R_\mathrm{dze}\geq2H_\mathrm{dze}$), in which RWI is not excited, no trapping of the planetary core occurs. Interestingly, the migration rate of the $10\,\mathrm{M_\oplus}$ planet exceeds that in the conventional Type I regime. This can be explained by the attenuated corotation torque caused by the smaller vortensity-related horseshoe drag compared to the unperturbed disc models.

For sharp dead zone edges ($R_\mathrm{dze}\leq1.5H_\mathrm{dze}$), a large-scale vortex forms due to RWI and the $10\,\mathrm{M_\oplus}$ planetary core is trapped. The torque exerted on the planet is highly oscillating. The trapping of the planetary core and the oscillating torque can be attributed to the interaction of high-magnitude oscillating corotation and stellar torques, that emerges due to the vortex periodic perturbation on horseshoe dynamics. However, when the planet gets through the density maximum, the vortex disintegrates, resulting in declined corotation and stellar torques. As a consequence, the planetary core feels a larger negative disc torque on average and eventually is ejected from the trap. Retrapping of the planetary core occurs if the width of the dead zone edge is sharp enough ($R_\mathrm{dze}\leq1.3H_\mathrm{dze}$), due to the redevelopment of the large-scale vortex. However, after a couple of retrapping, the planet finally leaves the trap and migrates towards the star with higher rate than in the Type I regime. For even sharper dead zone edge models ($R_\mathrm{dze}\leq1H_\mathrm{dze}$), the planetary core is ejected but remains close to the density maximum up to $200\times10^3$\,yr. This can be explained by the nearly vanishing average stellar torque and significant average corotation torque.

The discrepancy between the 1D and 2D simulations can be accounted for two assumptions made in 1D simulations. On one hand, the $10\,\mathrm{M_\oplus}$ planet significantly perturbs the disc due to low viscosity ($\alpha=10^{-4}$) in the dead zone. Namely, even the $10\,\mathrm{M_\oplus}$ planet is able to open a partial gap in the disc as the viscous condition of a gap opening is fulfilled. As a result, the horseshoe dynamics significantly changes compared to the unperturbed disc causing a more negative total disc torque. On the other hand, the neglect of high azimuthal non-axisymmetry, especially in sharp viscosity transitions, results in the emergence of an additional stellar torque due to the shift of the barycentre. Since neither the partial gap opening nor the barycentre shift is taken into account in 1D simulations, to model the migration of the $10\,\mathrm{M_\oplus}$ core of a giant planet in a low-viscosity regime ($\alpha=10^{-4}$) dead zone requires at least a two-dimensional approach.

Applying our results to planet formation via core accretion in discs having a dead zone, we can conclude the following. Initially, the gas accumulates near the dead zone edge. If the viscosity transition is sharp enough, the disc becomes RWI unstable and finally an $m=1$ mode large vortex develops. Inside of this vortex, the core accretion might be enhanced and planetary cores might form well within the disc lifetime. If the planetary core leaves the trap after reaching the mass limit for runaway gas accretion, the formation efficiency of a giant planet might be accelerated due to the mass accumulated inside the vortex. For this case, giant planets might form at the outer dead zone edge. If the planetary core leaves the trap before reaching the mass limit of runaway gas accretion, the rate of the inward migration of this core is faster than those in the conventional Type I regime. This might result in the formation of rocky planets without a significant gas envelope. If the viscosity transition is not sharp enough, the core is not trapped and subject to fast (faster than in the Type I regime) inward migration. For these latter two cases, the planetary core will be inevitably engulfed by the star within less time than in the conventional Type I scenario predicts or within the disc lifetime, unless there is no other stopping mechanism or slowing down of Type I migration. 

Our final conclusion is that the outer dead zone edge might be highly effective in the temporary trapping of the giant-planet core and forming a giant planet as the vortex acts as a gas reservoir. After the formation of a giant planet, the migration in the Type II regime is greatly reduced due to low viscosity in the dead zone. We note, however, that this scenario should be investigated in a more elaborated model, in which the viscous evolution of the density jump, the growth of the planetary core and runaway gas accretion are modelled.

\section*{Acknowledgments}

This project has been supported by the Lend\"ulet-2009 Young Researchers' Programme of the Hungarian Academy of Sciences and the HUMAN MB08C 81013 grant of the MAG Zrt and by City of Szombathely under agreement No. S-11-1027. We are grateful to F. Masset who provided the NVIDIA TESLA compatible version of {\small GFARGO} to us for helpful discussions regarding the modification of the {\small CUDA} kernel code of {\small GFARGO}. We are also grateful to C. P. Dullemond, W. Kley and A. Zsom for helpful discussions. We thank L. L. Kiss for carefully reading the manuscript.

\bibliographystyle{mn2e}
\bibliography{mn-jour,regaly}

\appendix

\section{Numerical convergence}
~~~~~~~~~~~~~~~~~~~~~~~~~~~~~~~~~~~~~~~~~~~~~~~~~~~~~~~~~~~~~~~~~~~~~~
\label{apx:numres}

Recently, \citet{Kleyetal2012} have investigated the effect of numerical resolution in 2D global disc simulations in which a low-mass planet ($2\,\mathrm{M_\oplus}$) orbits at a fixed radial distance in a disc with constant surface mass density. Their simulations used constant kinematic viscosity with the value of $\nu=10^{-8}$. They have found that close  to the planet ($|d|\leq4/3H_\mathrm{p}$, where $H_\mathrm{p}$ is the pressure scale height at the planetary orbit), the simulations are in the convergent domain with moderate numerical resolution, in a sense of capturing the shock in the spiral wakes induced by the planet. However, farther from the planet ($|d|\geq2H_\mathrm{p}$), where the spiral wake transforms to a shock wave, extremely high numerical resolution is required, e.g. 100 grid cells per $H_\mathrm{p}$ to satisfy numerical convergence. Note that in our standard resolution ($N_R=512$, $N_\phi=1024$), the resolution is only 15 grid cells per $H_\mathrm{p}$.

In order to be convinced that the results of our simulations with the above standard resolution are appropriate, we ran simulations where the numerical resolution was halved and doubled to $N_R=256,\,N_\phi=512$ (low resolution) and $N_R=1024,\,N_\phi=2048$ (high resolution), respectively. Note that the original version of {\small GFARGO} is capable to handle a limited number of radial zones due to the finite size of the internal cache of the NVIDIA TESLA C2050 card being 64kB. Therefore, the {\small CUDA} kernel code was modified and tuned to achieve simulations in higher resolution. 

In these simulations no viscosity reduction was applied, and the global viscosity was set to  $\alpha=10^{-4}$. The $10\,\mathrm{M_\oplus}$ planet was fixed in an orbit at $a=9.5$\,au, in a region where temporary trapping was observed in our standard resolution dead zone models. Emphasize that in these models the kinematic viscosity of the disc close to the planet, being $\nu=7.5\times10^{-7}$, is about two orders of magnitude larger than in the nearly inviscid simulations presented in \citet{Kleyetal2012}.

\begin{figure}
	\centering
	\includegraphics[width=\columnwidth]{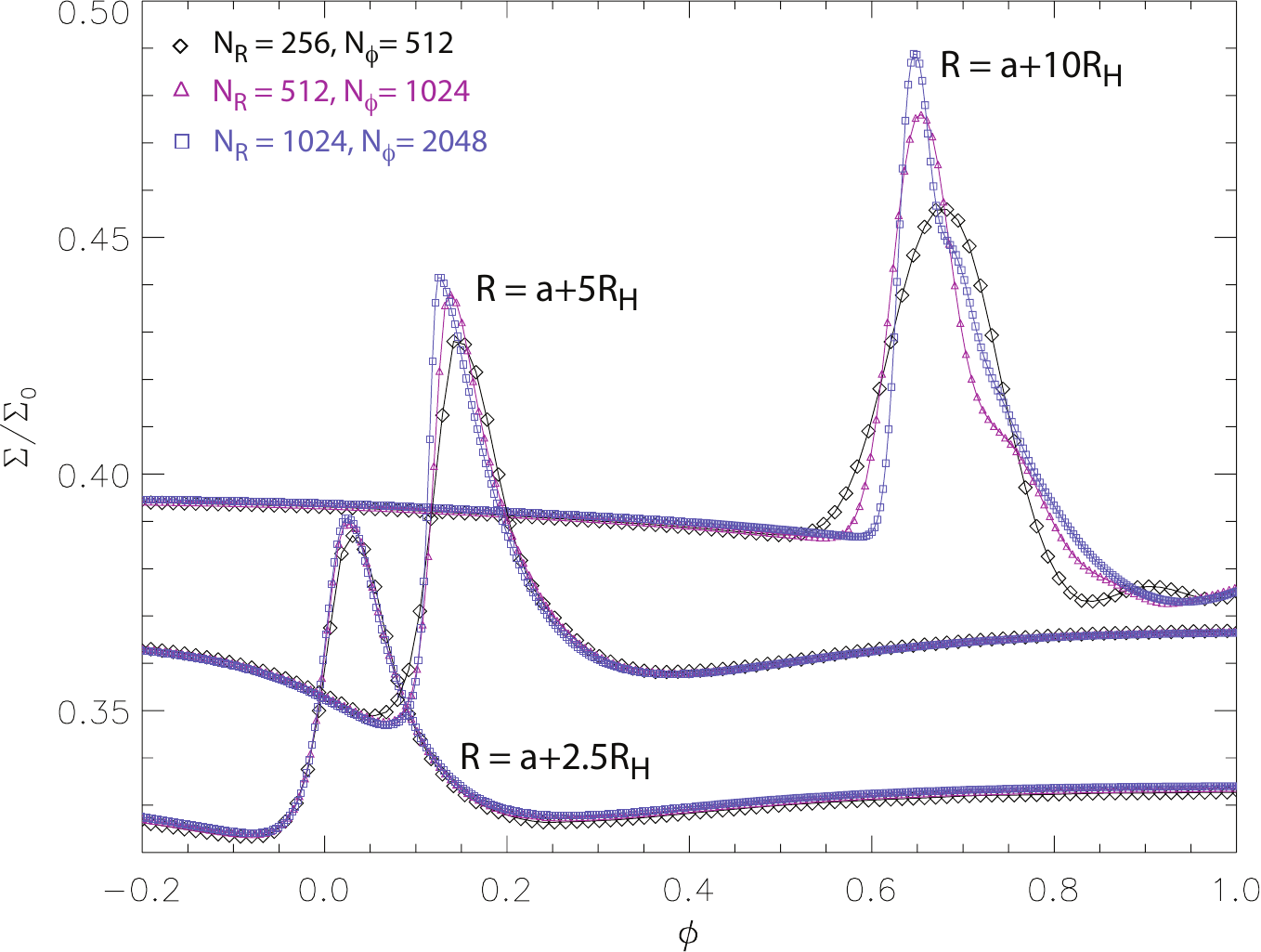}
	\caption{Snapshot of normalized azimuthal density profile of the outer spiral wave taken at $t=500$\,yr of simulation for the three different numerical resolutions (indicated on the figure) at a radial distance of $a+2.5R_\mathrm{H},\,a+5R_\mathrm{H}$ and $a+10R_\mathrm{H}$. The profiles are vertically shifted with an arbitrary value.}
	\label{fig:numres_shock}
\end{figure}

Since the spiral waves induced by the planet travel with the azimuthal velocity of planet, the wave angular velocity becomes supersonic inside and outside the planetary orbit at $R<a(-1+h)^2$ and $R>a(1+h)^2$, respectively. These distances correspond to $R>a+4.15R_\mathrm{H}$ and $R<a-4.52R_\mathrm{H}$ in a $10\,\mathrm{M_\oplus}$ planet, where $R_\mathrm{H}$ is the radius of the planetary Hill sphere. Fig.\,\ref{fig:numres_shock} shows snapshots of normalized azimuthal density profiles of the outer spiral wave taken after $t=500$\,yr of simulation (corresponding to about 17 orbits at $9.5\,$au) using low, high and standard numerical resolution at distances of $a+2.5R_\mathrm{H},\,a+5R_\mathrm{H}$ and $a+10R_\mathrm{H}$. The three simulations are perfectly matching at $R=a+2.5R_\mathrm{H}$. At larger distances ($R=a+5R_\mathrm{H}$), where the spiral wave transforms to shock, the shock is well captured by all simulations. However, at even larger distances ($R=a+10R_\mathrm{H}$) the shock is captured only for the standard and high-resolution simulations.

\begin{figure}
	\centering
	\includegraphics[width=\columnwidth]{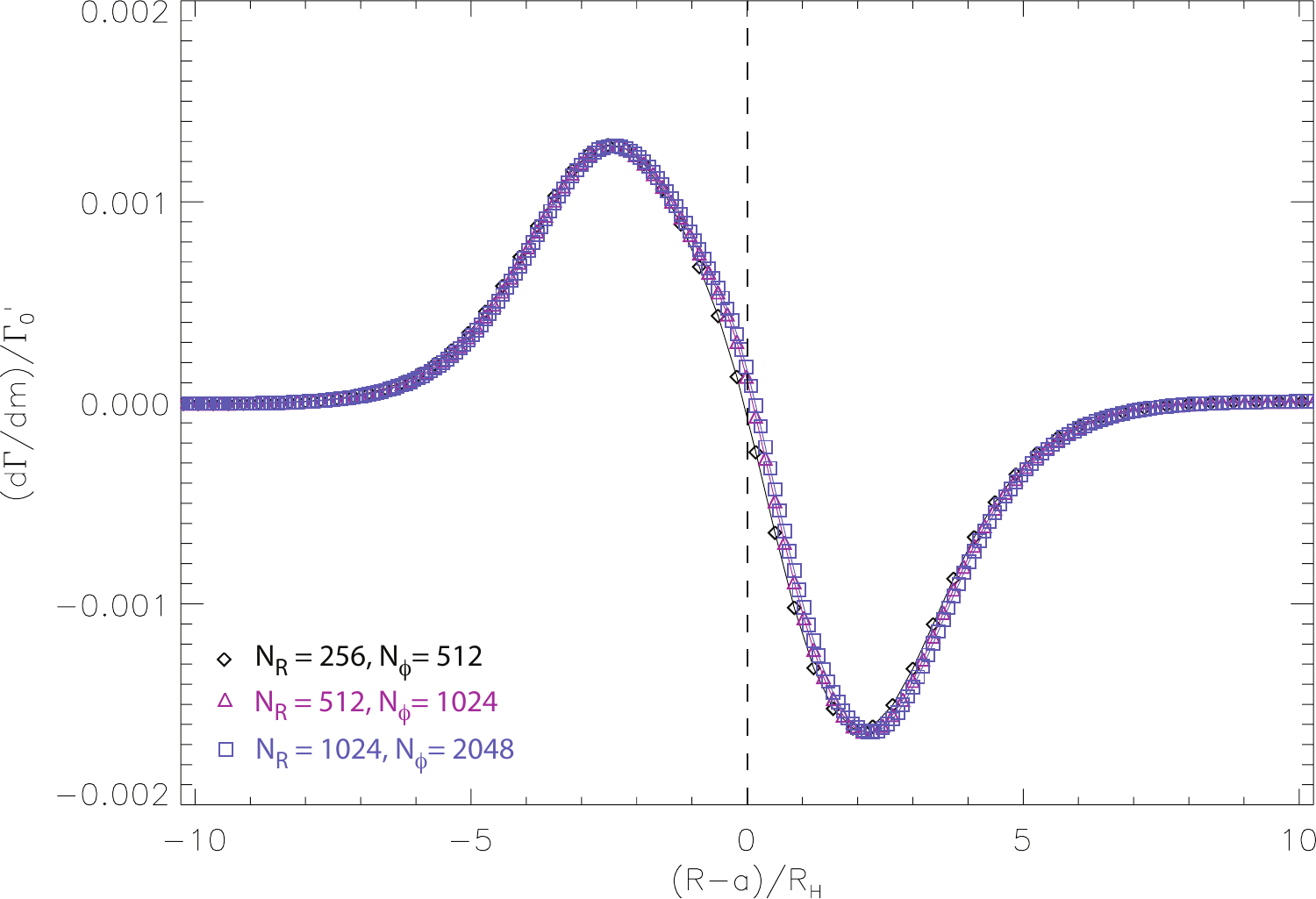}
	\includegraphics[width=\columnwidth]{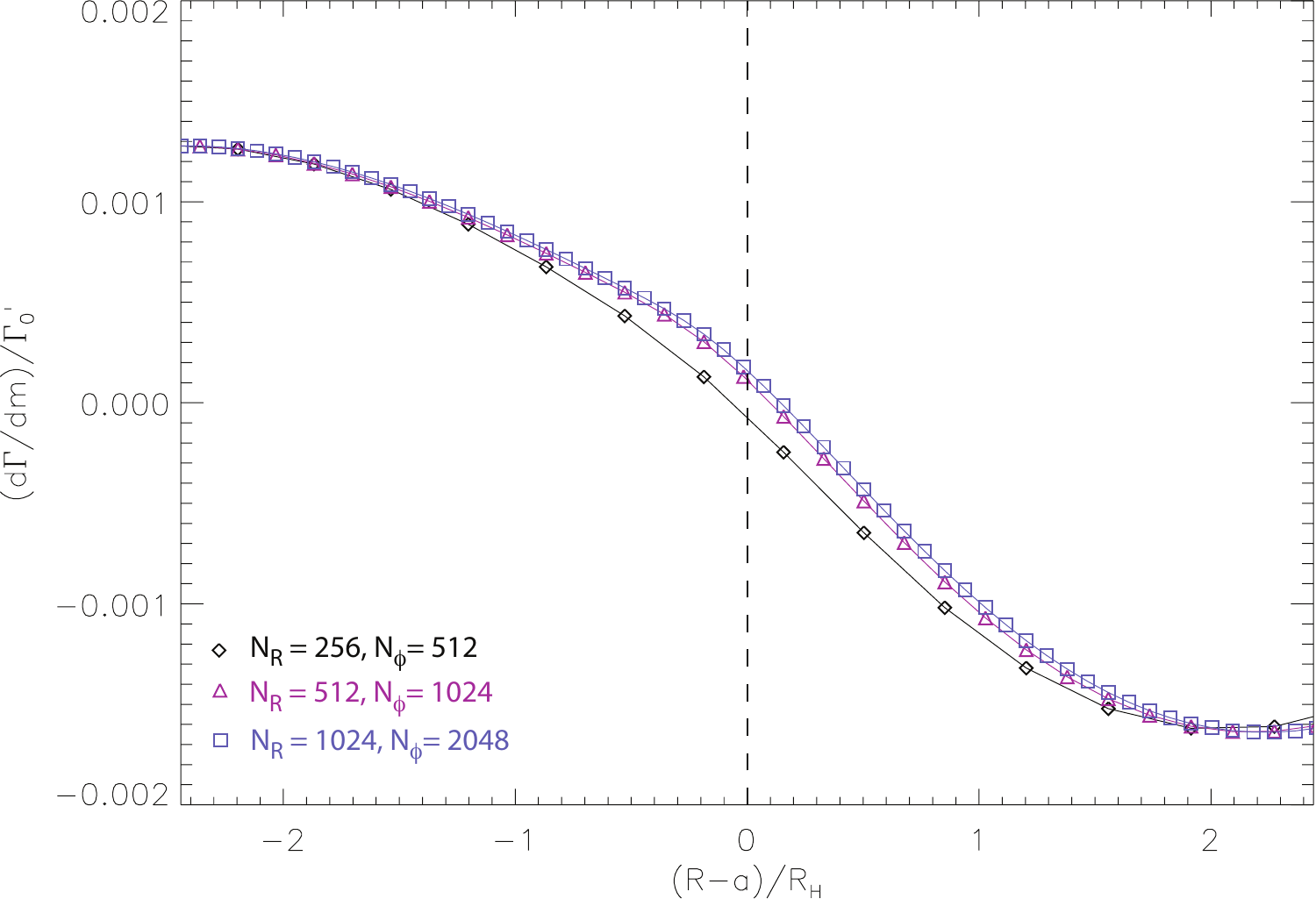}
	\caption{Radial torque density profile for models with different numerical resolutions (indicated in the figure) in the planet vicinity. The lower panel shows the region close to planet. The profiles are taken after $t=500$\,yr of simulation.}
	\label{fig:torque-numres}
\end{figure}

\begin{figure}
	\centering
	\includegraphics[width=\columnwidth]{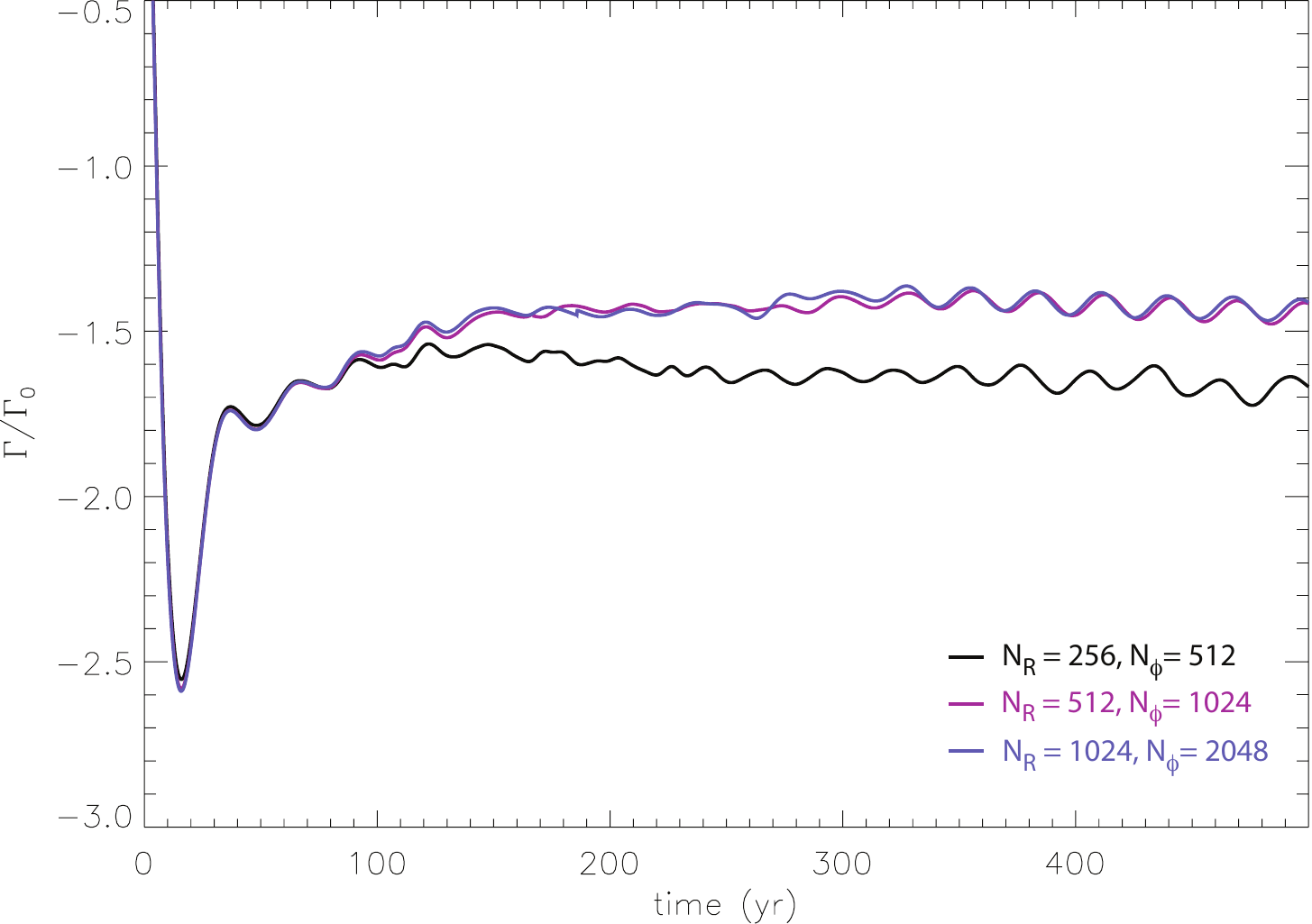}
	\caption{Evolution of the normalized total disc torque felt by the planet for the three different numerical resolution simulations.}
	\label{fig:t-numres}
\end{figure}

\begin{figure}
	\centering
	\includegraphics[width=\columnwidth]{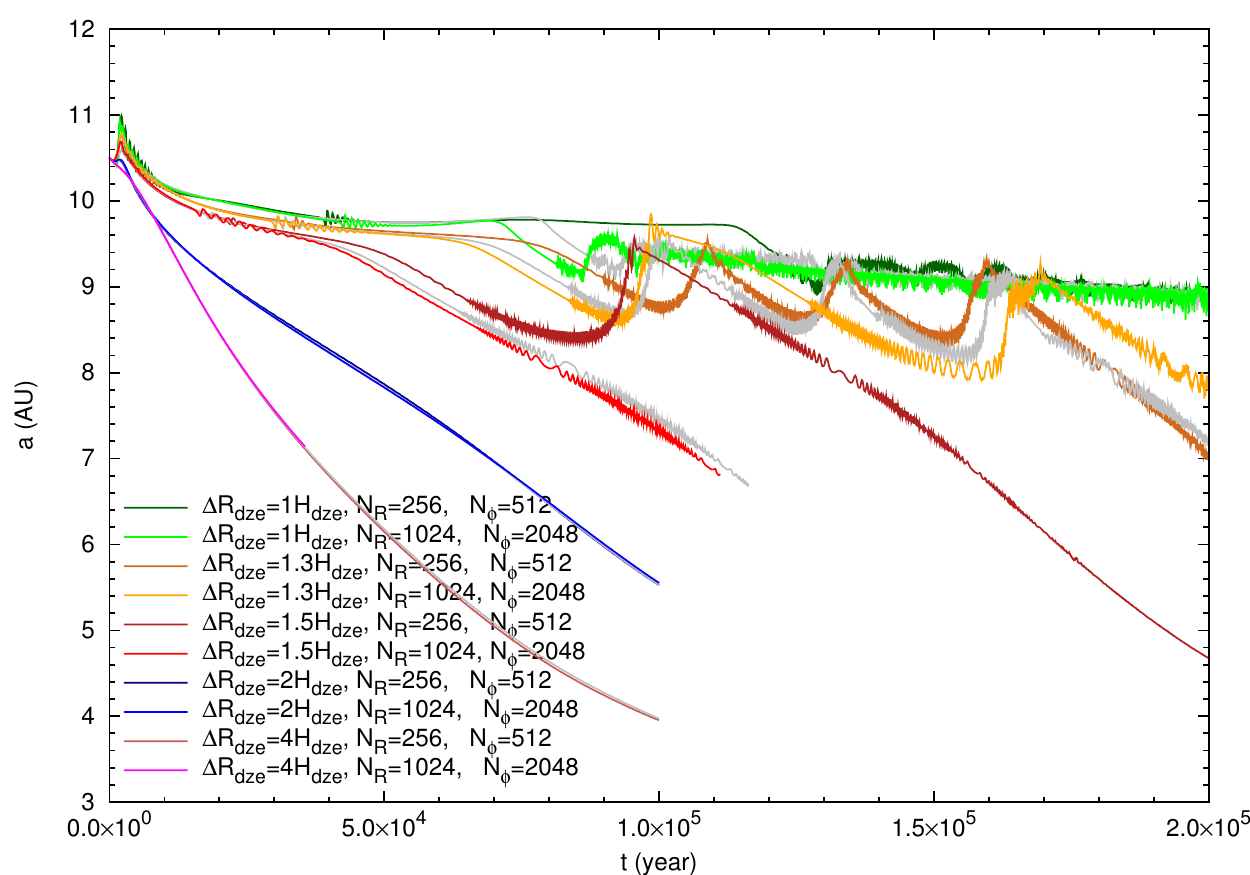}
	\caption{Migration history of a $10\,\mathrm{M_\oplus}$ planetary core in model $4H_\mathrm{dze}\leq\Delta R_\mathrm{dze}\leq1H_\mathrm{dze}$ with low ($N_R=256,\,N_\mathrm{\phi}=512$) and high ($N_R=1024,\,N_\mathrm{\phi}=2048$) numerical resolution. The standard medium resolution ($N_R=512,\,N_\mathrm{\phi}=1024$) results are plotted with grey curves.}
	\label{fig:mig-nr}
\end{figure}

Fig.\,\ref{fig:torque-numres} (upper panel) shows the radial torque density profile $(\mathrm{d}\Gamma(r)/\mathrm{d}m)/\Gamma_0'$ near the planet, where $\mathrm{d}\Gamma(r)$ is the torque exerted by a disc annulus located at $r$ with mass $\mathrm{d}m$ and $\Gamma_0'=\Gamma_0/\Sigma_0h^2$ is the normalization factor, with the three different numerical resolutions at the same time as in Fig.\,\ref{fig:numres_shock}. The torque density profile vanishes beyond $10R_\mathrm{H}$, meaning that the non-satisfactory capturing of shock waves at large radial distances ($R>5\textrm{--}10R_\mathrm{H}$) does not cause significant departure in the torque felt by the planet. However, the torque density for the smallest resolution simulation slightly differs from those in the standard and higher resolution simulations, shown in the lower panel of Fig.\,\ref{fig:torque-numres}.

The evolution of the normalized total disc torque for the three simulations (Fig.\,\ref{fig:t-numres}) also confirmed that the torques are perfectly matching for the two higher resolution models, while have a slightly larger magnitude for the low-resolution model. Thus, we conclude that simulations done with numerical resolution of $N_R=512$, $N_\phi=1024$ are in the numerically convergent domain in a sense that the torque felt by the planet converges.

In order to be convinced that the observed fast migration (for blunt viscosity reduction models) and temporary trapping (for sharp viscosity reduction models) of the $10\,\mathrm{M_\oplus}$ planetary core are a real physical phenomenon, we rerun simulations using different widths of the viscosity transition with small and high resolution. Fig.\,\ref{fig:mig-nr} shows our results. The migration history of the $10\,\mathrm{M_\oplus}$ planetary core is independent of the numerical resolution for blunt viscosity reduction ($\Delta R_\mathrm{dze}\geq2H_\mathrm{dze}$) models. However, for model $\Delta R_\mathrm{dze}=1.5H_\mathrm{dze}$, the low-resolution simulation shows failed recapturing of the planet, while the medium- and high-resolution simulations gives similar results. For even sharper viscosity reduction ($\Delta R_\mathrm{dze}\leq1.3H_\mathrm{dze}$), the ejection of the $10\,\mathrm{M_\oplus}$ planetary core occurs slightly earlier as the numerical resolution is increased. Nevertheless, the migration history of the medium- and high-resolution models is very similar. Thus, we conclude that with the standard numerical resolution of $N_R=512$ and $N_\mathrm{\Phi}=1024$, the simulations are in the numerically convergent domain. Therefore, for all models presented in this paper we use the standard (512$\times$1024) numerical resolution.

\section{Exclusion or inclusion of the planetary Hill sphere}
\label{apx:Hill}

According to \citet{Cridaetal2009b}, the circumplanetary disc plays an important role in planetary migration. If the disc self-gravity is neglected, three actions of the circumplanetary disc are missed. (1) The external perturbations (by star and the disc) make the circumplanetary disc non-axisymmetric, which exerts a direct torque on the planet. (2) The circumplanetary disc acts like a planet as it is bound to the planet, which enhances the perturbation caused solely by the orbiting planet. (3) The circumplanetary disc not feeling the gravity of the disc remains in a constant orbit; thus, the planet has to pull the circumplanetary disc, which eventually slows down its migration. \citet{Cridaetal2009b} conclude that removing the half of the material of the planetary Hill sphere in the computation of force felt by the planet from the disc could be appropriate to handle the effect of the circumplanetary disc. Although \citet{Cridaetal2009b} state that the inappropriate handle of a circumplanetary disc has a severe effect for planets with $q=M_\mathrm{p}/M_*>10^{-4}$ because the Roche lobe of low-mass planets disappears \citep{Massetetal2006a}, it is worth investigating the effect of the circumplanetary disc, since the planet migrates across such a regions, where the gas density abruptly increases, which results in filling up of the planetary Roche lobe, even in our $q=10^{-5}$ case.

Moreover, \citet{BaruteauMasset2008} found that the Type I migration rate of a low-mass planet slightly decreases (by a factor of 2) due to the shift of Lindblad resonance points, if the self-gravity of the disc is taken into account. We mention, however, that they investigated a disc with mass three times that in the minimum mass solar nebula model, which is three times higher than in our simulations. We also note that \citet{TerquemHeinemann2011} found that the atmosphere of a low mass planet ($5\,\mathrm{M_\oplus}$) extends all the way out to the Roche lobe; thus, the material present in the Roche lobe must be excluded from the torque calculation as it is bound to the planet.

\begin{figure}
	\centering
	\includegraphics[width=\columnwidth]{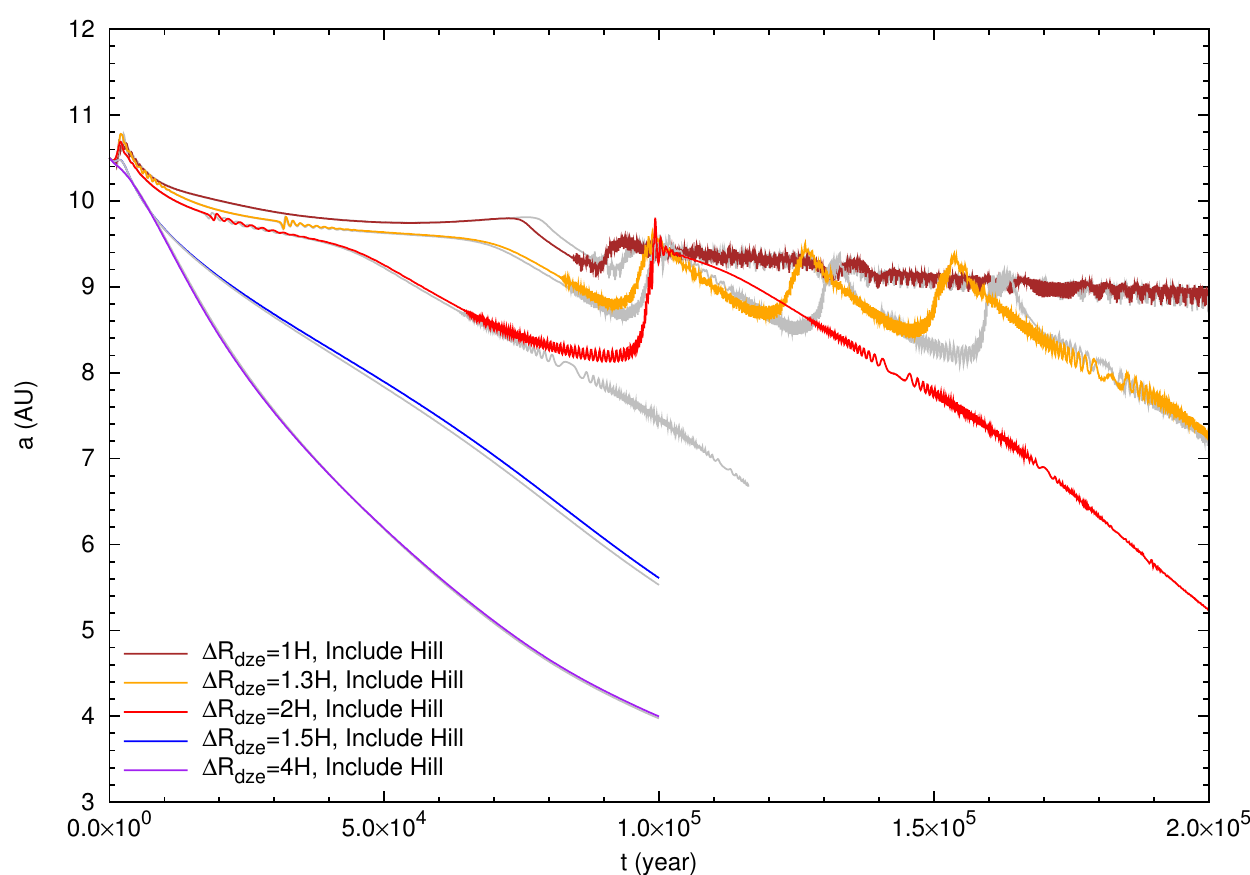}
	\caption{Migration history of a $10\,\mathrm{M_\oplus}$ planet for models $1H_\mathrm{dze}\leq\Delta R_\mathrm{dze}\leq4H_\mathrm{dze}$ without torque cut-off applied. The grey curves show the results with torque cut-off applied.}
	\label{fig:mig-Hill}
\end{figure}

In order to investigate the effect of the circumplanetary disc on the migration of the $10\,\mathrm{M_\oplus}$ planetary core, we ran simulations with and without torque cut-off. Applying the torque cut-off inside the Hill sphere, the torque of any zone (interior or exterior to the Roche lobe) is multiplied by $1-\exp(d/R_\mathrm{H})$ (where $d$ is the zone centre distance to the planet and $R_\mathrm{H}$ is the planetary Hill radius). Our results are presented in Fig.\,\ref{fig:mig-Hill}. 

There is no significant effect of the torque cut-off for models $\Delta R_\mathrm{dze}>1.5H_\mathrm{dze}$. However, the migration history is significantly different for model $\Delta R_\mathrm{dze}=1.5H_\mathrm{dze}$. The planetary core is retrapped for a short time at  $t\simeq100\times10^3$\,yr after the first ejection as the vortex reformed. For model $\Delta R_\mathrm{dze}=1.3H_\mathrm{dze}$, the ejection and retrapping occurs nearly identically with or without torque cut-off applied. For model $\Delta R_\mathrm{dze}=1H_\mathrm{dze}$, there is no significant difference in the planetary migration with or without torque cut-off applied. To summarize, the exclusion of the planetary Hill sphere in the torque calculation causes difference in the planetary migration only in certain thickness of the dead zone edge ($1.3H_\mathrm{dze}\leq\Delta R_\mathrm{dze}\leq1.5H_\mathrm{dze}$). Since we neglect the disc self-gravity in our 2D simulations, we presume that the exclusion of the planetary Hill sphere from the torque calculation gives the more physically reliable result.

\section{Smoothing of planetary potential}
\label{apx:Smooth}

\begin{figure}
	\centering
	\includegraphics[width=\columnwidth]{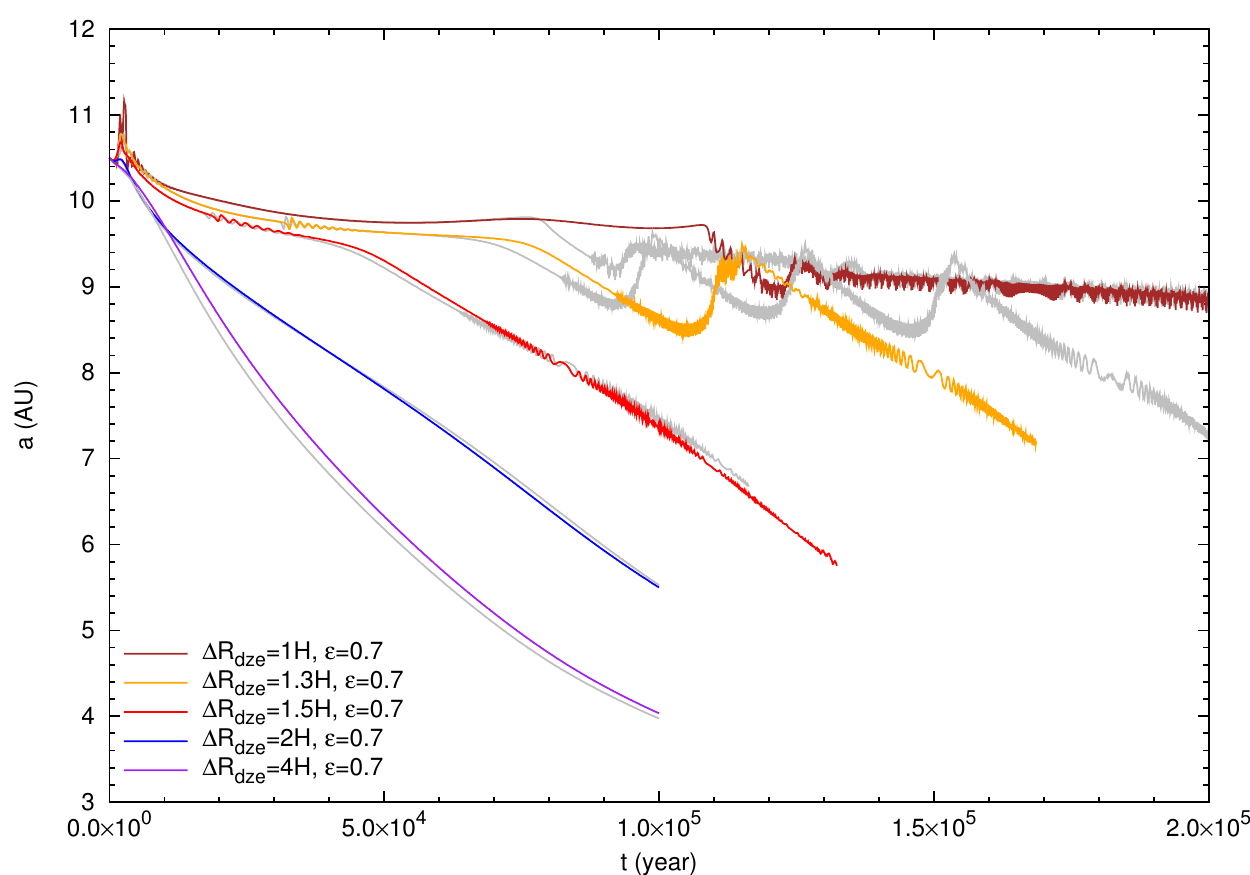}
	\caption{Migration history of a $10\,\mathrm{M_\oplus}$ planet in model $1H_\mathrm{dze}\leq\Delta R_\mathrm{dze}\leq4H_\mathrm{dze}$ with smoothing length $\epsilon=0.7$ and  torque cut-off applied. The grey curves show the results with $\epsilon=0.6$.}
	\label{fig:mig-thickness}
\end{figure}

We investigate the planet disc interactions by means of two-dimensional hydrodynamic simulations, in which the planet is treated as a point mass in the 2D simulations, placed in a grid cell. Thus, to regularize the planetary potential its smoothing is required. Additionally, the smoothing of planetary potential is also required to take into account the vertical thickness of the disc, which is otherwise neglected in 2D simulations. The gravitational potential due to the planet that acts on the disc is smoothed as
\begin{equation}
	\Phi_{\mathrm{p}(i,j)} = -\frac{GM_*M_\mathrm{p}}{\sqrt{(x_{i,j}-x_\mathrm{p})^2+(y_{i,j}-y_\mathrm{p})^2+(\epsilon H(a))^2}},
\end{equation}
where $x_{i,j}$ and $y_{i,j}$ are the Cartesian coordinates of the grid cell $i,j$, $x_\mathrm{p}$ and $y_\mathrm{p}$ are the Cartesian coordinates of the planet, while $\epsilon$ is the smoothing length, and $H(a)=ha$ is the disc thickness at the radial distance of the planet. 

\citet{Masset2002} found that the smoothing length being in the range $0.6\leq\epsilon\leq0.7$ will cause to underestimate and overestimate the differential Lindbald torque and the corotation torque, respectively. As a consequence, the total disc torque is underestimated because the differential Lindblad torque is negative, while the corotation torque is positive. However, \citet{Kleyetal2012} found that the choice of the smoothing parameter $\epsilon$ being in the range $0.6\leq\epsilon\leq0.7$ in 2D simulations gives reasonable good results compared to 3D simulations with a few Earth-mass planet embedded in the disc. Note that in the work of \citet{Kleyetal2012}, the disc has a constant surface mass density profile and spatially constant kinematic viscosity. Thus, it is worth investigating the effect of the choice of the smoothing length on the planetary migration.

In the simulations presented so far, the value of the smoothing length was set to $\epsilon=0.6$; thus, we performed additional 2D simulations using $\epsilon=0.7$ with torque cut-off applied. Our results are presented in Fig.\,\ref{fig:mig-thickness}. As one can see, the migration rate of the $10\,\mathrm{M_\oplus}$ planet is not affected significantly by the larger value of the smoothing length for models $\Delta R_\mathrm{dze}\leq1.5H_\mathrm{dze}$. However, the ejection and recapturing occurs only once for model $\Delta R_\mathrm{dze}=1.3H_\mathrm{dze}$ with larger smoothing length. Although the migration history is not altered by the larger smoothing length for model $\Delta R_\mathrm{dze}=1H_\mathrm{dze}$, the ejection occurs later by $\sim30\times10^3$\,yr. Thus, we conclude that the results are independent of the choice of the smoothing parameter being in the range $0.6\leq\epsilon\leq0.7$ for models $\Delta R_\mathrm{dze}\geq1.5H_\mathrm{dze}$; the observed temporary trapping and recapturing are robust phenomena for models $\Delta R_\mathrm{dze}\leq1.3H_\mathrm{dze}$.

\end{document}